\documentclass[11pt]{article}
\pdfoutput=1
\usepackage{jcapmod}

\usepackage[english]{babel}
\usepackage[per-mode=reciprocal, mode=math, group-digits=integer]{siunitx}[=v2]
\usepackage{slashed}
\usepackage[babel=true]{microtype}

\numberwithin{equation}{section}
\allowdisplaybreaks[1]

\renewcommand{\d}{\mathrm{d}}
\newcommand{\ee}{\mathrm{e}}
\newcommand{\ii}{\mathrm{i}}
\renewcommand{\k}{\vec{k}}
\newcommand{\x}{\vec{x}}
\newcommand{\fnl}{f_\mathrm{NL}}
\newcommand{\fnlalpha}{f_\mathrm{NL}^{\alpha, \slashed{\nu}}}
\newcommand{\fnlosc}{f_\mathrm{NL}^{\alpha, \nu}}
\newcommand{\Asin}{A_\mathrm{sin}}
\newcommand{\Acos}{A_\mathrm{cos}}
\newcommand{\kmax}{k_\mathrm{max}}
\newcommand{\kmin}{k_\mathrm{min}}
\renewcommand{\Re}{\mathrm{Re}}
\renewcommand{\Im}{\mathrm{Im}}

\DeclareSIUnit{\parsec}{pc}
\DeclareSIUnit{\Mpc}{\mega\parsec}
\DeclareSIUnit{\h}{\mathit{h}}
\DeclareSIUnit{\hPerMpc}{\h\per\Mpc}
\DeclareSIUnit{\MpcPerh}{\per\h\Mpc}

\DeclareRobustCommand{\SkipTocEntry}[4]{}

\begin{document}
	
\pagenumbering{roman}
\begin{titlepage}
	\baselineskip=15.5pt \thispagestyle{empty}
	
	\bigskip\
	
	\vspace{1cm}
	\begin{center}
		{\huge \sffamily \bfseries Extending the Cosmological Collider:}\\[8pt]
		{\LARGE \sffamily \bfseries New Scaling Regimes and Constraints from BOSS}
	\end{center}
	
	\vspace{0.2cm}
	\begin{center}
		{\large Daniel Green,$^{\bigstar}$ Jiashu Han\hskip1pt$^{\bigstar}$ and Benjamin Wallisch\hskip1pt$^{\spadesuit,\blacklozenge,\clubsuit}$}
	\end{center}
	
	\begin{center}
		\vskip8pt
		\textsl{$^\bigstar$ Department of Physics, University of California San Diego, La Jolla, CA 92093, USA}
		
		\vskip8pt
		\textsl{$^\spadesuit$ Oskar Klein Centre, Department of Physics, Stockholm University, 10691~Stockholm, SE}
		
		\vskip8pt
		\textsl{$^\blacklozenge$ Nordita, KTH Royal Institute of Technology and Stockholm University, 10691~Stockholm, SE}
		
		\vskip8pt
		\textsl{$^\clubsuit$ Texas Center for Cosmology and Astroparticle Physics, Weinberg Institute for Theoretical\\Physics, Department of Physics, The University of Texas at Austin, Austin, TX~78712, USA}
	\end{center}

	\vspace{1.2cm}
	\hrule \vspace{0.3cm}
	\noindent {\sffamily \bfseries Abstract}\\[0.1cm]
	Primordial non-Gaussianity generated by additional fields during inflation offers a compelling observational target. Heavy fields imprint characteristic oscillatory signals in non-Gaussian correlation functions of the inflaton, a process sometimes referred to as cosmological-collider physics. These distinct signatures are compelling windows into ultra-high-energy physics, but are often suppressed, making standard equilateral non-Gaussianity the most promising discovery channel in many scenarios. In this paper, we show that direct couplings between the inflaton and additional fields can lead to a wide variety of novel, observationally relevant signals which open new parameter regimes that simultaneously exhibit the characteristics of light and heavy fields. We identify these primordial signatures in the late-time observables of the large-scale structure of the Universe, where they most significantly modify the scale-dependent bias of the galaxy power spectrum to include an oscillatory modulation around a non-trivial power law. We explore the full range of parameters that phenomenologically arise in these models and study the sensitivity of current and future galaxy surveys, finding that this new class of primordial non-Gaussianity is particularly accessible in near-term surveys due to its oscillatory feature. Finally, we perform an analysis of existing data from the final release of the Baryon Oscillation Spectroscopic Survey~(BOSS~DR12). While we find no evidence for a signal, we demonstrate significant improvements in sensitivity over respective non-oscillatory scenarios and place the first constraints on this extended parameter space of oscillatory non-Gaussianity.
	\vskip10pt
	\hrule
	\vskip10pt
\end{titlepage}

\thispagestyle{empty}
\setcounter{page}{2}
\tableofcontents

\clearpage
\pagenumbering{arabic}
\setcounter{page}{1}
\section{Introduction}
\label{sec:introduction}

One of the central goals of cosmic surveys is to extract the physics of the early universe from the statistics of primordial fluctuations~\cite{Chang:2022lrw}. The expansion rate, particle content and interactions throughout cosmic history dictate the statistics of the primordial density perturbations, which are in turn encoded in cosmological maps~\cite{Green:2022bre}. These maps therefore contain information about the standard cosmological model, the nearly Gaussian fluctuations and their evolution after the hot big bang, and extensions of this model including additional primordial signatures such as isocurvature modes, features and non-Gaussianity~\cite{Achucarro:2022qrl}. A key challenge in cosmology therefore is to completely decode these maps in order to reveal all the physical laws that shaped our Universe.\medskip

Inflation provides a particularly compelling opportunity to translate maps of the Universe into insights about the particle content and interactions relevant at inflationary energy scales~\cite{Achucarro:2022qrl}. For additional fields weakly coupled to the inflaton, the statistics of non-Gaussian correlation functions encode the masses and spins of the excitations of these fields~\cite{Chen:2009we, Chen:2009zp, Baumann:2011nk, Assassi:2012zq, Noumi:2012vr, Arkani-Hamed:2015bza, Lee:2016vti, Chen:2016uwp, MoradinezhadDizgah:2017szk, Biagetti:2019bnp, Kumar:2019ebj, Bodas:2020yho, Lu:2021wxu, Chakraborty:2023qbp, Chakraborty:2023eoq, Goldstein:2024bky, Chakraborty:2025myb, Goldstein:2025eyj}. This probe of fundamental physics using maps of the Universe, often referred to as \textit{cosmological-collider physics}, is potentially sensitive to physics up to the scale of grand unified theories and beyond. Moreover, these processes introduce violations of the single-field consistency conditions~\cite{Maldacena:2002vr, Creminelli:2004yq} and are therefore distinguishable from effective self-interactions of the inflaton mediated by the massive~fields.

Despite the compelling theoretical motivation and broad space of signals, only a few observational analyses relevant to the cosmological collider have been performed with data from the distribution of galaxies~\cite{Green:2023uyz, Cabass:2024wob} and the cosmic microwave background~(CMB)~\cite{Sohn:2024xzd, Suman:2025vuf, Philcox:2025bbo, Suman:2025tpv}. Unfortunately, for conventional models like quasi-single-field inflation~\cite{Chen:2009zp}, the signals in the squeezed limit for massive fields are generally subdominant to the equilateral contribution to the bispectrum~\cite{Green:2023uyz}. As a result, there has been limited effort in extracting these exciting signatures from data. However, this conclusion applies only to models with approximate de~Sitter invariance. In the presence of stronger interactions between the additional fields and the inflaton, the soft limits are more complex~\cite{McAneny:2019epy, Green:2023ids} and the actual potential of current and future data is unknown. These signals of the \textit{extended cosmological collider} therefore strongly motivate that we revisit how fundamental physics is encoded in and extracted from data.\medskip

Isolating cosmological information about the history of the Universe in maps of the~CMB has been a remarkably successful enterprise~\cite{Planck:2018vyg, Planck:2018jri, Planck:2019kim, AtacamaCosmologyTelescope:2025blo, SPT-3G:2025bzu}. This is due to both the~CMB being an effectively linear system, which makes isolating primordial physics particularly straightforward, and the development of powerful statistical tools, which enable the data analysis. In other words, the fact that linear transfer functions directly relate primordial fluctuations to observed modes in the~CMB dramatically helps with the modeling of the signals. These insights into the~CMB are now standard practice~\cite{Babich:2004gb, Smith:2006ud} and have been applied to summary statistics for a wide range of primordial signals~\cite{Senatore:2009gt, Flauger:2009ab, Flauger:2010ja, Smith:2015uia, Philcox:2025wts}. These measurements have confirmed that fluctuations are predominantly adiabatic, nearly Gaussian and close to scale invariant, which has already ruled out broad classes of inflationary scenarios. In principle, there remain significant opportunities beyond correlation functions and searches for more complex signals directly in the maps have opened additional pathways to extracting primordial physics~\cite{Chang:2008gj, Feeney:2010dd, Munchmeyer:2019wlh, Kim:2023wuk, Philcox:2024jpd}. Ultimately, the two-dimensional CMB~sky provides access to roughly \num{e7}~independent modes and cosmic variance will therefore soon limit constraints from the primary~CMB.

Galaxy surveys can allow us to access vastly more information about the Universe than is available in the~CMB~\cite{Chang:2022lrw}. Since these surveys cover a similar range of scales, but in three dimensions, near-term surveys are expected to contain at least one to two orders of magnitude more modes~\cite{DESI:2022lza}. Unlike the~CMB, however, the large-scale structure~(LSS) of the Universe is generally nonlinear, and primordial information can additionally be mixed with late-time gravitational evolution and astrophysics~\cite{Green:2022hhj}. This not only moves information into higher-point correlation functions, but also complicates the modeling of the observed maps and inferred summary statistics. While considerable progress has been made through effective field theory approaches~\cite{Baumann:2010tm, Cabass:2022avo, Ivanov:2022mrd}, simulation-based forward modeling~\cite{Jasche:2012kq, Ramanah:2018eed, Villaescusa-Navarro:2021cni, Stopyra:2023yqm, Doeser:2023yzv, Bairagi:2025sux, McAlpine:2025uzh}, and their combination~\cite{Schmidt:2018bkr, Kokron:2021xgh, Pellejero-Ibanez:2022efv, LSSTDarkEnergyScience:2023qfp, Ivanov:2024xgb}, decoding these maps has proven to be quite challenging. This may explain why a much smaller space of possible primordial signals has been explored in LSS~analyses. Even looking forward, restricting to the quasi-linear modes for which the gravitational and astrophysical nonlinearities are under control often leaves the sensitivity of current and future galaxy surveys \mbox{comparable to that of the~CMB~today.}\medskip

One successful strategy for inferring primordial information from LSS~maps has been to focus on protected observables~\cite{Green:2022hhj}. Local primordial non-Gaussianity~(PNG) offers one such example since it leads to modifications to the power spectrum at large scales, due to the generation of scale-dependent bias~\cite{Dalal:2007cu, Meerburg:2019qqi, Achucarro:2022qrl}. The presence of an additional massless field beyond the inflaton is generically expected to give rise to local~PNG, making it a compelling observable. Moreover, the imprint of this signal in the form of a scale-dependent bias on large scales looks like an apparent violation of the equivalence principle during structure formation and cannot be produced locally with conventional gravity. While current bounds on the amplitude of local~PNG,~$\fnl^\mathrm{loc}$, which parametrizes the size of this signal,\footnote{We emphasize that any form of primordial non-Gaussianity will generate scale-dependent bias as we will also explicitly see throughout this paper. In this paragraph, we focus on its local type since the scale-dependent bias has the potential to provide the best measurement in this case~\cite{Meerburg:2019qqi, Achucarro:2022qrl}.} from galaxy surveys are weaker than those from the~CMB, SPHEREx~is expected to change this with a significant improvement~\cite{SPHEREx:2014bgr, Heinrich:2023qaa, Shiveshwarkar:2023afl}.

Primordial features provide a second compelling example since they imprint oscillatory patterns in the power~\cite{Flauger:2009ab} and higher-point spectra~\cite{Chen:2006xjb, Flauger:2010ja}: for example, sharp events during inflation or resonant interactions with heavy fields lead to a strong departure from scale invariance during inflation~(or its alternatives) producing distinct signals in the correlation functions~\cite{Slosar:2019gvt, Achucarro:2022qrl}. Late-time astrophysical and gravitational processes however cannot give rise to these high-frequency oscillations. This allows us to measure these features without careful modeling of structure formation, making them especially robust probes of the primordial universe~\cite{Beutler:2019ojk}~(cf.\ also~\cite{Vlah:2015zda, Vasudevan:2019ewf, Li:2021jvz, Chen:2020ckc, Ballardini:2024dto}). The Baryon Oscillation Spectroscopic Survey~(BOSS) has already provided the strongest constraints on such oscillatory features over a wide range of frequencies~\cite{Beutler:2019ojk, Ballardini:2022wzu, Mergulhao:2023ukp, Calderon:2025xod} and upcoming surveys will further extend these bounds to levels that CMB~observations will never be able to achieve~(cf.~e.g.~\cite{Sohn:2019rlq, Beutler:2019ojk, Euclid:2023shr}).\medskip

One major reason to be optimistic about the potential for discovering extra fields during inflation is that they can introduce a hybrid of these two robust signals~\cite{Chen:2009we, Chen:2009zp}. Specifically, if we introduce a nonzero interaction of the inflaton with a heavy scalar field with mass $m > 3H/2$, where~$H$ is the Hubble scale during inflation, one finds an oscillatory non-Gaussian signal. In the squeezed limit of the primordial bispectrum, it takes the form\hskip1pt\footnote{In a standard abuse of notation, $\lim_{k_1 \to 0}$ indicates that we take the squeezed limit $k_1 \ll k_2, k_3$ while keeping the leading term. The~(implicit) momentum conservation on the left-hand side of the equation then also forces $\k_2 = -\k_3$ so that~$k_1$ and~$k_2$ on the right-hand side are the wavenumbers of the short and long mode, respectively.\label{fn:limits}}
\begin{equation}
	\lim_{k_1 \to 0} \big\langle \zeta(\k_1) \zeta(\k_2) \zeta(\k_3) \big\rangle' = \lambda \fnl \left[ \ee^{\ii\phi} \left(\frac{k_1}{k_2}\right)^{\!\ii\nu} + \ee^{-\ii\phi} \left(\frac{k_1}{k_2}\right)^{\!-\ii\nu} \right] \!\left(\frac{k_1}{k_2}\right)^{\!3/2} P(k_1) P(k_2)\, ,	\label{eq:squeezed_bispectrum}
\end{equation}
where we employed the correlation function defined as $\langle \ldots \rangle = (2\pi)^3\hskip0.5pt\delta_D(\sum_i \k_i)\,\langle \ldots \rangle'$, with the Dirac delta function~$\delta_D$, $\phi$~is a calculable phase and $\nu = \sqrt{m^2/H^2 - 9/4}$. Since the non-Gaussian amplitude~$\fnl$ is set in the equilateral limit, we introduced~$\lambda$ to capture the relative amplitude in the squeezed configuration. Unfortunately, $\lambda$~is exponentially suppressed in simple models with heavy fields, $\lambda \propto \ee^{-\pi\nu}$~\cite{Noumi:2012vr}. Nevertheless, it is important that the signal in~\eqref{eq:squeezed_bispectrum} exhibits logarithmic oscillations in~$k_1/k_2$ with a frequency~$\nu$, which is set by the mass~$m$ of the field. These oscillations arise from the interference term between the wavefunction of the heavy field, which oscillates at a frequency set by its mass, and the massless mode of the inflaton~\cite{Arkani-Hamed:2015bza}.

\begin{figure}
	\centering
	\includegraphics{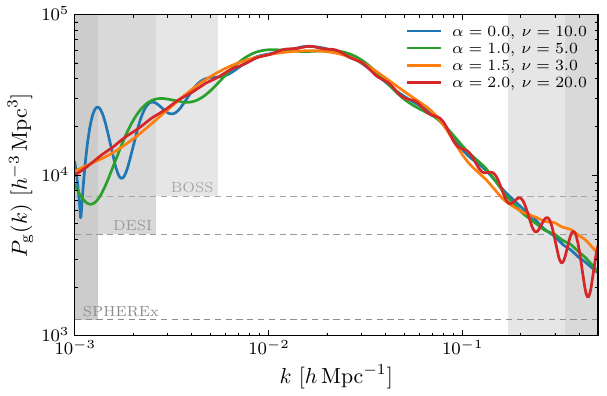}
	\caption{Illustration of the effect of oscillatory primordial non-Gaussianity on the linear galaxy power spectrum through the scale-dependent bias for different scaling dimensions~$\Delta = \alpha + \ii\nu$. In addition to the standard cosmological collider with $\alpha = 3/2$ as in~\eqref{eq:squeezed_bispectrum}, we also show examples for the wider range of values considered in this work. The employed non-Gaussian amplitudes are $\fnlosc = 3, 500, 2000\text{ and }2000$ for the shown cases~(with absorbed suppression factors; see especially Section~\ref{sec:forecasts} for its definition and details). We observe that the oscillatory imprint with logarithmic frequency~$\nu$ moves from large to small scales as we increase the scaling exponent~$\alpha$, which also sets the envelope of the oscillations. The horizontal dashed lines indicate the effective shot noise level for~BOSS, DESI and~SPHEREx after scaling them to the displayed redshift~$z = 0$ and linear bias~$b_1 = 1.6$. The gray-shaded regions on large scales indicate the wavenumbers below the minimum wavenumber of these surveys, $k < \kmin$, computed based on their entire spherical volume. The gray-shaded regions for large wavenumbers indicate the regimes where the scales exceed the nonlinear scale, $k > k_\mathrm{NL}$, for the maximum redshift of~BOSS and~DESI, respectively.}
	\label{fig:power_spectra}
\end{figure}

The squeezed bispectrum in~\eqref{eq:squeezed_bispectrum} is responsible for the robustness of the cosmological-collider signal: it induces a scale-dependent bias and imprints characteristic oscillations, as illustrated in Fig.~\ref{fig:power_spectra}.%
It therefore benefits from both forms of protection. These oscillations in addition to the smooth power-law bias are a particularly compelling theoretical and observational target for two reasons. First, the oscillation frequency~$\nu$ directly encodes the mass of the additional fields which could provide access to the particle spectrum during inflation. Second, this type of oscillations can generally neither arise without new degrees of freedom in the very early universe nor be produced by known astrophysical processes at late times. As a result, it is typically easier to robustly isolate this oscillatory signal in cosmological data.\medskip

Unfortunately, the amplitude of this compelling signal is doubly suppressed in most models. First, the squeezed limit is Boltzmann suppressed $\sim \ee^{-\pi\hskip1pt m/H}$ due to the high energy of the additional particles compared to the de~Sitter temperature. Second, the signal scales as~$(k_1/k_2)^{3/2}$. This means that it effectively vanishes as $k_1 \to 0$ which is a particularly significant observation since resolving an oscillation in the ratio of wavenumbers~$k_1/k_2$ requires taking this squeezed limit. Moreover, the suppression by at least~$(k_1/ k_2)^{3/2}$ in the squeezed limit is unavoidable for massive fields that effectively see a de~Sitter background, even in strongly interacting theories: the effective scaling dimension $\Delta = \alpha + \ii \nu$, with $\alpha \equiv \Re\Delta = 3/2$ and $\nu \equiv \Im\Delta$, in~\eqref{eq:squeezed_bispectrum} can receive corrections in interacting theories, but the optical theorem~(or, in other words, the positivity of probabilities) always requires $\alpha \geq 3/2$~\cite{Marolf:2010zp, Hogervorst:2021uvp, DiPietro:2021sjt, Lu:2021wxu, Loparco:2023rug, Green:2023ids, Cohen:2024anu, Pimentel:2025rds, Jiang:2025mlm}. The resulting suppression renders the oscillations encoded in~\eqref{eq:squeezed_bispectrum} difficult to observe in practice, unless an equilateral non-Gaussian signal has already been detected with high significance.

On the other hand, light fields with mass $m \leq 3H/2$ have a real-valued effective scaling dimension $0 \leq \Delta = \alpha \leq 3/2$, which alleviates both suppression factors. While this comes at the expense of generating non-oscillatory signals, these lighter fields have therefore been considered the more promising target for current and future LSS~observations~\cite{Gleyzes:2016tdh, Green:2023uyz}. At the same time, however, inflation does not occur in a pure de~Sitter space and allows for models that escape the discussed squeezed-limit suppression. In fact, due to the time evolution of the inflaton, the de~Sitter isometries are broken and the optical theorem no longer applies. In particular, explicit examples with Lorentz-violating or time-dependent mixing between fields have been shown to allow for complex scaling dimensions with $\Re\Delta = \alpha < 3/2$ and $\Im\Delta = \nu > 0$~\cite{McAneny:2019epy, Green:2023ids}. In such cases, high-frequency oscillations can appear in the squeezed limit without the mentioned exponential or power-law suppressions of the standard cosmological-collider scenario. In consequence, such models would be easier to search for in cosmological data, opening a new and observationally accessible window into the particle content and interactions of the inflationary epoch.\bigskip

In this paper, we explore the theoretical and observational implications of this broader class of models characterized by general complex-valued scaling dimension~$\Delta$. We first analyze the two-field model of~\cite{McAneny:2019epy} in Section~\ref{sec:models}, demonstrating that large frequencies~$\nu$ are achievable for $0 \leq \Re\Delta = \alpha \leq 3/2$. We in particular show that while the amplitude of the oscillatory signal in the galaxy power spectrum remains suppressed for vacuum fluctuations, this can be partially compensated by the enhanced observational sensitivity to these oscillations. We then deduce the observational signatures imprinted in galaxy power spectra, as exemplified in Fig.~\ref{fig:power_spectra} for a range of scaling exponents~$\alpha$ and frequencies~$\nu$, and assess the detectability of such signals through the scale-dependent bias in current and upcoming galaxy surveys in Section~\ref{sec:forecasts}. Using Fisher-matrix forecasts, we demonstrate and quantify the increased sensitivity of the galaxy power spectrum to these high-frequency oscillations in the squeezed limit. In Section~\ref{sec:analysis}, we subsequently place the first constraints on these bispectra for general complex values of~$\Delta$ from the BOSS~DR12 dataset. Finally, Section~\ref{sec:conclusions} contains our conclusions, and a set of appendices provide additional details on our forecasts~(Appendix~\ref{app:forecasting}) and our data analysis~(Appendix~\ref{app:analysis}).\medskip

Throughout the paper, we will use the following notation: A comoving vector~$\k$ is defined to have length $k \equiv |\k|$. We denote the $n$-point correlation function of operators~$\mathcal{O}_i$ without momentum-conserving Dirac delta function~$\delta_D$ as
\begin{equation}
	\big\langle \mathcal{O}_1(\k_1) \ldots \mathcal{O}_n(\k_n) \big\rangle = (2\pi)^3\hskip0.5pt \delta_D\!\left(\sum_{i=1}^n \k_i \right) \big\langle \mathcal{O}_1(\k_1) \ldots \mathcal{O}_n(\k_n) \big\rangle'\, .
\end{equation}
The inflationary adiabatic mode is~$\zeta$, the additional scalar fields are~$\sigma_i$ and the correlation function in a classical~$\sigma$ background is written as $\big\langle \zeta(\k_1) \ldots \zeta(\k_n) \big\rangle_\sigma$. This should not be confused with the squeezed limit of a correlator $\big\langle \mathcal{O}_1(\k_1) \mathcal{O}_2(\k_2) \big\rangle_{k_1 \ll k_2} \equiv \lim_{k_1 \ll k_2} \big\langle \mathcal{O}_1(\k_1) \mathcal{O}_2(\k_2) \big\rangle = \lim_{k_1 \to 0} \big\langle \mathcal{O}_1(\k_1) \mathcal{O}_2(\k_2) \big\rangle$~(see also footnote~\ref{fn:limits}). The scaling behavior of the additional fields~$\sigma_i$ coupled to the inflaton is parameterized by the complex dimension $\Delta = \alpha + \ii\nu$, with real-valued exponent~$\alpha$ and frequency~$\nu$, so that the power spectrum scales as~$k^{2\Delta - 3}$ in three spatial dimensions. For a single principal-series scalar~($m^2 > 9H^2/4$) in de~Sitter space, we therefore have $\alpha = 3/2$ and $\nu = \sqrt{m^2/H^2 - 9/4}$. We refer to the non-Gaussian amplitude for the oscillatory~($\nu \neq 0$) and non-oscillatory~($\nu = 0$) scenarios as~$\fnlosc$ and~$\fnlalpha$, respectively. Finally, we use the density perturbation~$\delta$, the smoothed density field~$\delta_M$ for~(halo) mass~$M$ and the galaxy overdensity~$\delta_\mathrm{g}$.

\section{Cosmological-Collider Models}
\label{sec:models}

The framework of the cosmological collider provides a natural origin for the distinct oscillatory and power-law signatures~\cite{Noumi:2012vr, Arkani-Hamed:2015bza, Lee:2016vti, Chen:2016uwp, MoradinezhadDizgah:2017szk, Kumar:2019ebj, Bodas:2020yho}. The simplest examples arise in quasi-single-field inflation~\cite{Chen:2009we, Chen:2009zp}, in which the inflaton interacts with an additional scalar of mass~$m$. For heavy fields, the signal is characterized by a frequency which is controlled only by this mass providing a direct connection between the inflationary particle spectrum and observable non-Gaussianity. However, the space of possible signals is large and we would consequently like to understand the viable parameter space in which to look for these primordial signals. This is why we will in this theoretical section characterize the range of frequencies~(\textsection\ref{sec:models_frequency}) and amplitudes~(\textsection\ref{sec:models_amplitude}) accessible in the two-field model of~\cite{McAneny:2019epy}, and evaluate the impact of the squeezed limit of the adiabatic modes~(\textsection\ref{sec:models_png}). While many aspects of our discussion were presented in~\cite{McAneny:2019epy}, our focus on the range of viable values of the scaling exponent~$\alpha$, oscillatory frequency~$\nu$ and PNG~amplitude~$\fnlosc$ is both new and relevant for observations.

\subsection{Frequency Range of Two-Field Models}
\label{sec:models_frequency}

The frequency of oscillatory signals in cosmological-collider scenarios is determined by the scaling behavior of the fields in the super-horizon limit~[cf.\ the frequency~$\nu$ in the exponents of~\eqref{eq:squeezed_bispectrum}]. We therefore start by analyzing this behavior in a simple model with two mixing fields which already exhibits a broad spectrum of scaling behaviors. To be specific, we consider two scalar fields~$\sigma_i$, $i = 1, 2$, which mix according to the following quadratic Lagrangian:
\begin{equation}
	\mathcal{L}_\sigma = -\frac{1}{2} \sum_{i=1}^2 \hskip-1pt\left(\partial_\mu \sigma_i\hskip1pt \partial^\mu \sigma_i - m_i^2 \sigma_i^2\right) + \rho\hskip-1pt \left(\dot\sigma_1 \sigma_2 - \dot\sigma_2 \sigma_1\right) ,	\label{eq:Lagrangian}
\end{equation}
with the masses~$m_i$ of the two fields, the mixing parameter~$\rho$ and dots denoting derivatives with respect to physical time~$t$. The last term is responsible for the mixing and explicitly breaks the de~Sitter isometries. This is particularly significant because the allowed space of scaling behaviors~(values of~$\Delta$) is highly constrained by symmetries and unitarity in a de~Sitter background. Once de~Sitter invariance is broken, much wider ranges of~$\Delta$ are~(at least in principle) allowed. As we will see, the relative sizes of the masses~$m_i$, the mixing~$\rho$ and the Hubble scale~$H$ will determine whether the scaling dimensions are real or complex and what the size of the imaginary part is, which sets the characteristic frequencies that we are interested in.\medskip

We start by computing the scaling dimension of the model~\eqref{eq:Lagrangian} as a function of~$m_i$, $\rho$ and~$H$. The equations of motion are given by
\begin{equation}
	\begin{aligned}
		\ddot\sigma_1 + 3 H \dot\sigma_1 + \left(\frac{k^2}{a^2} + m_1^2\right)\hskip-1pt \sigma_1 + \rho\hskip-1pt \left(2 \dot\sigma_2 + 3 H \sigma_2\right) &= 0\, ,	\\
		\ddot\sigma_2 + 3 H \dot\sigma_2 + \left(\frac{k^2}{a^2} + m_2^2\right)\hskip-1pt \sigma_2 - \rho\hskip-1pt \left(2 \dot\sigma_1 + 3 H \sigma_1\right) &= 0\, .
	\end{aligned}	\label{eq:eom}
\end{equation}
Since the Fourier modes are described by the time evolution in the super-horizon limit, $k \ll aH$, we make the ansatz $\vec\sigma = (\sigma_1, \sigma_2) = a(t)^{-\Delta} \vec v$. We consequently have to solve
\begin{equation}
	\begin{pmatrix}
		\Delta (\Delta - 3) H^2 + m_1^2	& H \rho\hskip1pt (-2\Delta + 3)				\\[2pt]
		H \rho\hskip1pt (2 \Delta - 3)	& \Delta (\Delta - 3) H^2 + m_2^2\hskip2pt{}
	\end{pmatrix} \hskip-1pt\vec{v} = 0\, .	\label{eq:eom-matrix}
\end{equation}
The four solutions to this equation are given by
\begin{equation}
	\Delta_\pm = \frac{3}{2} - \sqrt{\frac{9}{4} + a_\pm} \quad \text{and} \quad \bar\Delta_\pm = 3 - \Delta_\pm\, ,	\label{eq:delta_pm}
\end{equation}
where we introduced
\begin{equation}
	a_\pm \equiv -\frac{1}{2} \frac{m_1^2 + m_2^2 + 4\rho^2}{H^2} \pm 2 \sqrt{\frac{1}{2} \frac{\rho^2}{H^2} \left(\frac{m_1^2 + m_2^2 + 2\rho^2}{H^2} - \frac{9}{2}\right) + \frac{1}{16} \frac{(m^2_1 - m^2_2)^2}{H^4}}\, .	\label{eq:a_pm}
\end{equation}
We therefore see that $\Delta = \alpha + \ii\nu$ is in general complex. Now, we are interested in the ranges of~$\alpha$ and~$\nu$ that are possible within this model, in particular whether small~$\alpha$ and large~$\nu$ may occur concurrently.

To gain further insight by simplifying the expressions of these four solutions, it is helpful to define
\begin{equation}
	x \equiv \frac{1}{2} \frac{m_1^2 + m_2^2}{H^2}\, ,	\qquad	y \equiv \frac{1}{16} \frac{(m_1^2 - m_2^2)^2}{H^2 \rho^2}\, ,	\qquad	z \equiv \frac{\rho^2}{H^2}\, ,
\end{equation}
so that we have
\begin{equation}
	a_\pm = -x - 2z \pm 2\sqrt{z} \sqrt{x + y + z - \frac{9}{4}}\, .
\end{equation}
Given the assumption of real fields, the parameters~$x$, $y$ and~$z$ are also real, and $y, z > 0$. If we demand that both squared masses are positive, $m_i^2 > 0$, we also have $x > 0$. This limits~$a_\pm$ to be real-valued, except at small~$x$, $y$ and~$z$ such that $x + y + z < 9/4$. While this does not eliminate complex dimensions~$\Delta$ since~$a_\pm$ can be negative, it does not lead to novel phenomenology.

In contrast, for tachyonic masses, $m_i^2 < 0$,\footnote{The use of negative squared masses for the fields in the Lagrangian~\eqref{eq:Lagrangian} does not imply any non-attractor behavior in the solutions we will discuss. Tachyonic matter without mixing~($\rho = 0$) has long been known to have unique phenomenology~(see e.g.~\cite{McCulloch:2024hiz}), due in part to the fact that the background is unstable. We however emphasize the novel signals are stable and distinct from the usual differences \mbox{between attractor and non-attractor phenomenology.}} we find a much wider range of viable values of~$\Delta$. Considering the special case for which $y = 0$ and $x + z = 0$, i.e.\ $\rho^2 = -m_1^2 = -m_2^2 > 0$, we get $a_\pm = -z \pm 3\ii\sqrt{z}$, which implies the following four complex-valued solutions:
\begin{equation}
	\Delta \in \left\{\frac{3}{2} - \sqrt{\hskip-1pt\left(\frac{3}{2} \pm \ii\sqrt{z}\right)^{\!2}}, \frac{3}{2} + \sqrt{\hskip-1pt\left(\frac{3}{2} \pm \ii\sqrt{z}\right)^{\!2}}\right\} = \left\{\pm \ii \frac{\rho}{H}, 3 \pm \ii \frac{\rho}{H}\right\} .
\end{equation}
As a result, allowing $m_i^2< 0$ leads to solutions where $\Delta = 0 \pm \ii \nu$. In other words, we found at least one region of the parameter space~$\{m_i, \rho\}$ of the two-field model~\eqref{eq:Lagrangian} in which the frequency~$\nu$ may be large while the real-valued scaling exponent~$\alpha$ vanishes. Small deviations from our special case will generally preserve this behavior with a combination of potentially large frequency~$\nu$ and scaling exponent~$0 \leq \alpha \ll 3/2$. We reiterate that this does not violate the optical theorem since the mixing~$\rho$ between the two fields breaks the de~Sitter isometries. This is therefore a phenomenologically compelling region of viable parameter space {that has so far been under-explored.}

\subsection{The Amplitude of Primordial Fluctuations}
\label{sec:models_amplitude}

To establish whether this phenomenologically compelling region of parameter space is also observationally interesting, we need to compute the expected amplitude of the oscillatory signal, which depends on how efficiently the massive modes are excited during inflation. We therefore need to calculate the normalization of the fluctuations produced during inflation, focusing on the regime where $\alpha \approx 0$ and $\nu \gg 1$. While the scaling exponent~$\alpha$ and the frequency~$\nu$ are determined by the long-wavelength behavior, as we saw above, calculating the amplitude requires matching to the sub-horizon limit. This means that the full dynamics of the coupled system must be solved. In practice, solving for the mode functions analytically is difficult in this case, but a numerical solution is fortunately sufficient to determine the amplitude. In the following, we will set up the differential equations to be solved and discuss how we set their initial conditions using the WKB~approximation, numerically solve these equations and validate our procedure.\medskip

In order to calculate the amplitude of the oscillatory signal, we return to the quadratic equations of motion~\eqref{eq:eom}. To make solving these coupled differential equations numerically more straightforward, we define the dimensionless fields $\tilde\sigma_i \equiv k^{3/2} \sigma_i / H$, and employ the dimensionless quantities $u \equiv k \tau$, $\tilde{m}_i \equiv m_i/H$ and $\tilde\rho \equiv \rho/H$ so that
\begin{equation}
	\begin{aligned}
		\tilde\sigma_1'' - \frac{2}{u} \tilde\sigma_1' + \left(1 + \frac{\tilde{m}_1^2}{u^2}\right)\hskip-1pt \tilde\sigma_1 + \tilde\rho\hskip-1pt \left(-\frac{2}{u}\tilde\sigma_2' + \frac{3}{u^2} \tilde\sigma_2 \right)	&= 0\, ,	\\
		\tilde\sigma_2'' - \frac{2}{u} \tilde\sigma_2' + \left(1 + \frac{\tilde{m}_2^2}{u^2}\right)\hskip-1pt \tilde\sigma_2 - \tilde\rho\hskip-1pt \left(-\frac{2}{u}\tilde\sigma_1' + \frac{3}{u^2} \tilde\sigma_1 \right)	&= 0\, ,
	\end{aligned}	\label{eq:odes_mixing}
\end{equation}
where primes denote derivatives with respect to~$u$. Following~\cite{Assassi:2013gxa}, these equations can be solved numerically by starting at a time $u \ll -1$ when the mode functions are well inside the horizon. Since we took the Fourier transform, it is in principle straightforward to numerically compute the solutions in terms of~$u$ using any integrator for ordinary differential equations.

The non-trivial step in determining the amplitude of the vacuum fluctuations is imposing numerically that the modes are in the Bunch-Davies vacuum. We implement this as an initial condition for our numerical analysis as follows: in the early-time limit, $u = k \tau \to -\infty$~($a(t) \to 0$), the WKB~approximation in physical time~$t$ should correspond to a massless field with dispersion relation $\omega(t) \to k a^{-1}(t) + O(a^0)$. The WKB~calculation incorporates the first sub-leading correction to~\eqref{eq:odes_mixing}, i.e.\ terms of~$O(u^{-1})$ that are suppressed by~$(k \tau)^{-1}$, but neglects the second-order terms,~$O(u^{-2} = (k \tau)^{-2})$. Instead of the $\sigma_{1,2}$~basis used in~\eqref{eq:eom-matrix}, the WKB~solutions for the initial mode functions are more conveniently expressed in a different basis, $\sigma_{A,B}$, which are defined with respect to the original basis by
\begin{equation}
	\sigma_1 = \frac{\sigma_A^{(+)} - \sigma_A^{(-)}}{\sqrt{2}\hskip0.5pt\ii}\, ,	\qquad \sigma_2 = \frac{\sigma_B^{(+)} + \sigma_B^{(-)}}{\sqrt{2}}\, .
\end{equation}
The dimensionless WKB~mode functions $\tilde\sigma_{A,B} \equiv k^{3/2} \sigma_{A,B}/H$ are then given by
\begin{equation}
	\tilde\sigma_A^{(\pm)}(u) = \mp \frac{\ii u}{2} \hskip1pt\ee^{-\ii [u \pm \rho \log(-u)]}\, ,	\qquad	\tilde\sigma_B^{(\pm)}(u) = -\frac{u}{2} \hskip1pt\ee^{-\ii [u \pm \rho \log(-u)]}\, .	\label{eq:wkb-mf}
\end{equation}
A priori, one might have expected the modes to be defined in the $\sigma_{1,2}$~basis since the mixing term is subdominant when the WKB~approximation holds. However, if we include only the canonical kinetic term, we can rotate to any linear combination of~$\sigma_{1,2}$ and the action will be unchanged. The benefit of using the $\tilde\sigma_{A,B}$~basis is that it diagonalizes the~(sub-leading) mixing term~$\rho\log(-u)$ in the limit where the WKB~approximation is exact, $u \to -\infty$. Keeping the mixing term in the WKB~approximation allows us to match the numerical result to the analytic WKB~solution at smaller values of~$|u|$. This, in turn, improves the accuracy of our results as we do not have to integrate the equations for as long to reach the super-horizon regime, $|u| \ll 1$.\medskip

Having determined the analytic initial conditions that hold when $|u|^2 \gg m_i^2/H^2$,\footnote{Note that in this representation, the mass term is time dependent~(rather than the physical frequency) and is therefore relevant to the WKB~approximation.} we can now solve the differential equations numerically starting from the WKB~regime. The analytic form of the WKB~solution is sufficient to calculate the initial values and first derivatives of~$\sigma_{A,B}^{(\pm)}$. Since we have to additionally determine the correct normalizations, we match to the Bunch-Davies solution by isolating the positive frequency modes in the WKB~limit and canonically normalize the fields according to $q_i = a \sigma_i = a k^{-3/2} H \tilde\sigma_i$, $i = 1, 2$. We can then solve for~$\tilde\sigma_i$ numerically given these initial conditions. If needed, their~(equal-time) power spectra may subsequently be obtained according to
\begin{equation}
	H^{-2} k^3 P_{\sigma_1} = \big|\tilde\sigma_A^{(+)}\big|^2 + \big|\tilde\sigma_A^{(-)}\big|^2\, ,	\qquad H^{-2} k^3 P_{\sigma_2} = \big|\tilde\sigma_B^{(+)}\big|^2 + \big|\tilde\sigma_B^{(-)}\big|^2\, ,
\end{equation}
and we naturally have $P_{\sigma_1} = P_{\sigma_2}$ for the case of no mixing and equal masses, $\rho = 0$ and $m_1 = m_2$.\medskip

\begin{figure}
	\centering
	\includegraphics{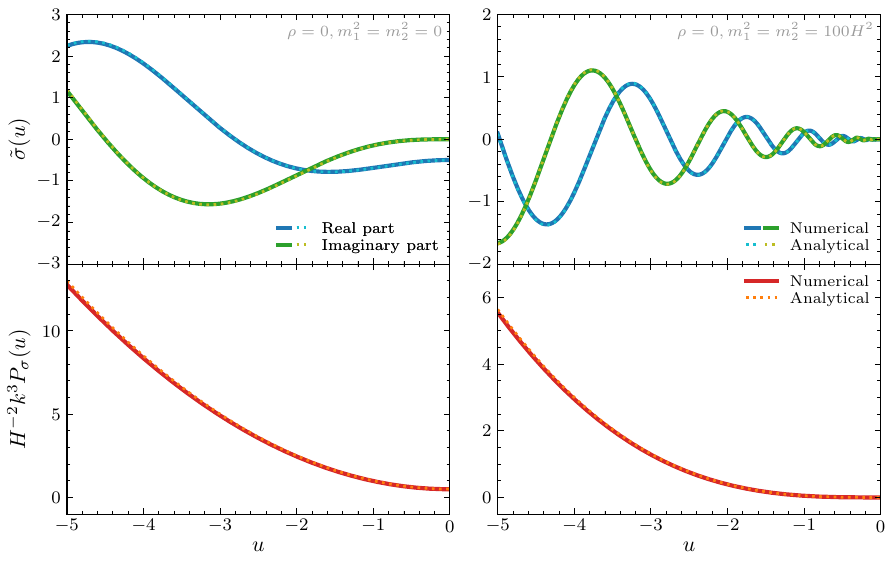}
	\caption{Comparison of the numerical and analytical solutions for the dimensionless mode function~$\tilde\sigma(u)$~(\textit{top}) and its respective power spectrum $H^{-2} k^3 P_{\sigma}(u)$~(\textit{bottom}) as a function of $u = k \tau$ for the case of no mixing, $\rho = 0$, and equal masses of the two fields, $m_1 = m_2$. We display both the real and imaginary parts of~$\tilde\sigma$, and note that $\tilde\sigma \equiv \tilde\sigma_A^{(+)} = -\tilde\sigma_A^{(-)} = \ii\tilde\sigma_B^{(\pm)}$~[cf.~\eqref{eq:wkb-mf} since $\sigma_1 = \sigma_2$]. The left and right panels illustrate the massless scenario and an example for the massive case with $m = 10H$, respectively. The numerical solution is computed using the Dormand-Prince method with initial step size $\Delta u = 0.01$ and initial point $u_0 = -5000$ to minimize the phase error, and exhibits very good agreement with the analytic result.}
	\label{fig:numerical_vs_analytic_mf}
\end{figure}

As a first step, we consider the special case of a principal-series field~$\sigma$ with no mixing, $m^2 > 9 H^2/4$ and $\rho = 0$, since it can serve as an important point of comparison between the analytic and numerical results. In this particular case, the mode functions have an analytic solution,
\begin{equation}
	\tilde\sigma(u) = \frac{\sqrt{\pi}}{2} \ee^{\frac{\pi \nu}{2}} (-u)^{3/2} H_{\ii\nu}^{(2)}(-u)\, ,
\end{equation}
where~$H_n^{(2)}(x)$ are Hankel functions of the second kind. In the limit $u \to 0$, we can expand the solution to find~(see e.g.~\cite{Cohen:2024anu})
\begin{equation}
	\tilde\sigma(u \to 0) \to \ii \frac{(-u)^{3/2}}{2\sqrt{\pi}} \left[\Gamma(-\ii\nu) \left(-\frac{\ii}{2} u\right)^{\ii\nu} + \Gamma(\ii\nu) \left(-\frac{\ii}{2} u\right)^{-\ii\nu}\right] ,
\end{equation}
with the gamma function~$\Gamma$. Since $\ii^{\pm\ii\nu} = \ee^{\mp \nu \pi/2}$ and the asymptotic behavior of the gamma function is $|\Gamma(\ii \nu)| \to \sqrt{2\pi} \nu^{-1/2} \ee^{-|\nu|\pi/2}$ as $\nu \to \infty$, the two terms of the mode function have an exponential hierarchy for large~$\nu$,
\begin{equation}
	\tilde\sigma(u \to 0) \to \ii \frac{(-u)^{3/2}}{\sqrt{2\nu}} \left[\ee^{-\pi\nu} \left(-\frac{1}{2} u\right)^{\ii\nu} + \left(-\frac{1}{2} u\right)^{-\ii\nu}\right] .	\label{eq:modes_fncs_long}
\end{equation}
As a result, the power spectrum of the original field~$\sigma$ in the super-horizon and large-frequency limits, $u \to 0$ and $\nu \gg 1$, is
\begin{equation}
	P_\sigma(k, \tau) \approx \frac{H^2 \tau^3}{2\nu} \left[1+ 2 \ee^{-\pi\nu} \cos\!\left(2\nu \log(k \tau)\right) + \ldots\right] .
\end{equation}
We see that the leading term is analytic in~$k$ and, consequently, is a pure contact term. The leading oscillatory~(non-analytic) behavior in~$k$ is therefore suppressed by $\ee^{-\pi\nu}$. This behavior highlights an important feature of the principal-series field which is that the wavefunctions are highly oscillatory as $u \to 0$ while the power spectrum is mostly analytic. This is because the oscillations in the power spectrum arise from the interference between the two oscillatory terms of the wavefunctions. We additionally note that the exponential suppression of the massive power spectrum makes it also useful as a test of our numerical implementation in a regime with known analytic behavior. In Fig.~\ref{fig:numerical_vs_analytic_mf},%
we show this comparison between the analytic and numerical results for both the mode function~$\tilde\sigma(u)$ and its power spectrum~$P_\sigma(u)$, both of which generally agree with one another within about two percent or (much)~less. This therefore serves as a validation of our numerical approach that employs the Dormand-Prince method to solve the ordinary differential equations~\eqref{eq:odes_mixing} for which we set the initial conditions at $u_0 = -5000$ and use an initial step size of $\Delta u = 0.01$.\medskip

Having established that the accuracy of our numerical methods for massive fields is sufficient, we will now re-introduce $\rho \neq 0$ to find the solutions in this novel parameter regime. As discussed in~\textsection\ref{sec:models_frequency}, the mode function in the $u \to 0$ limit is a linear combination of four power-law solutions, with their dimensions being two complex-conjugate pairs. In general, the dimensionless mode functions~$\tilde\sigma_A^{(\pm)}(u)$ and~$\tilde\sigma_B^{(\pm)}(u)$ can consequently be parameterized as
\begin{equation}
	\tilde\sigma_{A,B}^{(\pm)}(u) = a_{A,B}^{(\pm)} (-u)^{\Delta} + b_{A,B}^{(\pm)} (-u)^{\Delta^*} + c_{A,B}^{(\pm)} (-u)^{\bar\Delta} + d_{A,B}^{(\pm)} (-u)^{\bar\Delta^*}\, ,	\label{eq:sig_powers}
\end{equation}
where~$\Delta^*$ and~$\bar\Delta^*$ are the complex conjugates of~$\Delta$ and~$\bar\Delta$, respectively, which are defined in~\eqref{eq:delta_pm}, and the coefficients~$a_{A,B}^{(\pm)}, b_{A,B}^{(\pm)}, c_{A,B}^{(\pm)}$ and~$d_{A,B}^{(\pm)}$ are all complex. In the limit $u \to 0$, the two solutions with smaller~$\alpha$ become dominant, which allows us to drop two out of the four terms in~\eqref{eq:sig_powers}.\footnote{For $\alpha = 3/2$, some of the scaling dimensions become degenerate, i.e.\ $\Delta^* = \bar\Delta$, which requires a more careful treatment of the solutions and fitting all four coefficients in~\eqref{eq:sig_powers}.} Our expectation therefore is that our numerical solutions for $|u| \ll 1$ can be used to determine these coefficients.

\begin{figure}
	\centering
	\includegraphics{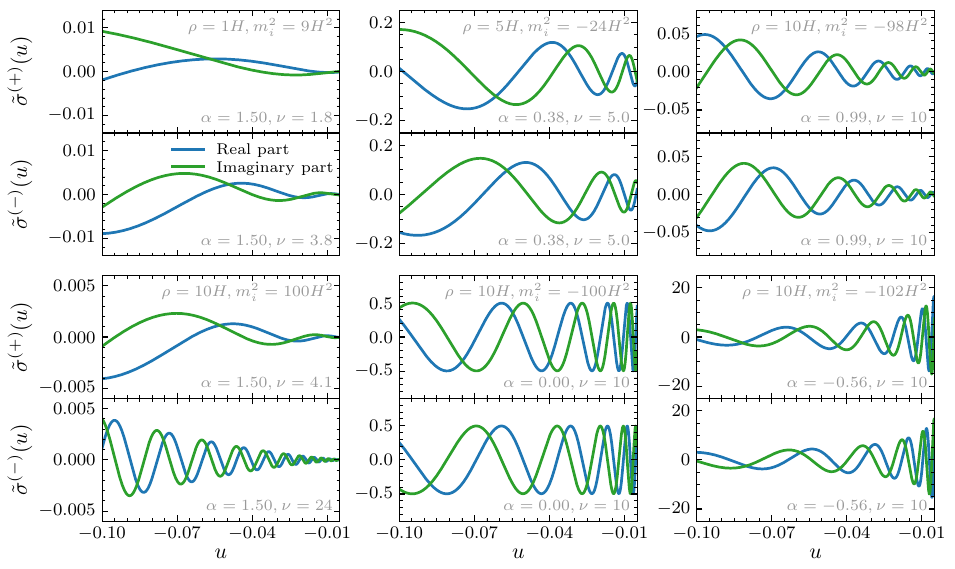}
	\caption{Numerical solutions for the mode functions~$\tilde\sigma^{(\pm)}(u) \equiv \pm\tilde\sigma_A^{(\pm)}(u) = \ii\tilde\sigma_B^{(\pm)}(u)$ for six different cases of non-zero mixing, $\rho \neq 0$. We show the real~(blue) and imaginary~(green) parts for a few different scenarios, taking the masses to be equal for simplicity. Fitting the parametrization~\eqref{eq:sig_powers} to these curves over the shown range of~$u$ leads to excellent agreement, with differences being at most at the sub-percent level. We explicitly highlight the dominant scaling solution as parameterized by $\Delta = \alpha + \ii\nu$ in each panel. We observe that the oscillatory mode functions are not only non-zero in all cases, but may in fact have an appreciably large amplitude, especially for tachyonic masses in the new regime of interest with small scaling exponent~$\alpha$ and large frequency~$\nu$.}
	\label{fig:mode-functions}
\end{figure}

We fit the power laws in~\eqref{eq:sig_powers} to the numerical mode functions and show several cases in Fig.~\ref{fig:mode-functions},%
including their dominant power laws. We find that the parametrization~\eqref{eq:sig_powers} provides excellent fits to the numerical solutions, with the relative differences being at most at the sub-percent level. The central take-away from these solutions consequently is that the mode functions corresponding to the Bunch-Davies vacuum populate the long-wavelength scaling solutions associated with~$\Delta$ and~$\Delta^*$ for $\alpha \to 0$ and $\nu > 1$. Concretely, our analytic calculations confirm the existence of these solutions in the super-horizon limit, but we need the numerical computations to confirm these solutions are realized with a sufficient amplitude to dominate the super-horizon evolution. We additionally see that we can generate positive and negative scaling exponents in this two-field model and in particular observe that the novel regime where $\alpha \approx 0$ and $\nu \gg 1$ is indeed generated from vacuum fluctuations with an amplitude of~$O(1)$. Figure~\ref{fig:mode-functions} also illustrates the sensitivity of~$\alpha$ to the precise values of~$\rho$ and~$m^2$. In this regard, there is nothing special about the case $\alpha = 0$, but we see instead that~$\alpha$ can take on a \mbox{wide range of positive and negative values for a narrow range of~$\rho$ and~$m_i^2$.}

While the power spectrum provides a useful benchmark for numerical results, the observational signatures will be driven by the unequal-time two-point functions. As a result, it is primarily the mode functions that directly control observational predictions. To avoid distractions, we therefore do not show the (equal-time) power spectra of the numerical mode functions in Fig.~\ref{fig:mode-functions}. Importantly, we have now explicitly seen that the mode functions can be reliably determined numerically and indeed exhibit the large oscillatory signals that we anticipated analytically. From here, we will see that the amplitude of the observational imprints are driven more by the coupling to the inflaton, rather than the intrinsic size of the statistics of the field~$\sigma$.

\subsection{Non-Gaussianity and the Squeezed Limit}
\label{sec:models_png}

Having established the existence of a phenomenologically viable parameter space for $\Delta = \alpha + \ii\nu$ with general~$\alpha$ and~$\nu$, particularly small~$\alpha$ and large~$\nu$, we want to know the amplitude of any observational signatures. In general, the signal of primordial non-Gaussianity in the scale-dependent bias of galaxies arises from the squeezed and collapsed limits of the primordial correlation functions of the adiabatic mode~$\zeta$. Since the correlations are induced entirely when the massive field is outside the horizon, these limits are fortunately particularly straightforward to theoretically understand and calculate which we do in the following.\medskip

Suppose we have a cubic interaction between our massive fields~$\sigma_i$ and the adiabatic inflaton mode~$\zeta$, such as given by the interaction Hamiltonian $\mathcal{H}_\mathrm{int} = \kappa M_\mathrm{pl}^2 \dot{H} \dot\zeta^2 (a \sigma_1 + b \sigma_2) \equiv \kappa M_\mathrm{pl}^2 \dot{H} \dot\zeta^2 \sigma$, with coupling constant~$\kappa$, Planck mass~$M_\mathrm{pl}$, and parameters~$a$ and~$b$ governing the relative coupling strengths of the massive fields to the inflaton. At a very basic level, this gives rise to the non-Gaussian signal because it means that the power spectrum of~$\zeta$ is modulated by the long-wavelength~$\sigma$. In this regime, the long-wavelength modes are effectively classical and we can therefore think about this as a purely statistical correlation of the form~\cite{Assassi:2012zq, Mirbabayi:2015hva}
\begin{equation}
	\left\langle\! \sigma(\k_1, \tau_2)\hskip1pt \big\langle \zeta(\k_2, \tau_1) \zeta(\k_3, \tau_2) \big\rangle'_\sigma \right\rangle'_{k_1 \ll k_2 \approx k_3} \approx \langle \sigma(k_1, \tau_1) \sigma(-\k_1, \tau_2)\rangle' \frac{\partial}{\partial\sigma} P_\zeta(k_2)\, ,
\end{equation}
where $\langle \zeta(\k_2) \zeta(\k_3) \rangle_\sigma$ is the two-point correlator in a classical $\sigma$~background. The derivative on the right-hand side of this expression is easily calculated by treating~$\sigma$ as a background field and the two-point correlator of~$\sigma$ is computed by the methods in~\textsection\ref{sec:models_amplitude}. We however have not specified~$\tau_1$ and~$\tau_2$ which play crucial roles in determining the signal when $\Delta \neq 0$.\medskip

In practice, it is the squeezed bispectrum of the adiabatic modes that is the most observationally relevant.\footnote{The collapsed limit of the trispectrum is the leading signal in some models~\cite{Chen:2009we, Baumann:2011nk, Baumann:2012bc}. The generalization to that case is straightforward, but we focus on the bispectrum to simplify the presentation.} The role of the field~$\sigma$ in this case is to enhance the correlations between adiabatic modes of different wavelengths. We can write the squeezed limit in the suggestive form
\begin{equation}
	\big\langle \zeta(\k_1) \zeta(\k_2) \zeta(\k_3) \big\rangle_{k_1 \ll k_2 \approx k_3} \approx \big\langle \langle \zeta(\k_1) \rangle_\sigma \langle \zeta(\k_2) \zeta(\k_3) \rangle_\sigma \big\rangle\, .			\label{eq:sqbi}
\end{equation}
In order to generate a large three-point function, we must introduce an additional interaction, a Hamiltonian mixing term $\mathcal{H}_\mathrm{mix} = \kappa' M_\mathrm{pl}^2 \dot{H} \dot\zeta \sigma$.\footnote{In principle, there is no reason that the linear combination of~$\sigma_i$ in the mixing term matches the combination in the cubic interaction. To simplify the notation, we assume here that they are both the same $\sigma = a\sigma_1+ b\sigma_2$ without impacting our results.} Note that we have not yet specified in this expression at which time these background fields are evaluated. The correct answer of course is that we integrate over time, but understanding the scaling requires identifying the times that dominate this expression.

Interactions of~$\zeta$ always involve derivatives and, therefore, exponentially decay in time~$t$ outside of the horizon. Similarly, due to the $\ii\epsilon$~prescription~(the interacting Bunch-Davies vacuum), fluctuations inside the horizon are also exponentially suppressed. As a result, the dominant contributions to the correlations of~$\zeta$ are associated with the fields at the time of horizon crossing of the corresponding mode. This implies that we should evaluate the squeezed limits at the times when~$\zeta(\k_i)$ is crossing the horizon, i.e.
\begin{equation}
	\big\langle \zeta(\k_1) \zeta(\k_2) \zeta(\k_3) \big\rangle_{k_1 \ll k_2, k_3}' \approx \Re\!\left[\hskip-1pt \big\langle\hskip-0.5pt \sigma(\k_1, (a H)_1) \sigma(-\k_1, (a H)_2) \hskip-0.5pt\big\rangle' \hskip1pt\frac{\partial}{\partial\sigma} P_\zeta(\k_2) \hskip1pt \frac{\partial}{\partial\sigma} \zeta(\k_1) \right] ,	\label{eq:squeezed_bi}
\end{equation}
where we defined the time of horizon crossing via $(a H)_i = k_i$ and~$\frac{\partial\zeta}{\partial\sigma}$ is the linear response of~$\zeta$ to~$\sigma$. The unequal-time correlators of~$\sigma$ can be evaluated using the leading power laws in the mode expansion,
\begin{equation}
	\sigma(\k, a H)\hskip1pt k^{3/2} = c_+ \left(\frac{k}{a H}\right)^{\!\alpha + \ii\nu} + c_- \left(\frac{k}{a H}\right)^{\!\alpha - \ii\nu}\, ,
\end{equation}
with the coefficients~$c_\pm$ having the potential of being~$O(1)$ as explicitly shown in the numerical calculations of~\textsection\ref{sec:models_amplitude}, so that
\begin{equation}
	\big\langle \sigma(\k_1, (a H)_1) \sigma(-\k_1, (a H)_2) \big\rangle' \propto \frac{k_1^{\alpha-3}}{k_2^\alpha} (c_+ + c_-) \left[c_+^* \left(\frac{k_1}{k_2}\right)^{\!-\ii\nu} + c_-^* \left(\frac{k_1}{k_2}\right)^{\!\ii\nu} \right] .	\label{eq:notequaltime}
\end{equation}
Since $(a H)_1 = k_1$, we see that the correlator scales as~$(k_1/k_2)^\Delta$. Moreover, there is no possibility of canceling the oscillatory behavior unlike in the calculation at equal times. In this sense, the squeezed limit of the primordial bispectrum necessarily takes the form
\begin{equation}
	\big\langle \zeta(\k_1) \zeta(\k_2) \zeta(\k_3) \big\rangle_{k_1 \ll k_2, k_3}' \propto \frac{1}{k_1^3 k_2^3}\hskip1pt \frac{\partial}{\partial\sigma} \zeta(\k_1)\hskip1pt \!\left(\frac{k_1}{k_2}\right)^{\!\alpha} \cos\!\left[\nu \log(k_1/k_2) + \varphi\right] ,
\end{equation}
where~$\varphi$ is a~(model-dependent) phase.

While the amplitude of the oscillations in the unequal-time correlation function of~$\sigma$ in~\eqref{eq:notequaltime} is unsuppressed, the same conclusion cannot be reached for the bispectrum. Specifically, the correlation between~$\zeta$ and~$\sigma$ is an integrated rather than instantaneous quantity and must be evaluated with care. This means that we should actually consider
\begin{equation}
	\frac{\partial\zeta}{\partial\sigma} \propto \Im\!\left\{k_1^3 \zeta(\k_1, \tau_0) \int\! \frac{\d\tau}{(-\tau)^4}\, \dot\zeta(k_1, \tau) \!\left[ c_+ (-k_1 \tau)^{\alpha + \ii\nu} + c_- (-k_1 \tau)^{\alpha - \ii\nu} \right] \!\right\} .
\end{equation}
This integral can be evaluated by a Wick rotation~\cite{Chen:2012ge} noting that
\begin{equation}
	\int_0^\infty\! \d\bar\tau\, \ee^{-\bar\tau} \bar\tau^{\alpha - 1 \pm \ii\nu} = \Gamma(\alpha - 1 \pm \ii\nu) \propto \nu^{\alpha - 3/2} \ee^{-\nu\pi/2}\, ,
\end{equation}
where $\bar\tau = -k_1 \tau$. Consequently, the mixing term alone is expected to exponentially suppress the amplitude at large frequencies, $\nu \gg 1$. We note that the precise form of the bispectrum is model dependent, including on both the definition of~$\sigma$ and the specific interactions. For an explicit calculation of the bispectrum and the trispectrum in one such model, we refer to~\cite{McAneny:2019epy}, which confirms the above scaling. As a result, we expect the scaling of our signal to take the model-independent form
\begin{equation}
	\big\langle \zeta(\k_1) \zeta(\k_2) \zeta(\k_3) \big\rangle_{k_1 \ll k_2, k_3}' \propto \frac{\nu^{\alpha - 3/2} \ee^{-\nu\pi/2}}{k_1^3 k_2^3} \left(\frac{k_1}{k_2}\right)^{\!\alpha} \cos\!\left[\nu \log(k_1/k_2) + \varphi\right] ,	\label{eq:squeezed_bispectrum_final}
\end{equation}
with the mentioned exponential suppression with frequency,~$\ee^{-\nu\pi/2}$.

The origin of this exponential suppression has a relatively simple physical origin. In order to measure the two-point function~\eqref{eq:notequaltime} of the field~$\sigma$, we would have to have a detector with perfect time resolution. This requires an injection of energy sufficient to produce physical particles of any mass~\cite{Green:2022fwg}, as we would expect from the uncertainty principle. When the signal is not localized in time, there is correspondingly no energy injection and the signal is Boltzmann suppressed~\cite{Arkani-Hamed:2015bza}. In this particular model, the same behavior holds, although the ``energy'' associated with these models is determined by~$\nu$, but does not have the familiar interpretation in terms of the \mbox{mass of a particle.}\medskip

We have focused on the behavior in the soft limit because it plays an outsized role in the description of the large-scale structure of the Universe. As we can see from~\eqref{eq:sqbi}, the long-distance behavior of these soft correlators is controlled by~$\sigma(k)$ outside the horizon and is largely independent of the details of how the short-wavelength $\zeta$~fluctuations behave. Moreover, these signals violate the single-field consistency relations~\cite{Maldacena:2002vr, Creminelli:2004yq, Creminelli:2012ed} and, therefore, cannot be mimicked by loop corrections to correlators, either in the early~\cite{Assassi:2012zq, Hinterbichler:2012nm, Goldberger:2013rsa, dePutter:2016moa, Hui:2018cag} or late universe~\cite{Creminelli:2013mca, Creminelli:2013poa, Horn:2014rta, Esposito:2019jkb}. This implies that the exponential suppression with~$\nu$, the power-law behavior with~$\alpha$ and the logarithmic oscillations with frequency~$\nu$ can generally be expected to be \mbox{present in late-time cosmological~observations.}\medskip

While this suppression is typical, it is specific to models with an approximate time-translation symmetry that is broken at the scale~$H$ by the expansion rate. The fields in our model effectively carry energy $E \approx H\nu \gg H$ in the high-frequency regime and are consequently Boltzmann suppressed in an approximate de~Sitter background. However, the inflaton background carries an effective energy of $\dot \phi^{1/2} \approx 60H$~\cite{Baumann:2011su} which is why direct couplings to the inflaton can overcome this suppression. Explicit examples have realized this idea~(see e.g.~\cite{Flauger:2010ja, Behbahani:2012be, Flauger:2016idt, Bodas:2020yho}) and, therefore, suggest our scaling behavior could arise without the exponential suppression.\footnote{Some classes of models with broken time translations produce additional oscillatory signals in the power spectrum~\cite{Behbahani:2011it}. There are physical reasons these corrections can be subdominant in other examples, including chemical-potential-like couplings~\cite{Behbahani:2012be}.}\medskip

For the rest of this paper and in line with this phenomenological approach, we will consider a range of scaling exponents of $\alpha \in [-1, 3]$. While the focus of this section was on models with $\alpha \in [0, 3/2]$, this broader range for~$\alpha$ allows for connections with a wider range of models~\cite{Jazayeri:2025vlv}. Concretely, for $\nu = 0$, some relevant values of~$\alpha$ include: the tachyonic collider~($\alpha = -1$)~\cite{McCulloch:2024hiz}, local~PNG~($\alpha = 0$), the conventional cosmological collider~($\alpha = 3/2$)~\cite{Noumi:2012vr}, equilateral and orthogonal~PNG~($\alpha = 2$), and inflationary loops or composite operators~($\alpha = 3$)~\cite{Green:2013rd}.

\section{Oscillations in the Galaxy Power Spectrum}
\label{sec:forecasts}

We have established a novel type of oscillatory behavior in the squeezed limit of inflationary correlators in concrete multi-field models. We therefore expect that these primordial signatures are imprinted on the late-time distribution of matter as characteristic modulations in the large-scale clustering of halos and galaxies. Structure formation should translate these violations of the single-field consistency relations into a distinct scale-dependent bias. In this section, we will explicitly determine the relationship between the primordial statistics of the adiabatic mode~$\zeta$, which we determined in the previous section, and the observable statistics of galaxies, which are accessible to cosmological surveys of the large-scale structure of the Universe. We first review how the primordial bispectrum in the squeezed limit affects the halo power spectrum by modifying its bias~(\textsection\ref{sec:bias}). We then translate this modification into the scale-dependent and oscillatory behavior of the galaxy power spectrum~(\textsection\ref{sec:spectrum}). Finally, we describe the Fisher forecasts and their results for the sensitivity of current and future surveys to \mbox{this class of oscillatory primordial non-Gaussianity~(\textsection\ref{sec:Fisher}).}\medskip

In the following, we primarily focus on the origin of the scale-dependent bias from primordial non-Gaussianity~\cite{Dalal:2007cu}. Due to the novel parameter regime found in the models in Section~\ref{sec:models}, the relationship between the primordial signal and the galaxy power spectrum does not follow directly from many conventional discussions which is why we discuss it in detail. In contrast, the nonlinear biases due to gravity and galaxy formation will be the same as those that appear in any number of analyses of LSS~data and have been thoroughly reviewed in~\cite{Desjacques:2016bnm}.

\subsection{Oscillations in the Scale-Dependent Bias}
\label{sec:bias}

The essential feature of the models studied in this paper is the complex scaling dimension~$\Delta$ which encodes the time evolution of the additional fields. Our goal is to infer this parameter from the observed distribution of galaxies. The map from initial conditions to the statistics of galaxies, i.e.~galaxy biasing, is a well-established subject with many comprehensive reviews~\cite{Bernardeau:2001qr, McDonald:2009dh, Desjacques:2016bnm}. Having said that, the complex scaling behavior of the primordial bispectrum~\eqref{eq:squeezed_bispectrum_final} introduces a novel challenge in describing the signals accurately. This section will review some of the well-known tools in modeling galaxies with the goal of making the derivation and assumptions of our signal transparent to all readers. To some degree, the results mirror the scenario in which~$\Delta$ is real~\cite{Baumann:2012bc, Gleyzes:2016tdh, Green:2023uyz}. As we have however already seen when calculating the squeezed limit of the primordial signal in~\textsection\ref{sec:models_png}, high-frequency oscillations demand more care when determining the normalization of the signal.\medskip

In order to quantify the impact of primordial non-Gaussianity on the statistics of galaxies, we need a model that relates initial conditions to galaxy abundances. The essential input is the idea that galaxy formation is effectively local on cosmological scales. Specifically, we often assume that the density contrast of galaxies,~$\delta_\mathrm{g}(\x)$, is determined by some unknown local function,
\begin{equation}
	\delta_\mathrm{g}(\x) = F_\mathrm{g}(\delta(\x), \nabla_i \nabla_j \Phi(\x), \nabla^2 \delta(\x), \ldots) \, .	\label{eq:bias_general}
\end{equation}
Here, we have allowed not only for the density perturbation~$\delta(\x)$ and its derivatives to appear in this expression, but also for the tidal tensor,~$\nabla_i \nabla_j \Phi(\x)$, with the Newtonian potential~$\Phi$, since it is allowed by the equivalence principle. In principle, Taylor expanding this expression tells us that any local operator allowed by the equivalence principle can appear in the relationship between~$\delta_\mathrm{g}$ and~$\delta$.

While this is a compelling physical argument, it presents a challenge when it comes to the initial conditions. The above expression is only local on scales larger than the scale of the halos,~$k_\mathrm{halo}$, or the nonlinear scale,~$k_\mathrm{NL}$. Since this is better expressed in terms of Fourier modes~$\k$, we are considering local operators
\begin{equation}
	\mathcal{O}(\k) \equiv \int \!\d^3x\, \ee^{\ii \k \cdot \x}\, \mathcal{O}(\x)\, ,
\end{equation}
with $k < k_\mathrm{NL}$. However, for non-Gaussian correlators, this constraint does not mean we are neglecting short-wavelength modes. For example, using~\eqref{eq:squeezed_bi}, we see that
\begin{equation}
	\big\langle \delta^2(\k) \delta(\k') \big\rangle' = \int \!\frac{\d^3k_s}{(2\pi)^3}\, \big\langle \delta(\k_s) \delta(\k - \k_s) \delta(\k') \big\rangle' \propto \big\langle \sigma(\k, (a H)_s)\hskip1pt \sigma(\k',a H) \big\rangle'\, .
\end{equation}
In this sense, in a theory with primordial non-Gaussianity, $\delta^2(\k)$~behaves at a practical level like a proxy for~$\sigma(\k)$ as much as it encodes the local density squared.

The unusual dependence on initial conditions is related to the fact that the long-distance description of the large-scale structure is local in space, but nonlocal in time. One consequence is that~$\sigma(\x, t_\star)$ effectively plays a role in the clustering of matter at a time $t \gg t_\star$. In addition, the biasing model in~\eqref{eq:bias_general} defines the halos in terms of these locations in the initial conditions~(Lagrangian coordinates), which are then mapped to biases in Eulerian coordinates through the motion of the underlying structures through gravitational evolution.\medskip

In order to provide a concrete model for the formation of halos~(and, therefore, the statistics of halos and galaxies), we will use the barrier-crossing formalism~\cite{Press:1973iz}. The key input is the smoothed density field,
\begin{equation}
	\delta_M(x) = \int\! \frac{\d^3 k}{(2\pi)^3}\, \ee^{-\ii \k \cdot \x}\, W_M(k)\, \delta(\k)\, ,	\qquad	W_M(k) = 3 \frac{\sin(k R_M) - k R_M \cos(k R_M)}{(k R_M)^3} \, ,
\end{equation}
where~$R_M$ is the Lagrangian radius corresponding to the halo mass~$M$ and $W_M$~is the Fourier transform of the associated top-hat window function. An important simplifying feature of using the smoothed density field is that it eliminates short modes from the description of the bias and moves the squeezed~PNG into the scale dependence of the bias.

The barrier-crossing model defines the number density of halos in terms of the linear density above a critical threshold~$\delta_c$. Following~\cite{Baumann:2012bc}, it is first useful to define the Edgeworth expansion of the probability distribution of the matter overdensity. To determine the power spectrum of galaxies, we will need the two-point probability distribution
\begin{equation}
	p(\nu, \nu^\prime) = \exp \!\left(\!\kappa_{1,1} \frac{\partial^2}{\partial\nu\hskip1pt \partial\nu^\prime} + \sum_{m + n \geq 3} \frac{(-1)^{m + n}}{m! n!} \kappa_{m, n} \frac{\partial^{m + n}}{\partial\nu^m\hskip1pt \partial\!\left(\nu^\prime\right)^n}\!\right) p_g(\nu)\hskip1pt p_g(\nu^\prime)\, ,
\end{equation}
where $\nu^{(\prime)} = \delta\!\left(\x^{(\prime)}\right)\! / \sigma_M$, $\sigma_M$~is the root-mean-square amplitude of~$\delta_M$ and $p_g(\nu)$~is the normal distribution. The coefficients of this expansion are the cumulants
\begin{equation}
	\kappa_{n, m}(r = |\x - \x'|, M, \bar{M}) = \frac{\big\langle [\delta_M^n](\x) [\delta_{\bar{M}}^m](\x') \big\rangle'}{\sigma_M^n \sigma_{\bar{M}}^m}\, ,
\end{equation}
where the notation~$[\mathcal{O}]$ means the operator~$\mathcal{O}$ minus all self-contractions~(or, equivalently, that we only evaluate the connected correlator). In cosmological applications, we frequently also need to determine the cross-correlation between the galaxies and the matter overdensity,
\begin{equation}
	p(\nu | \delta(\x')) = \exp \!\left[\sum_{n \geq 1} \frac{(-1)^{n}}{n!} \kappa_{\hat{1}, n} \frac{\partial^n}{\partial\nu^n}\right] p_g(\nu) \, ,
\end{equation}
with the cumulants relevant here being defined as
\begin{equation}
	\kappa_{\hat{n}, m}(r = |\x - \x'|, M) = \frac{\big\langle [\delta^n](\x) [\delta_M^m](\x') \big\rangle'}{\sigma^n \sigma_M^m} \, .
\end{equation}
This is, in effect, a definition of the linear bias. In order to therefore determine the scale-dependent bias, we need to calculate
\begin{equation}
	\kappa_{\hat{1}, n}(k, M) \equiv \int \!\d^3x\, \ee^{\ii \k \cdot \x}\, \kappa_{\hat{1}, n}(r, M) = \int \!\d^3x\, \ee^{\ii \k \cdot \x}\, \frac{\big\langle \delta^{(1)}(\x) [\delta_M^n](\x') \big\rangle'}{\hat\sigma \sigma_M^n}\, ,
\end{equation}
where $\delta^{(1)}$~is the linear matter overdensity and $\hat\sigma$~is its root-mean-square amplitude.

In addition, we now define $\delta^{(1)}(\k, z) = k^2 \mathcal{T}(k, z) \Phi(\k)$, with $\mathcal{T}(k, z) = \frac{2 T(k) D(z)}{3 H_0^2 \Omega_m}$, where $T(k)$~is the transfer function defined such that $T(k \to 0) \to 1$, $D(z)$~is the linear growth factor normalized as $D(z) = 1/(1 + z)$ during matter domination, $H_0$~is the Hubble constant and $\Omega_m$~is the matter density. Although measuring the matter-galaxy cross-correlation is not possible with~BOSS~(or other galaxy surveys) directly, the calculation of the cross-correlation is important for distinguishing the stochastic and deterministic scale-dependent biases.\medskip

We can determine the relevant deterministic bias parameters by taking the \mbox{$k \to 0$}~limit of~$\kappa_{\hat{1}, n}$. We will only need the first two cumulants,
\begin{align}
	\kappa_{\hat{1}, 1}(k, M) \stackrel{k \to 0}{\longrightarrow}	&\ \frac{P_\mathrm{mm}(k)}{\hat\sigma \sigma_M}\, ,		\\
	\kappa_{\hat{1}, 2}(k, M) \stackrel{k \to 0}{\longrightarrow}	&\ \hat\sigma^{-1} \sigma_M^{-2} \int_{\vec{q}_1} \int_{\vec{q}_2} k^6\hskip1pt W_M(q_1) W_M(q_2)\hskip1pt \mathcal{T}(q_1) \mathcal{T}(q_2) \mathcal{T}(k)\hskip1pt \big\langle [\Phi](\k) [\Phi](\vec{q}_1) [\Phi](\vec{q}_2) \big\rangle	\nonumber	\\
																			&= \frac{4\lambda \fnlosc}{\hat\sigma \sigma_M^2} \frac{P_\mathrm{mm}(k)}{k^2 \mathcal{T}(k)} (k R_M)^\Delta \int_{\vec{q}} W_M(q)^2 P_\mathrm{mm}(q) (q R_M)^{-\Delta}\, ,
\end{align}
where $\int_{\vec{q}} \equiv \int\!\frac{\d^3 q}{(2\pi)^3}$ and $P_\mathrm{mm}$~is the standard~(Gaussian) matter autocorrelation power spectrum. As we will explain in~\textsection\ref{sec:spectrum}, we will need the real part of this expression for complex-valued exponents~$\Delta$. We note that we reintroduced here the relative normalization of the squeezed limit in terms of the parameter~$\lambda$ which is defined in~\eqref{eq:squeezed_bispectrum}. To simplify the equations, we will hereafter denote the integral on the right-hand side as
\begin{equation}
	\sigma_M^2(\Delta) \equiv \int_{\vec{q}} W_M(q)^2 P_\mathrm{mm}(q) (q R_M)^{-\Delta} \, .	\label{eq:sigma2_M_integral}
\end{equation}
This integral isolates the bispectrum shapes that contribute to the long-distance scale-dependent bias and are more easily measured.

The leading contribution to the halo-matter power spectrum is then given by
\begin{align}
	P_\mathrm{mh}(k) \stackrel{k \to 0}{\longrightarrow}	&\left[b_\mathrm{G} + \frac{1}{2} \delta_c b_\mathrm{G}^L \frac{\kappa_{\hat{1}, 2}(k, M)}{\kappa_{\hat{1}, 1}(k, M) \sigma_M} + \ldots \right] P_\mathrm{mm}(k)	\nonumber	\\
																	&= \left[b_\mathrm{G} + 2 \lambda \fnlosc \delta_c (b_\mathrm{G} - 1) \frac{(k R_M)^\Delta}{k^2 \mathcal{T}(k)} \frac{\sigma^2_M(\Delta)}{\sigma^2_M} + \ldots \right] P_\mathrm{mm}(k)\, ,
\end{align}
where~$b_\mathrm{G}$ is the linear bias in the limit of purely Gaussian initial conditions and the additional terms denoted by~$\ldots$ are of~$O(\fnl^2)$, with the contributions of higher-order biases vanishing in the \mbox{$k \to 0$}~limit. For later convenience, we will define the second term in the bracket as the non-Gaussian bias
\begin{equation}
	b_\mathrm{NG}(k; M, \Delta) \equiv 2 \lambda \fnlosc \delta_c (b_\mathrm{G} - 1) \frac{(k R_M)^\Delta}{k^2 \mathcal{T}(k)} \frac{\sigma^2_M(\Delta)}{\sigma^2_M} \propto \frac{k^\Delta}{k^2 \mathcal{T}(k)}\, .	\label{eq:b_ng}
\end{equation}
The essential feature of this expression is that the scale-dependent term scales as $k^{\Delta - 2}$ at small wavenumbers~$k$ since $\mathcal{T}(k \to 0) \to \mathrm{const}$. In addition, the amplitude of the signal is controlled by~$\sigma_M^2(\Delta)$.

What we actually observe from galaxy surveys is the halo-halo correlation, which involves an additional contribution from the primordial trispectrum whose amplitude is~$O(\fnl^2)$ and should be sub-dominant compared to the term linear in~$\fnl$. For simplicity, we will approximate this contribution using~$b_\mathrm{NG}^2$, so that the halo-halo correlation in Fourier space is
\begin{equation}
	P_\mathrm{hh}(k) \stackrel{k \to 0}{\longrightarrow} \left[b_\mathrm{G} + b_\mathrm{NG}(k; M, \Delta) + \ldots \right]^2 P_\mathrm{mm}(k)\, .	\label{eq:phh}
\end{equation}
Here, we are working in the small $\fnl$~approximation and are ignoring the stochastic halo biases that are of~$O(\fnl^2)$. Some of these terms contribute to the stochastic bias, which is given a more rigorous treatment in~\cite{Baumann:2012bc}.\medskip

The key take-away from this result is that the shape and behavior of the scale-dependent bias are determined by the scaling dimension~$\Delta$. For instance, the scale-dependent bias has a power-law behavior in~$k$ when the field coupled to the inflaton is light so that $\Delta = \alpha$ is real. On the other hand, when the scaling dimension is complex, $\Delta = \alpha + \ii\nu$, we can rewrite $(k R_M)^\Delta = (k R_M)^\alpha \ee^{\ii \nu \log(k R_M)}$ which shows that the imaginary component manifests itself in the power spectrum as an oscillatory feature in~$\log k$. It is also useful to note that this oscillation arises from a scale-invariant initial condition due to the breaking of scale invariance associated with the size of the halo.

\subsection{Modeling the Galaxy Power Spectrum}
\label{sec:spectrum}

Having derived the contribution to the halo-halo power spectrum from a generic primordial non-Gaussian signal, we can now discuss how to parameterize the galaxy-galaxy power spectrum in order to extract this signal. Our goal is to measure the amplitude of the non-Gaussian signal in a model-independent way while accounting for uncertainties in galaxy biasing. We will therefore discuss the new non-Gaussian contributions in detail and mention the standard nonlinear biases at the end.\medskip

We parameterize the theoretical galaxy power spectrum by separating the scale-dependent bias contribution into a smooth power-law piece,~$S(k)$, and an oscillatory modulation,~$O(k)$, such that~\eqref{eq:phh} is written as\hskip1pt\footnote{Since we are only considering the galaxy auto-power spectrum and no cross-spectra, we are denoting it by~$P_\mathrm{g} \equiv P_\mathrm{gg}$ for simplicity.}
\begin{equation}
	P_\mathrm{g}(k) = \left[b_1 + 3 \fnlosc b_\phi |\Sigma_M^2(\Delta)| S(k; M, \Delta) O(k; M, \Delta) + \ldots\right]^2 P_\mathrm{mm}(k) \, ,	\label{eq:Pg}
\end{equation}
where $b_\phi = 2 \delta_c (b_1 - 1)$ following the universality relation,\footnote{A simple relation between the non-Gaussian bias~$b_\phi$ and the linear Gaussian bias~$b_\mathrm{G}$ can be derived in halo-bias theory~\cite{Desjacques:2016bnm}. For local~PNG, $b_\phi = 2 \delta_c (b_\mathrm{G} - p)$ can be obtained, where $\delta_c \approx 1.686$ is the critical overdensity for spherical collapse and $p = 1$ when assuming a universal halo mass function. While this relationship was originally derived for dark matter halos, it is usually also applied to galaxies by using their linear bias~$b_1$ instead of~$b_\mathrm{G}$.} $\Sigma_M^2(\Delta) \equiv \sigma_M^2(\Delta)/\sigma_M^2$ and we have taken $\lambda = 3$ to match the normalization for the equilateral template used especially in~\cite{Schmidt:2010gw, Gleyzes:2016tdh, Green:2023uyz} when $\Delta = 2$.\footnote{Even for equilateral non-Gaussianity, the normalization in the squeezed limit is model dependent and only has $\lambda = 3$ for the factorizable template. The normalization obeys a non-trivial soft theorem that depends on the specific interaction~\cite{Creminelli:2013cga} and is exponentially sensitive to the frequency~$\nu$ which we implicitly absorb into the PNG~amplitude~$\fnlosc$.} To make a distinction between the bias for halos and galaxies, we switched the notation from~$b_\mathrm{G}$ to the standard notation for the linear galaxy bias as~$b_1$. We note that the form and parameter values of the universality relation may change for different galaxy populations or types of primordial non-Gaussianity, for instance, and should therefore be used with care~(cf.~\cite{Desjacques:2008vf, Pillepich:2008ka, Grossi:2009an, Scoccimarro:2011pz, Baldauf:2015vio, Biagetti:2016ywx, Barreira:2020kvh, Barreira:2020ekm, Barreira:2021ukk, Coulton:2022rir, Barreira:2023rxn, Sullivan:2023qjr, Hadzhiyska:2025rez, Sharma:2025xss, Perez:2026mjt}). Having said that, we will generally apply and use this particular form of the relation throughout this work.\footnote{It is interesting to note in this context that when we marginalized over the universality relation with a broad Gaussian prior for non-oscillatory~PNG in~\cite{Green:2023uyz}, we found the same qualitative behavior with some degradation in constraining power.}

\begin{figure}
	\centering
	\includegraphics{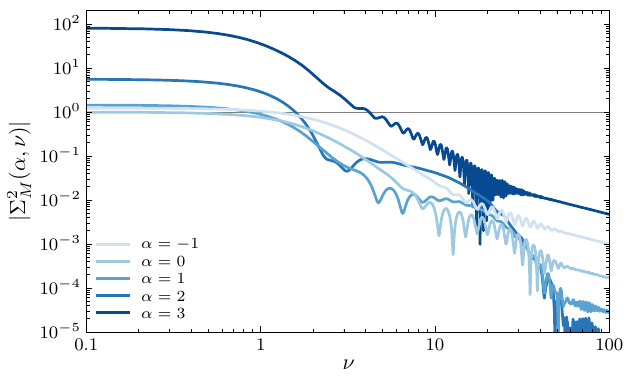}
	\caption{Suppression of the overall amplitude of the oscillatory PNG~signal in the galaxy power spectrum as captured by~$|\Sigma_M^2(\alpha, \nu)|$ in~\eqref{eq:Pg} for a halo mass of $M = \num{e13} M_\odot$. For all values of the scaling exponent~$\alpha$, $|\Sigma_M^2|$~sharply decreases at large frequencies~$\nu$ since we integrate over the highly oscillatory signal. It approaches a constant for small frequencies which is~(close to) unity for~(around) $\alpha = 0$, as expected, but is~$O(100)$ for $\alpha = 3$ since the integrand in~\eqref{eq:sigma2_M_integral} peaks around the scale of matter-radiation equality and is additionally enhanced by $(q R_M)^{-\alpha} \gg 1$ at that scale. We note that we absorb this suppression factor into the PNG~amplitude~$\fnlosc$ for the rest of our discussion, forecasts and data analysis. This is because we are mostly interested in investigating the sensitivity to a potential signal in this new PNG~class and this suppression factor can easily be reinstated to get specific model predictions and results.}
	\label{fig:sigma2}
\end{figure}

The numerical results for~$|\Sigma^2_M(\Delta)| = |\Sigma^2_M(\alpha, \nu)|$ are shown in Fig.~\ref{fig:sigma2}%
as a function of frequency~$\nu$ for the range of scaling exponents~$\alpha$ considered in this work. We note that~$\Sigma_M(\Delta)$ may be complex valued since~$\Delta$ may be complex. We therefore factored out the absolute value~$|\Sigma^2_M(\Delta)|$ in~\eqref{eq:Pg} so that the term~$S(k; M, \Delta) O(k; M, \Delta)$ carries only the phase information of~$\sigma^2_M(\Delta)$. The overall magnitude of~$|\Sigma^2_M(\Delta)|$, especially for small~$\nu$, depends on the range of wavenumbers in which the integrand in~\eqref{eq:sigma2_M_integral} peaks. For $\alpha = 0$, this is the case around \SI{0.5}{\hPerMpc} which shifts to larger~(smaller) scales for positive~(negative)~$\alpha$, reaching approximately the scale of matter-radiation equality for $\alpha = 3$. In the latter case, the peak is additionally enhanced since $(q R_M)^{-\alpha} \gg 1$ for those scales. For $\nu > O(1)$, we see a large suppression of the signal due to the integration over the rapidly oscillating correlator.

Comparing~\eqref{eq:Pg} with~\eqref{eq:b_ng} and~\eqref{eq:phh}, it remains to define the two terms~$S(k; M, \Delta)$ and~$O(k; M, \Delta)$. We take the former to be the power-law part of the scale-dependent bias, which only depends on the real part of~$\Delta$ and the halo mass~$M$,
\begin{equation}
	S(k; M, \Delta) = S(k; M, \alpha) = \frac{(k R_M)^{\alpha}}{k^2 \mathcal{T}(k)}\, .
\end{equation}
For complex~$\Delta$, i.e.\ if $\nu \neq 0$, the latter describes the oscillatory part of the scale-dependent bias, which only depends on the frequency~$\nu$~(and the halo mass~$M$),
\begin{equation}
	\fnlosc O(k; M, \Delta) = \fnlosc O(k; M, \nu) = \Acos \cos[\nu \log(k R_M)] + \Asin \sin[\nu \log(k R_M)]\, ,	\label{eq:A-cos-sin}
\end{equation}
i.e.\ we parameterize the oscillations in the amplitude parametrization with two amplitudes~$\Acos$ and~$\Asin$. For notational convenience, we defined these two parameters to effectively absorb the PNG~parameter~$\fnlosc$, but we will only use this as an intermediate step and state our results again in terms of~$\fnlosc$. In addition, it is implied that $\fnlosc O(k) = \fnlalpha$ for $\nu = 0$ and a change of the pivot scale in the logarithmic oscillations, which we took as~$R_M$, would only lead to a physically irrelevant shift in the phase.

We note that it would have been equally possible to describe the oscillations in terms of one amplitude and a phase, i.e.\ in the phase parametrization. While this can be a conceptually simpler way to understand the signal, it can complicate a parameter inference, whether for a forecast or a data analysis. This same challenge is present in searches for primordial features in the power spectrum and we refer to~\cite{Beutler:2019ojk} for a detailed discussion of the merits and shortcomings of these parametrizations. In short, the posterior distributions for~$\Acos$ and~$\Asin$ are both Gaussians that are mostly uncorrelated at large~$\nu$ and peak at zero for a vanishing signal. This contrasts with the phase parametrization, with a phase $\varphi \in [0, 2\pi)$, for which the phase will be uniformly distributed in the absence of a signal, while the amplitude will have a positive prior. In this regard, Fisher forecasts in the phase parametrization may not accurately capture these non-Gaussian priors and can therefore be a weaker approximation to the results of Markov chain Monte Carlo analyses like those used in Section~\ref{sec:analysis}.\medskip

Throughout the rest of this paper, we adopt two approximations for the theoretical non-Gaussian signal or, equivalently, the normalization of~$\fnlosc$. First, we know that large-scale gravitational nonlinearities exponentially damp any oscillatory signal in the late universe~\cite{Beutler:2019ojk}~(see also~\cite{Eisenstein:2006nj, Eisenstein:2006nk, Crocce:2007dt, Vlah:2015zda, Vasudevan:2019ewf, Li:2021jvz, Chen:2020ckc, Ballardini:2024dto}). While it is important to take this damping into account in general, we are able to neglect it in the following because we make conservative scale cuts given the rest of our modeling~(and, more generally, the original signal can be reconstructed on smaller scales to some level). This means that the damping effect remains small over the range of wavenumbers that we are taking into account~(see e.g.\ Fig.~19 of~\cite{Beutler:2019ojk} for an illustration). More generally, our work is of a more exploratory nature and with results shown on a logarithmic scale for the PNG~amplitude which means that corrections to the amplitude at the $O(10\%)$~level will not have a visible impact. Second, we saw in Fig.~\ref{fig:sigma2} that~$|\Sigma_M^2(\alpha, \nu)|$ is suppressing the non-Gaussian oscillations by many orders of magnitude for large frequencies~$\nu$. In our forecasts and data analysis, we are however mostly interested in investigating the sensitivity of the surveys to a potential signal. This is why we set this term to unity, $|\Sigma_M^2(\alpha, \nu)| \to 1$, or alternatively absorb it into a redefinition of the PNG~parameter, $\fnlosc \to \fnlosc/|\Sigma_M^2(\alpha, \nu)|$.\footnote{We therefore effectively remove an $O(1)$~number for small and vanishing~$\nu$~(and scaling exponents around zero), cf.~Fig.~\ref{fig:sigma2}, which we had implicitly done in~\cite{Green:2023uyz} following~\cite{Gleyzes:2016tdh}. For $\alpha \gtrsim 2$ and small~$\nu$, we however remove a larger factor. For the two-field model of Section~\ref{sec:models}, this is consequently the second exponential factor in~$\nu$ that we absorb into the PNG~amplitude. More generally speaking, our convention isolates the survey sensitivity to the phenomenological signature of the combined power law and logarithmic oscillations, and model-dependent suppression factors can be reinstated when mapping inferred constraints to specific inflationary scenarios.} As discussed in Section~\ref{sec:models}, the primordial signal itself may be exponentially suppressed or enhanced in a given inflationary model so that these factors can be reinstated when projecting the inferred constraints on the model space or specific scenarios.\medskip

\begin{figure}
	\centering
	\includegraphics{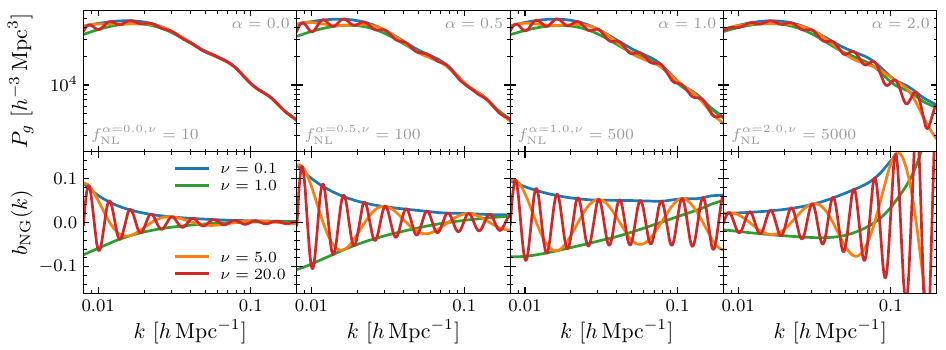}
	\caption{Illustration of the effect of the oscillatory scale-dependent bias on the linear BOSS~galaxy power spectrum at $z = 0.35$, with the phase chosen so that $\Asin = 0$. The top panels display the power spectra~$P_g(k)$ for different values of the scaling exponent~$\alpha$ and frequency~$\nu$, while the bottom panels show the non-Gaussian bias~$b_\mathrm{NG}(k)$. We observe how the non-Gaussian signature moves from the largest to the smallest scales as we increase the scaling exponent~$\alpha$ from the leftmost to the rightmost panels. (Note that the PNG~amplitude~$\fnlosc$ also increases between the panels and we have already absorbed the suppression factor~$|\Sigma_M^2|$ into its normalization as explained in the main text.) We can also see how the logarithmic oscillations with frequency~$\nu$ are modulated by the power-law contribution, which is governed by~$\alpha$, and that we would only measure part of an oscillation for small frequencies given the limited range of wavenumbers~(see main text).}
	\label{fig:shape-of-signal}
\end{figure}

The shape of the signal is shown in Fig.~\ref{fig:shape-of-signal}%
for a range of~(real-valued) scaling exponents~$\alpha$ and frequencies~$\nu$~(see also Fig.~\ref{fig:power_spectra}). We note that the oscillatory signal looks the same on all scales since these are logarithmic oscillations on a logarithmic wavenumber axis. We can already anticipate degeneracies with the smooth broadband galaxy power spectrum, particularly for small frequencies, because only a fraction of a period is covered in the finite range of~$k$. On the other hand, the oscillatory behavior of the signal, especially for high frequencies, will clearly break these degeneracies. At the same time, the signal-to-noise ratio is not uniform in~$k$ both for sample variance due to the finite cosmic volume and the finite number density of objects in a realistic cosmological survey.\footnote{The wavenumber dependence of the signal-to-noise ratio depends primarily on the shape of the scale-dependent bias and power spectrum relative to the sample variance and shot noise. This means that shot noise being independent of~$k$ does not simplify our understanding of the constraints.} As a result, the leading constraints may arise from limited ranges in~$k$ for which the oscillations have the largest signal-to-noise ratio. As can be anticipated from Figures~\ref{fig:power_spectra} and~\ref{fig:shape-of-signal}, this directly depends on the value of~$\alpha$ which will also be evident in our forecasts below.\medskip

Our treatment of the non-Gaussian signal as a complex power law is a simplification that is valid for small~$\fnlosc$. Nonlinear corrections will generate additional frequencies, $n \nu$ for integer~$n$, that are suppressed by~$\fnl^n$. This phenomenon arises for any nonlinear amplifier and is suppressed by the amplitude of the oscillatory signal~(and not just the nonlinear scale). Additionally, there are stochastic bias terms proportional to the collapsed trispectrum that can contain oscillating and power-law behavior that we have neglected. This should not qualitatively affect our results given the existing bounds on primordial non-Gaussianity and features which allow us to anticipate that the potential non-Gaussian signal at any given frequency is small and uncorrelated across frequencies. This is also why we can analyze one pair of~$(\alpha, \nu)$ at a time. Having said that, the constraints between frequencies will themselves be correlated since we cannot isolate each logarithmic frequency in our analysis ahead of time~\mbox{(unlike for a time-series analysis with white noise).}\medskip

In addition to the linear and scale-dependent biases, we also have to model nonlinear corrections to the galaxy spectrum. We will follow the same bias expansion as~\cite{Green:2023uyz} which includes operators up to third order in~$\delta_m$. At the field level, the bias parametrization is given by
\begin{equation}
	\begin{split}
		\delta_\mathrm{g}(\x) =\	&b_1 \delta_\mathrm{m}(\x) + b_{\nabla^2} R_M^2 \nabla^2\delta_\mathrm{m}(\x) + b_{\delta^2} \!\left(\delta_\mathrm{m}^2(\x) - \left\langle \delta_\mathrm{m}^2 \right\rangle\right) + b_{s^2} \!\left(s^2(\x) - \left\langle s^2\right\rangle\right)	\\
									&\!+ b_{\Pi\Pi^{[2]}} \left.\operatorname{Tr}\!\left[\Pi^{[1]} \Pi^{[2]}\right]\right|^{(3)}\!(\x) + \ldots\, ,
	\end{split}	\label{eq:delta_g}
\end{equation}
where $s_{ij} = \left(\nabla_i \nabla_j/\nabla^2 - \delta_{ij}/3\right)\! \delta_\mathrm{m}(\x)$ is the tidal tensor and the tensors~$\Pi^{[n]}_{ij}$ are defined in~\cite{Desjacques:2016bnm}, with the superscript~(3) indicating that the operator includes terms up to third order in perturbation theory. We additionally include redshift-space distortions and higher-derivative terms through a change to the linear bias,
\begin{equation}\label{eq:bz}
	b_1(z) \to b(k, \mu, z) = b_1(z) + f(z) \mu^2 + \sum_{n>0} b_{k^{2n}}(z) \hskip1pt(k R_M)^{2n}\, ,
\end{equation}
with the cosine~$\mu$ between the wavevector~$\k$ and the line of sight, and the growth rate $f(a) = \d\hskip-1pt\log D(a)/\d\hskip-1pt\log a$, where~$D(a)$ is the linear growth factor, while keeping~$b_\phi = b_\phi(z)$ as is.

While the details of our modeling, forecasting and eventual analysis are largely the same as in~\cite{Green:2023uyz}, it is noteworthy that in the regime $\nu > 1$, the detailed description of the nonlinear galaxy power spectrum has a limited impact on the sensitivity to our non-Gaussian signal. This is an expected property of oscillatory features~\cite{Eisenstein:2006nj, Beutler:2019ojk}. Since these details do not significantly affect our main results, we refer to~\textsection2.2 and Appendix~A.1 of~\cite{Green:2023uyz} for a complete description of the nonlinear galaxy power spectrum with Gaussian initial conditions, including the observational Alcock-Paczynski effect and the theoretical one-loop corrections.\medskip

The key take-away of this section is that we describe our non-Gaussian oscillatory signal in the galaxy power spectrum at small wavenumbers~$k$ by
\begin{equation}
	P_\mathrm{g}(k) = \left[b_1 + 3 b_\phi\, \frac{(k R_M)^{\alpha}}{k^2 \mathcal{T}(k)} \left( \Acos \cos[\nu \log(k R_M)] + \Asin \sin[\nu \log(k R_M)] \right)\right]^2 P_\mathrm{mm}(k) \, ,	\label{eq:Pgfinal}
\end{equation}
where the PNG~amplitude~$\fnlosc$ is determined from the oscillatory amplitudes~$\Acos$ and~$\Asin$ by~\eqref{eq:A-cos-sin}. This expression receives corrections from higher-order biasing given by~\eqref{eq:delta_g} and~\eqref{eq:bz} that are important and taken into account at larger wavenumbers and in redshift space.

\subsection{Forecasts for Current and Future Surveys}
\label{sec:Fisher}

Having translated the phenomenological two-field model of Section~\ref{sec:models} into a specific, but generalized observational signal as a combination of power-law and oscillatory~PNG in the galaxy power spectrum, we are now ready to examine the signal in realistic surveys. Our forecasting methodology generally follows the detailed description in~\cite{Green:2023uyz}, with some additional aspects explained in Appendix~\ref{app:forecasting}. In the following, we will summarize our approach based on the Fisher information matrix, provide our forecasted results for current and future surveys, and discuss their main implications and differences to the case of power-law~PNG~($\nu = 0$), \mbox{which we comprehensively studied in~\cite{Green:2023uyz}.}\medskip

We compute the standard Fisher matrix for four galaxy surveys: BOSS~\cite{BOSS:2016wmc},~which is completed, the Dark Energy Spectroscopic Survey~(DESI)~\cite{DESI:2016fyo}\footnote{Since DESI and Euclid~\cite{Euclid:2021icp} are expected to have very similar constraining power for our non-Gaussian signal~(cf.~\cite{Green:2023uyz}), we only include one of these current surveys explicitly as part of our reported results.} and the Spectro-Photometer for the History of the Universe, Epoch of Reionization and Ices Explorer~(SPHEREx)~\cite{SPHEREx:2014bgr}, which are currently taking data, and the billion-object survey, which we introduced in~\cite{Green:2023uyz} to investigate the potential capabilities of a more futuristic survey. The former are modeled to roughly replicate the specifications of the existing surveys once completed, while the latter assumes one billion spectroscopic objects split into two target samples with different linear galaxy biases and distributed with a constant number density over 10~bins in the redshift range~$0 \leq z \leq 5$ covering half the sky. This could therefore be thought of as a spectroscopic follow-up survey to the Vera Rubin Observatory's Legacy Survey of Space and Time~(LSST~\cite{LSSTScience:2009jmu}; cf.~the MUltiplexed Survey Telescope~\cite{Zhao:2024alp} and Spec-S5~\cite{DESI:2022lza} as steps in this direction) and/or an estimation for a \SI{21}{cm}~or other line intensity mapping survey over these redshifts~(see e.g.~\cite{Camera:2014bwa, Meerburg:2016zdz, Li:2017jnt, MoradinezhadDizgah:2018lac, CosmicVisions21cm:2018rfq, PUMA:2019jwd, Karagiannis:2020dpq, Castorina:2020zhz}). Our detailed specifications for these surveys, including the values of the linear bias and number density in each redshift bin, can be found in Appendix~A of~\cite{Green:2023uyz}.

We model the galaxy power spectrum including our non-Gaussian signatures as described~in~\textsection\ref{sec:spectrum}, cf.~\eqref{eq:Pgfinal}. Unless stated otherwise, we vary the cosmological parameters~$\{\omega_b, \omega_c, \theta_s, \log(\num{e10} A_\mathrm{s}), n_\mathrm{s}\}$ and oscillatory PNG~parameters~$\{\Acos, \Asin\}$ defined in~\eqref{eq:A-cos-sin} for a given survey, while the bias parameters~$\{b_1, b_{k^2}, b_{k^4}, b_{\delta^2}, b_{s^2}, b_{\Pi\Pi^{[2]}}\}$ are independently varied in each redshift bin.\footnote{The $\Lambda\mathrm{CDM}$~parameters are the physical baryon density~$\omega_b$, the physical cold dark matter density~$\omega_c$, the angular size~$\theta_s$ of the sound horizon at recombination, the~(logarithm of the) primordial scalar amplitude~$A_\mathrm{s}$ and the scalar spectral index~$n_\mathrm{s}$. The bias parameters that we include are the linear bias~$b_1$, the gradient biases up to quartic order~$b_{k^2}$ and $b_{k^4}$, the tidal bias~$b_{s^2}$ and the evolution bias~$b_{\Pi\Pi^{[2]}}$ defined in~\eqref{eq:delta_g} or explicitly in terms of the theoretical galaxy power spectrum in~(2.9) of~\cite{Green:2023uyz}.} The fiducial cosmology is the same $\Lambda\mathrm{CDM}$~cosmology with Gaussian initial conditions as in~\cite{Green:2023uyz},\footnote{The fiducial values of the linear biases are provided in Appendix~A of~\cite{Green:2023uyz} for all surveys together with their other specifications. The fiducial values of the nonlinear bias parameters are set to zero except for~BOSS, for which we assumed fiducial values for~$b_{s^2}$ and~$b_{\Pi\Pi^{[2]}}$ based on the Lagrangian local-in-matter-density biasing model~\cite{Desjacques:2016bnm} and for~$b_{\delta^2}$ based on the halo-simulation fit of~\cite{Lazeyras:2015lgp} to align with previous and current analyses. As expected from~\cite{Green:2023uyz}, this different choice in the fiducial values for~BOSS has only minimal effects on the forecasted constraints.} i.e.\ $\Acos = \Asin = 0$, and we also impose a marginalized Planck-like CMB~Fisher matrix as a $\Lambda\mathrm{CDM}$~prior.\footnote{For~BOSS, we impose the same Gaussian priors on the biases~$b_{\delta^2}$, $b_{s^2}$ and~$b_{\Pi\Pi^{[2]}}$ as~\cite{Philcox:2021kcw}. For all other surveys, we take the priors to be uniform for all biases. Similar to the findings in~\cite{Green:2023uyz}, imposing priors on the loop biases results in $O(1)$~improvements in the forecasted constraints for small~$\nu$, but leads to only minor changes of around a few percent for $\nu \gtrsim 10$.} We set the minimum wavenumber~$\kmin$ of a given redshift bin by its volume assuming a spherical geometry. The maximum wavenumber~$\kmax$ is taken to be the minimum of the nonlinear scale~$k_\mathrm{NL} = \frac{\pi}{2 R_\mathrm{NL}}$, with the radius~$R_\mathrm{NL}$ at which the variance of linear fluctuations is halved at a given~$z$, and the wavenumber~$k_\mathrm{halo} \approx \SI{0.19}{\hPerMpc}$ associated with the Lagrangian scale $R_M \approx \SI{2.66}{\MpcPerh}$ corresponding to the minimum halo mass of the surveyed population which we generally assume to be~$\num{e13} M_\odot$. Our choices for both~$\kmin$ and~$\kmax$ are conservative. In this work, we implement two extensions of our forecasting methodology compared to~\cite{Green:2023uyz} which are both necessitated by the oscillations on top of the previously studied smooth PNG~signal, which we discuss in the following and refer to Appendix~\ref{app:forecasting} for further details: using bandpowers and reporting 95\%~upper limits on the PNG~amplitude.

Surveys can only observe a finite cosmic volume which implies that we can only access a limited number of discrete wavenumbers~$k_i$. For sufficiently small spacing~$\Delta k$ between these wavenumbers, we can approximate a smooth enough (theoretical)~power spectrum~$P(k)$, such as the $\Lambda\mathrm{CDM}$~power spectrum with Gaussian or power-law non-Gaussian initial conditions, in a given band~$[k_i - \Delta/2, k_i + \Delta/2]$ by taking the value at its midpoint,~$P(k_i)$. On the other hand, the logarithmic oscillations~\eqref{eq:A-cos-sin} lead to significant variations within bands depending on the frequency~$\nu$ and wavenumber~$k_i$. This is why we compute the finite-size bandpowers~$P_i$ as linear averages of the galaxy power spectrum within these bands taking $\Delta k = \kmin$. We refer to Appendix~\ref{app:forecasting} for additional details, and to Fig.~4 of~\cite{Beutler:2019ojk} for an illustration of the effect on the observed signal and the reason why we do not additionally need to take the survey window function into account.

As a result of our forecasts and analyses, we would like to report constraints on the overall non-Gaussian signal. Given its oscillatory nature and our use of the amplitude parametrization with the two amplitudes~$\Acos$ and~$\Asin$ instead of the phase parametrization with an amplitude and a phase for good reasons~[see our discussion below~\eqref{eq:A-cos-sin}], this requires more care than in the case of simple power-law~PNG. Motivated by the phase parametrization, we define the non-Gaussian amplitude of our signal~\eqref{eq:Pg} as
\begin{equation}
	A\fnlosc \equiv \sqrt{\Acos^2 + \Asin^2}\, ,	\label{eq:png-amplitude}
\end{equation}
which is positive semi-definite. Phenomenologically, there is no preferred phase and the values of the amplitudes are unknown. We therefore do not assume any prior knowledge of~$\Acos$ or~$\Asin$ and the forecasted constraints on~$A\fnlosc$ should be defined so that they are independent of the fiducial values for~$(\Acos, \Asin)$ or, equivalently, the fiducial phase. First, as noted, we take the fiducial values for both amplitudes to be zero, i.e.\ the fiducial cosmology has Gaussian initial conditions. Second, following~\cite{Beutler:2019ojk}, we describe the sensitivity of the surveys in terms of the 95\%~exclusion limit of~$A\fnlosc$ which is a convenient estimate of the constraining power that however does not quite articulate the statistical power at small frequencies due to the strong degeneracy between the two amplitudes~$\Acos$ and~$\Asin$. Given the definition~\eqref{eq:png-amplitude}, this means in practice that we compute the radius of the circle in the marginalized $(\Acos, \Asin)$~parameter plane that contains 95\%~of the forecasted probability. In practice, we perform this calculation by Monte Carlo integration of the two-dimensional Gaussian posterior distribution for~$(\Acos, \Asin)$ around their origin given their marginalized Fisher matrix.\footnote{While the inherent scatter of the inferred mean values around the truth in an observational measurement was also taken into account in the forecasts of~\cite{Beutler:2019ojk} by additionally sampling the fiducial cosmology, we keep the fiducial values fixed to the origin in this work.} This way of computing the survey sensitivity to our oscillatory PNG~signal is both robust to changes of the fiducial values and implicitly marginalizes over the phase information~(see Appendix~\ref{app:forecasting} for more details). We additionally note that the sensitivity metric to compare to for the non-oscillatory power-law non-Gaussianity of~\cite{Green:2023uyz} with $\nu = 0$ is the 95\%~limit on the absolute value of its amplitude,~$|\fnlalpha|$ given the positive semi-definite nature~of~\eqref{eq:png-amplitude}.\medskip

\begin{figure}
	\centering
	\includegraphics{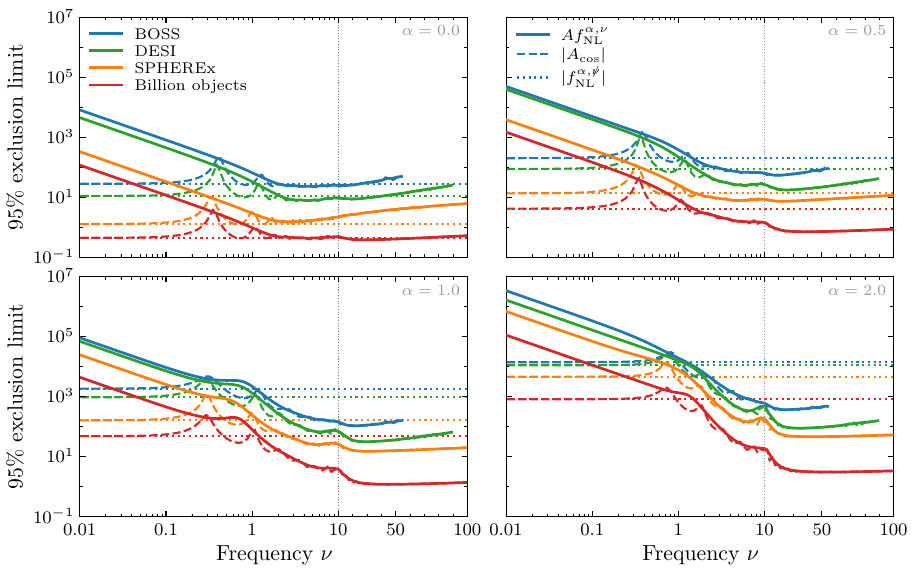}\vspace{-5pt}
	\caption{Forecasted constraints on the amplitude of primordial non-Gaussianity in terms of its 95\%~exclusion limit as a function of the oscillation frequency~$\nu$ for four values of the scaling exponent~$\alpha$. We note the logarithmic-linear scaling of the frequency axis as separated by the vertical dotted line. We show the Fisher-based limits on the overall oscillatory amplitude~$A\fnlosc$~(solid), the oscillatory cosine amplitude~$|\Acos|$~with fixed $\Asin = 0$~(dashed) and non-oscillatory power-law amplitude~$|\fnlalpha|$~(dotted) for the current and future surveys~BOSS~(blue), DESI~(green; Euclid has a very similar forecasted sensitivity), SPHEREx~(orange) and the billion-object survey~(red) described in the main text. The fiducial cosmology has Gaussian initial conditions, and we marginalized over the galaxy biases and $\Lambda\mathrm{CDM}$~parameters, with priors on the latter from a Planck-like CMB~experiment. For small frequencies~$\nu$, the constraints on the overall oscillatory amplitude~$A\fnlosc$ are worse than those for the power-law case with $\nu = 0$ denoted by~$|\fnlalpha|$ due to the implicit marginalization over the phase exhibiting the phase degeneracy, while we recover the $\nu = 0$ result when the phase is fixed to zero by setting $\Asin = 0$. On the other hand, the forecasted limits are much more similar for different values of~$\alpha$ at large~$\nu$ since the main constraining power comes from the imprinted oscillations. The frequency range available to the experiments is limited by the probed volume since it sets the minimum wavenumber and bandwidths which we conservatively estimate for each redshift bin~(see Fig.~\ref{fig:bandpower} in Appendix~\ref{app:forecasting} for a visualization). We also note that we compare the PNG~amplitude~$A\fnlosc$ to the absolute values of~$\Acos$ and~$\fnlalpha$ since~$A\fnlosc$ is a positive semi-definite quantity.\vspace{-4pt}}
	\label{fig:sigma-fnl-experiments}
\end{figure}

\begin{figure}
	\centering
	\includegraphics{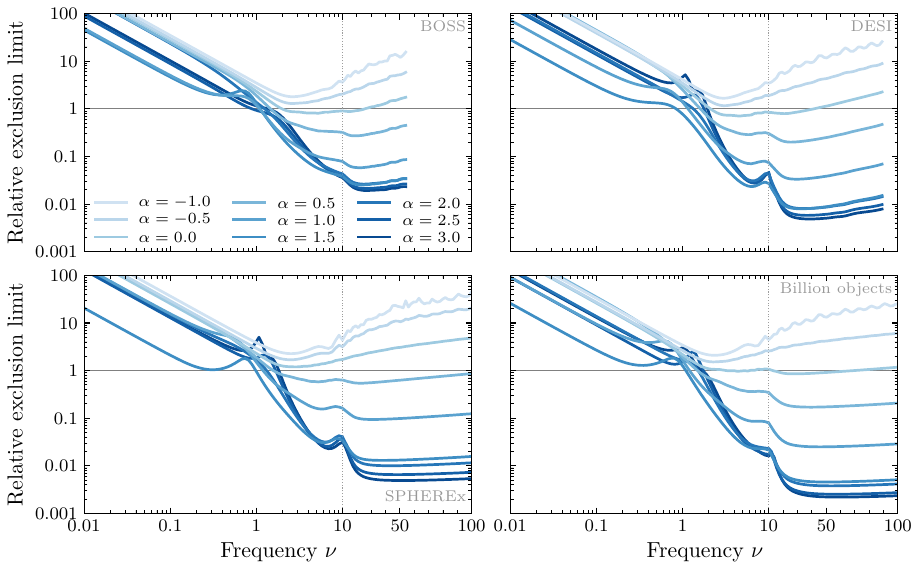}
	\caption{Ratio of the 95\%~exclusion limits for the oscillatory PNG~amplitude~$A\fnlosc$ with $\nu \neq 0$ and the~(absolute value of the) power-law PNG~amplitude~$|\fnlalpha|$ with $\nu = 0$ as a function of frequency~$\nu$ for a range of scaling exponents~$\alpha$. We note the logarithmic-linear scaling of the frequency axis as separated by the vertical dotted line. The four panels show the relative improvement~(or degradation) of the oscillatory signature for the four current and future surveys~BOSS, DESI, SPHEREx and the billion-object survey. While the forecasted limits are degraded by the inclusion of the oscillations for small~$\nu$ and $\alpha \lesssim 0$ due to the phase degeneracy and the signal peaking at the lowest wavenumbers, respectively, we observe a significant improvement in particular for large frequencies and scaling exponents since the oscillatory signal breaks the degeneracies with the broadband galaxy power spectrum.}
	\label{fig:sigma-fnl-experiments_relative}
\end{figure}

We present the forecasted 95\%~exclusion limits on the amplitude~$A\fnlosc$ of our oscillatory PNG~signal of~\eqref{eq:Pgfinal} in Fig.~\ref{fig:sigma-fnl-experiments}.%
For all displayed experiments~(BOSS, DESI, SPHEREx and the billion-object survey), we can observe a few general trends and regimes as a function of the scaling exponent~$\alpha$ and frequency~$\nu$. First, we can generally see that the constraints at small and intermediate frequencies change by several orders of magnitude with increasing~$\alpha$, but remain within about one order of magnitude at large frequencies indicating the impact of the oscillations. More generally speaking, we expect that oscillatory features in a power spectrum make the primordial signal easier to identify and distinguish from nonlinear physics. As a result, we might anticipate that introducing $\nu > 0$ will strengthen our constraints for all frequencies, but the actual situation is more nuanced as can be seen in Fig.~\ref{fig:sigma-fnl-experiments} when comparing the solid to the dotted line with the 95\%~limit on~$|\fnlalpha|$, which is the absolute value of the power-law PNG~amplitude considered in~\cite{Green:2023uyz}. We additionally present their ratio in Fig.~\ref{fig:sigma-fnl-experiments_relative}.%
Taken together, we can see distinct behavior at small, intermediate and large frequencies, and for positive and negative scaling exponents.

For scaling exponents $\alpha \lesssim 0$, the overall constraining power on the PNG~amplitude~$\fnlosc$ is most stringent in absolute terms, reminiscent of the light-field case of~\cite{Green:2023uyz}, which we have now extended to negative values of~$\alpha$. At the same time, there is effectively no improvement or even a severe degradation of the sensitivity by more than an order of magnitude even for large frequencies compared to the $\nu = 0$ case. The reason is that the PNG~signal peaks at small wavenumbers $k \to 0$, leaving little room for our oscillations to be visible in the observed power spectra~(see Fig.~\ref{fig:shape-of-signal}) and providing little benefit. On the other hand, due to the power-law term~$S(k; M, \alpha)$ in~\eqref{eq:Pg}, the signal moves to larger wavenumbers as the scaling exponent increases and peaks at~$\kmax$ for the largest~$\alpha$ under consideration. While this degrades the overall constraints on~$\fnlosc$, the imprinted oscillations can as a result dramatically improve the relative PNG~sensitivity by up to two orders of magnitude for~BOSS and even almost three orders of magnitude for the billion-object survey for $\nu \gtrsim 30$. So, as for primordial features~\cite{Beutler:2019ojk}, introducing oscillations allows the PNG~signal to be measured at $k \sim k_\mathrm{NL}$ while being robust to nonlinearities. In fact, as mentioned above, our treatment is conservative since we only consider linear wavenumbers, i.e.\ we limit the potential reach of the observations that would be possible following~\cite{Beutler:2019ojk} in this regime.\footnote{The range of considered wavenumbers could be significantly extended to all observed modes of a survey for the oscillatory imprint with the additional modeling as for primordial features in~\cite{Beutler:2019ojk}. While we left this combined analysis to future work, we can already anticipate that it would further improve the sensitivity considerably in this regime of~$(\alpha, \nu)$.} Even with our limited range of wavenumbers, we however clearly see the improvement over $\nu = 0$ for $\alpha \gtrsim 0$ and $\nu \gtrsim 1$ precisely due to the fact that the oscillations break the degeneracies of the power-law term with biasing and short-distance physics more generally.

\begin{figure}
	\centering
	\includegraphics{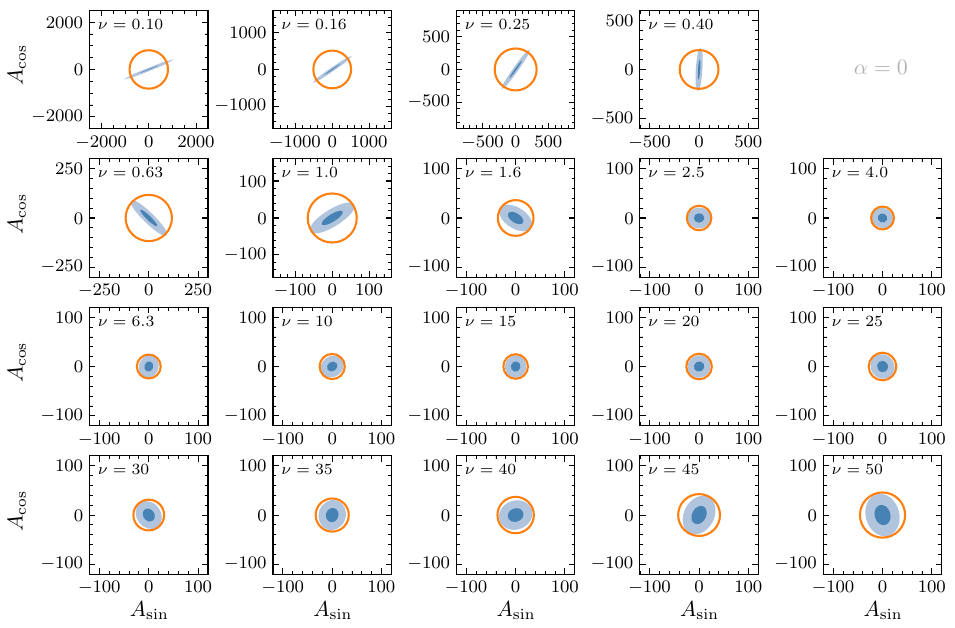}
	\caption{Forecasted posterior distributions of~$(\Acos, \Asin)$ for~BOSS, a scaling exponent of $\alpha = 0$ and different frequencies~$\nu$. The standard two-dimensional~$1\sigma$ and $2\sigma$~limits around the fiducial value of zero are included as the blue contours. The orange circle encloses~95\% of the probability, with its radius being the corresponding limit on the overall oscillatory PNG~amplitude~$A\fnlosc$. While the sine and cosine amplitudes are strongly degenerate for small~$\nu$, and the degeneracy line rotates as we increase the frequency, the degeneracy effectively vanishes for $\nu \gtrsim 2$ as evidenced by the~(nearly) circular posterior contours. We note that the light-blue $2\sigma$~contours and the orange circle do not coincide even for the frequencies for which~$\Acos$ and~$\Asin$ are essentially independent since the former~(two-dimensional~$2\sigma$) contours enclose approximately~86.4\% of the probability and not~95\% as in the case of the latter to display the~(one-dimensional) limit on the PNG~amplitude.}
	\label{fig:acos-asin_delta-0}
\end{figure}

At small frequencies, $\nu \lesssim 1$, the forecasts weaken significantly, effectively independently of the experimental specifications. We can see this both when comparing the limits for $\nu \neq 0$ and $\nu = 0$ in Fig.~\ref{fig:sigma-fnl-experiments} or in their ratio in Fig.~\ref{fig:sigma-fnl-experiments_relative}. This is due to marginalizing over the phase. Concretely, without seeing multiple periods of the oscillation at a large signal-to-noise ratio, changing the phase allows to escape the more stringent bounds. Put differently, the amplitude parameter~$\Asin$ becomes nearly unconstrained at $\nu \to 0$, which makes constraining the overall amplitude more difficult. The role of this degeneracy is clearly demonstrated in Fig.~\ref{fig:acos-asin_delta-0},%
in which the forecasted contours in the $(\Acos, \Asin)$~plane are shown as a function of~$\nu$ for $\alpha = 0$.\footnote{Note that the light-blue two-dimensional $2\sigma$~contours enclose about 86.4\%~of the probability while the orange circles visualize the respective (one-dimensional)~95\%~limits.} We see that the degeneracy line rotates as we decrease the frequency. As $\nu\to 0$, the $\Acos$~direction remains as constraining as the non-oscillatory scenario~(see also the dashed line in Fig.~\ref{fig:sigma-fnl-experiments} for the 95\%~exclusion limit on~$|\Acos|$ which is essentially the same as the limit on~$\fnlalpha$ for $\nu \lesssim 0.1$), while constraints along the $\Asin$~direction degrade rapidly. Since we also display the circle indicating the 95\%~limit on~$A\fnlosc$, we can also directly see how this degeneracy \mbox{impacts the sensitivity to the overall PNG~amplitude.}

When we increase the frequency, we can observe in Fig.~\ref{fig:acos-asin_delta-0} that the contours are nearly circular for $\nu \gtrsim 2$ which indicates that this phase degeneracy is effectively broken at these larger frequencies. We note that the same rotation and vanishing of the degeneracy line with increasing~$\nu$ is not only present for $\alpha = 0$ as shown, but all scaling exponents under consideration. This therefore also explains why we then start being able to see the improvements in the forecasted limits for $\alpha \gtrsim 0$ since the power-law term moves the signal to the largest wavenumbers in the analysis where the oscillations really help to distinguish between the primordial signature and the nonlinear galaxy power spectrum with Gaussian initial conditions. Comparing the limits on~$A\fnlosc$ and~$|\fnlalpha|$ in Fig.~\ref{fig:sigma-fnl-experiments} or their ratio in Fig.~\ref{fig:sigma-fnl-experiments_relative} at fixed~$\alpha$ confirms our expectation that oscillations make it easier to measure this signal at large~$k$.

Additional features of Figures~\ref{fig:sigma-fnl-experiments} and~\ref{fig:sigma-fnl-experiments_relative} are the bumps around $\nu \approx 1\text{ and }10$, with their significance depending on the scaling exponent. These bumps can be attributed to degeneracies with cosmological and bias parameters since the oscillatory signal may become degenerate with the broadband and/or the standard baryon acoustic oscillations. We further discuss this in Appendix~\ref{app:forecasting}, in particular around Fig.~\ref{fig:kmax-ratio}, which shows that these features depend on the marginalization over both the bias and $\Lambda\mathrm{CDM}$~parameters. We additionally note that we cut off the frequency range for~BOSS and~DESI since the oscillations for frequencies exceeding the effective Nyquist frequency, which is set by the employed conservative bandwidths for each redshift bin,\footnote{The bandwidths employed in our forecasts are inferred from a spherical volume separately for each redshift bin which results in overly cautious estimates. This also means that the bandwidths for~BOSS are larger in our forecasts than in the data analysis presented in Section~\ref{sec:analysis}. Compared to~\cite{Beutler:2019ojk}, the smaller maximum wavenumber further limits the effective reach in the logarithmic frequency~$\nu$.\label{fn:bandwidth}} are aliased for all wavenumbers~(see~Fig.~\ref{fig:bandpower} in Appendix~\ref{app:forecasting} for a visualization). This implies that the forecasts for~BOSS and~DESI should not be trusted for $\nu \gtrsim 55\text{ and }90$, respectively, while the other surveys cover a larger cosmic volume with correspondingly larger Nyquist frequencies far beyond the frequencies considered here.\medskip

The potential reach of~SPHEREx, which will have results within the next few years, is of particular interest in Figures~\ref{fig:sigma-fnl-experiments} and~\ref{fig:sigma-fnl-experiments_relative}. When $\alpha \approx 0$, the improvement from~BOSS and~DESI to~SPHEREx is more than an order of magnitude and persists across the full range for frequencies. This improvement is largely compatible with the reach in~$\fnl^\mathrm{loc}$ since the higher-frequency results do not meaningfully exceed the $\nu = 0$ sensitivity. In contrast, for $\alpha = O(1)$, the relative improvement is the same within a factor of a few for small~$\nu$, but we see substantial improvements at higher frequencies relative to the $\nu = 0$ forecasts. Both results are compatible with the idea that~SPHEREx has a larger advantage when the signal peaks at low~$k$, while the benefit of large frequencies is pronounced for larger~$\alpha$ when the signal is pushed to higher wavenumbers.\medskip

\begin{figure}
	\centering
	\includegraphics{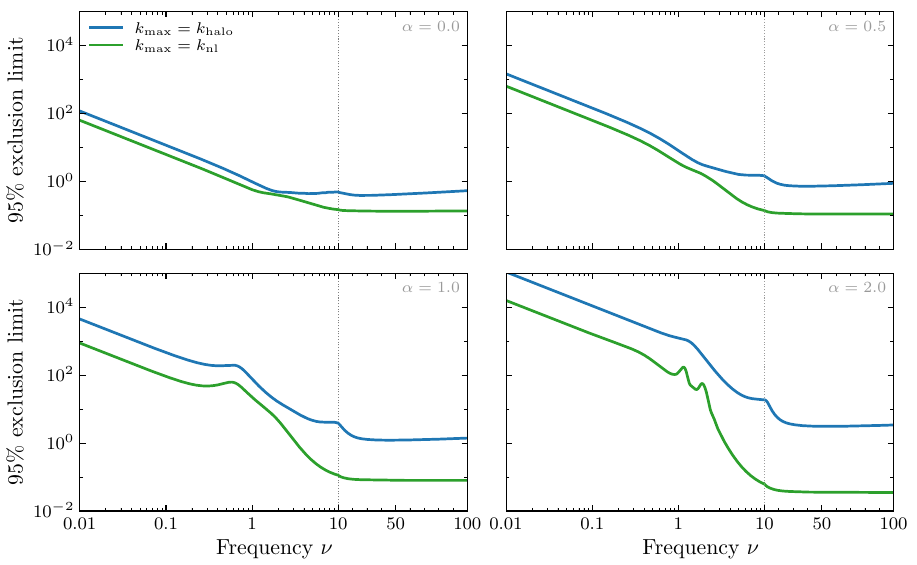}
	\caption{Dependence of the 95\%~exclusion limit of the PNG~amplitude~$\fnlosc$ on the maximum wavenumber~$\kmax$ of the billion-object survey for different values of the scaling exponent~$\alpha$. We compare the two cases for which we set the maximum wavenumber to correspond to the Lagrangian scale of the minimum halo mass, $\kmax = k_\mathrm{halo}$, and to the nonlinear scale, $\kmax = k_\mathrm{NL}(z)$. The latter is in particular much more optimistic at higher redshifts given our modeling of the~(Gaussian) galaxy power spectrum. While the impact on the forecasted constraints for small frequencies is limited, as expected, we see decisive improvements when increasing~$\kmax$ to~$k_\mathrm{NL}$ at high frequencies, especially for large~$\alpha$, since many more oscillatory periods are being captured by the survey.}
	\label{fig:kmax}
\end{figure}

The relative importance of the smallest modes can also be seen through the dependence on the maximum wavenumber~$\kmax$ as shown in Fig.~\ref{fig:kmax}.%
For all values of~$\alpha$, we see that for $\nu \gtrsim 10$, there is a significant improvement in the forecasted limits with $\kmax = k_\mathrm{NL} > k_\mathrm{halo}$. For $\nu < 1$, there is a more pronounced improvement with larger~$\kmax$ only for $\alpha = O(1)$. This reinforces the conclusion that (i)~$\nu > 1$ gains sensitivity from small-scale modes since the oscillations are not degenerate with biasing and (ii)~$\alpha \lesssim 0.5$ gets more constraining power from $k \ll k_\mathrm{NL}$ and is therefore less sensitive to the choice of~$\kmax$. As confirmed in additional forecasts based on the billion-object survey~(as in~\cite{Green:2023uyz}), this same pattern is important for understanding the relative importance of the survey volume~(or the minimum wavenumber~$\kmin$) and the number density~(or~$\kmax$) when designing future surveys for these kinds of power-law and/or oscillatory signals.\medskip

Of course, larger and deeper surveys have the potential to find ever weaker PNG~signals of this kind. We however see in Fig.~\ref{fig:sigma-fnl-experiments_relative} that the shapes of the curves illustrating the relative limits as a function of~$\nu$ are very similar across the four displayed surveys. As a result, the impact of survey design on the $\nu = 0$ forecasts, which were discussed at length in~\cite{Green:2023uyz}, will be effectively the same for all values of~$\nu$. In this regard, the large improvement in going from~DESI to~SPHEREx can again be attributed to the higher biases and the slightly larger number densities of the latter. Furthermore, we again expect a multi-tracer analysis from~LSST to offer improved sensitivity over~SPHEREx. Having said that, it would be worthwhile to understand how these non-Gaussian oscillations are affected by systematics. While the raw sensitivity of~SPHEREx and~LSST offer the potential for large improvements over spectroscopic surveys like~DESI and~Euclid, the scale-dependent bias from local non-Gaussianity can be particularly sensitive to large-angle effects and~(catastrophic) photometric redshift errors among other potential systematic biases~(see e.g.~\cite{Pullen:2012rd, Foglieni:2023xca, Martinez-Carrillo:2021lcn, Castorina:2020blr, Shiveshwarkar:2023xjv}). An oscillatory signal can potentially evade this potential sensitivity to systematic errors, making these kinds of improvements more robust than for the scale-dependent bias from light fields parametrized by~$\fnlalpha$.\medskip

Ultimately, inferring constraints on the PNG~signal in terms of~$\fnlosc$ for a given~$\alpha$ and~$\nu$ is a determination of the strength of the coupling of~$\sigma$ and~$\zeta$ for a fixed quadratic action for~$\sigma$~(i.e.\ its masses and mixing). For $\alpha \approx 1$ and $\nu \approx 1$, the improvement in the limits on~$\fnlosc$ relative to $\nu = 0$ from Fig.~\ref{fig:sigma-fnl-experiments_relative} is substantially larger than the effect due to the LSS~suppression factor~$|\Sigma_M^2|$ in Fig.~\ref{fig:sigma2}~(as well as the primordial factor~$\ee^{-\pi\nu}$). This is the regime where our new analysis is expected to be more sensitive to the coupling to additional fields than the models with no oscillations. This might appear to be a small region of parameters space, but we note that there is good reason for dimensionful parameters to be set by the Hubble scale~$H$~\cite{Baumann:2011nk}. Moreover, as we discussed in~\textsection\ref{sec:models_png}, there are phenomenological ways to boost the primordial amplitude to further overcome the suppression factors over a wider frequency range. Either way, it is evident that the improved PNG~sensitivity is a consequence of model building and analysis, which highlights the discovery potential that is driven by theoretical work in this area.\medskip

To summarize, galaxy surveys are more sensitive to oscillatory signals with $\nu \gtrsim 1$ when $\alpha = O(1)$. In these situations, the signal-to-noise ratio is dominated by large-$k$ modes and is sufficiently distinct from nonlinear corrections to be robust to marginalization over bias parameters. For scaling exponents~$\alpha$ near or smaller than zero, the signal is peaked at smaller wavenumbers which results in less improvement in sensitivity relative to the non-oscillatory scenario, $\nu = 0$. These models are not degenerate with more conventional PNG~shapes, which means that an analysis of~BOSS or DESI~data could find a non-zero signal for non-zero~$\alpha$ and~$\nu$ while being consistent with previous analyses.\footnote{The results must also be consistent with the existing constraints on primordial features from~BOSS~\cite{Beutler:2019ojk, Ballardini:2022wzu, Mergulhao:2023ukp, Calderon:2025xod}.}

\section{Constraints from BOSS~Galaxy Clustering}
\label{sec:analysis}

Having established the expected sensitivities of current and future LSS~surveys to our broad class of oscillatory non-Gaussianity in the previous section, we now apply its phenomenological template to observations and derive constraints from observed galaxy-clustering data. We analyze BOSS~DR12~data based on the galaxy power spectrum. The methodology, which we summarize in~\textsection\ref{sec:analysis_details}, closely follows that of recent analyses of standard PNG~templates~\cite{Cabass:2022wjy, DAmico:2022gki, Cabass:2022ymb, Chudaykin:2025vdh} and our previous analysis of non-oscillatory models~\cite{Green:2023uyz}, with extensions to address the presence of the frequency-dependent oscillations. Our inferred limits on the PNG~amplitude~$A\fnlosc$ are presented in~\textsection\ref{sec:analysis_limits} and exhibit the same behavior as seen in the forecasts. This includes confirming the significance of an oscillatory component in reducing the sensitivity to nonlinear astrophysical effects even for frequencies $\nu = O(1)$. Since we find no evidence for a non-zero oscillatory PNG~signal, we place upper limits on the explored $(\alpha, \nu)$~parameter space. The results of this section suggest that this type of analysis should be straightforward to apply to current and future data, for example from~DESI~(e.g.\ as an extension of~\cite{Chudaykin:2025vdh}) and~SPHEREx.

\subsection{Analysis Pipeline for BOSS~DR12~Data}
\label{sec:analysis_details}

To extract constraints on oscillatory~PNG, we analyze the BOSS~DR12~\cite{BOSS:2016wmc} galaxy power spectrum following the same methodology used in~\cite{Green:2023uyz}, which was based on~\cite{Cabass:2022wjy, Cabass:2022ymb}~(see also~\cite{DAmico:2022gki}) with a few different~(generally more conservative) choices, after extending it to account for the scale-dependent oscillatory modulation predicted by the class of models studied in this work. In particular, we modify the theoretical prediction for the galaxy power spectrum to include the scale- and frequency-dependent bias induced by primordial oscillations in the bispectrum. As discussed in more depth after~\eqref{eq:A-cos-sin}, we use the amplitude instead of the phase parametrization since the former results in nearly Gaussian posterior distributions for the oscillatory parameters while the phase is relatively poorly constrained in the latter in the absence of a signal. Compared to~\cite{Green:2023uyz}, we therefore extend our theoretical model of the galaxy power spectrum~$P_g(k)$ to include the oscillatory amplitudes~$\Acos$ and~$\Asin$ as follows:\footnote{We also refer to~\eqref{eq:Pgfinal} and remember that we absorbed the overall PNG~amplitude~$\fnlosc$ into the amplitudes~$\Acos$ and~$\Asin$ for notational convenience while ultimately still reporting the results of our analysis in terms of~$\fnlosc$.}
\begin{equation}
	\begin{split}
		b_\mathrm{NG}^\Delta&(k, z) = 3 \fnl^\Delta \frac{b_\phi(z)}{k^2 \mathcal{T}(k, z)}(k R_M)^\Delta																						\\
							&\to\ b_\mathrm{NG}^{\alpha, \nu}(k, z) = 3 \frac{b_\phi(z)}{k^2\mathcal{T}(k, z)}(k R_M)^\alpha \{\Acos \cos[\nu \log(k R_M)] + \Asin \sin[\nu \log(k R_M)]\}\, ,
	\end{split}
\end{equation}
where we took $R_M \approx \SI{2.66}{\MpcPerh}$ corresponding to the Lagrangian radius of a halo with a mass of~$\num{e13} M_\odot$, as in our Fisher forecasts presented in~\textsection\ref{sec:Fisher}. Let us in particular reiterate that we effectively absorbed the suppression factor~$|\Sigma_M^2(\alpha, \nu)|$ of~\eqref{eq:Pg} into the PNG~amplitudes.\medskip

The dataset, the observables and covariances, and the nonlinear and bias modeling are identical to the analysis presented in~\cite{Green:2023uyz}. In short, we use the BOSS~DR12 dataset~\cite{BOSS:2016wmc} with about 1.2~million galaxies which are split into two redshift bins, $z \in [0.2, 0.5]$ and~$[0.5, 0.75]$, and two galactic caps. The monopole, the quadrupole and hexadecapole of the galaxy power spectrum~$P^{(\ell)}_g(k)$ are estimated using a quadratic window-function-free estimator~\cite{Philcox:2020vbm} while the employed covariance matrices are computed from the MultiDark-Patchy mocks~\cite{Kitaura:2015uqa}. We limit the range of wavenumbers to $k \in \SIrange{0.01}{0.13}{\hPerMpc}\text{ and }\SIrange{0.01}{0.16}{\hPerMpc}$ for the low- and high-redshift bin, respectively, with bandpowers $\Delta k = \SI{0.005}{\hPerMpc}$. The theoretical power spectra account for redshift-space distortions, nonlinear evolution and the Alcock–Paczynski effect, include parameters from the bias expansion and EFTofLSS~description following~\cite{Ivanov:2019pdj, DAmico:2019fhj}~(cf.~\cite{Baumann:2010tm, Carrasco:2012cv, Pajer:2013jj, Carrasco:2013mua, Desjacques:2016bnm} and~\cite{Cabass:2022avo, Ivanov:2022mrd} for recent reviews) and are calculated using a modified version of CLASS-PT~\cite{Chudaykin:2020aoj} which in turn is based on~CLASS~\cite{Blas:2011rf}.

The parameter inference is performed via Markov chain Monte Carlo using the Metropolis-Hastings algorithm implemented in MontePython~\cite{Audren:2012wb, Brinckmann:2018cvx}, with all chains converging with a Gelman-Rubin criterion of $R-1 < 0.01$~(usually much smaller) for each varied parameter. While the cosmological parameters are fixed to the Planck~2018 best-fit $\Lambda\mathrm{CDM}$~values~(with the sum of neutrino masses $\sum m_\nu = \SI{0.06}{eV}$)~\cite{Planck:2018vyg},\footnote{We saw in the forecasts~(see Fig.~\ref{fig:prior-impact} in Appendix~\ref{app:forecasting}) that fixing the five $\Lambda\mathrm{CDM}$~parameters has essentially the same effect on the constraints on~$\fnlosc$ as imposing a $\Lambda\mathrm{CDM}$~prior consistent with Planck~2018. For~BOSS, the main difference of fixing these parameters is a slight overestimation of the constraining power around $\nu \approx 10$ which is broadly negligible on our logarithmic scale for~$\fnlosc$ and given the main focus of this work. We therefore effectively infer the limits on the PNG~amplitude from the BOSS~data while imposing a Planck prior on the $\Lambda\mathrm{CDM}$~parameters.} the bias and EFT~parameters are sampled independently for each redshift bin and galactic cap as in~\cite{Philcox:2020zyp, Philcox:2021kcw, Cabass:2022wjy, Green:2023uyz}. On the other hand, the oscillatory non-Gaussian parameters~$\Acos$ and~$\Asin$ are shared across all subsets, with the~(real-valued) scaling exponent~$\alpha = \Re\Delta$ and frequency~$\nu = \Im\Delta$ being fixed for individual runs and scanned over a linear and logarithmic-linear grid with nine and 19~values, respectively: $\alpha \in [-1, 3]$ and $\nu \in [0.1, 10] \cup [10, 50]$. As discussed above, these ranges are motivated by theoretical and observational considerations, respectively: we interpolate between the benchmark scaling exponents~$\alpha$ for the tachyonic collider~($\alpha = -1$), local~PNG~($\alpha = 0$), the conventional cosmological collider~($\alpha = 3/2$), equilateral and orthogonal~PNG~($\alpha = 2$) and originating from inflationary loops or composite operators~($\alpha = 3$); we take $\nu \geq 0.1$ since we saw in our forecasts that the constraints further weaken for smaller frequencies given the phase degeneracy and $\nu \leq 50$ to approximately remain within the effective logarithmic Nyquist~frequency of the BOSS~DR12~volume~(cf.~\cite{Beutler:2019ojk} and Fig.~\ref{fig:bandpower} in Appendix~\ref{app:forecasting}). We impose flat priors on the non-Gaussian and the linear bias parameters, and Gaussian priors on higher-order bias, the non-Gaussian bias and nuisance parameters. We refer to~\cite{Green:2023uyz} for further information and details on the data and choices underlying our analysis.

To infer the value of the PNG~amplitude~$\fnlosc$, rather than the product~$b_\phi \fnlosc$, we marginalize over~$b_\phi$ with a Gaussian prior separately in each redshift bin. This prior has a mean value that is set by the universality relation, $b_\phi = 2 \delta_c (b_1 - 1)$, and a large standard deviation of five times this mean~(see~\cite{Green:2023uyz} for additional details). This procedure alleviates degeneracies between~$\fnlosc$ and~$b_1$ which can be severe if we assumed that the universality relation is exact, especially for $\alpha \sim 1$. Although the constraint on~$\fnlosc$ is sensitive to the prior on~$b_\phi$, the degeneracy is further reduced by the fact that~$b_\phi$ varies in each redshift bin while there is only a single primordial amplitude~$\fnlosc$~(or rather~$\Acos$ and~$\Asin$) for the entire survey.\medskip

\begin{figure}
	\centering
	\includegraphics{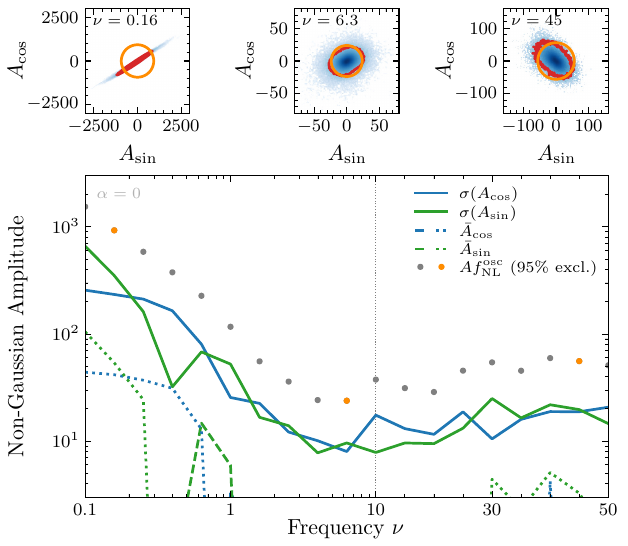}
	\caption{Constraints on the non-Gaussian amplitude(s) inferred from BOSS~DR12 for a scaling exponent of $\alpha = 0$ as a function of frequency~$\nu$. We note that the frequency axis is displayed on a logarithmic-linear scale. This figure illustrates the relationship between the MCMC~samples and the inferred mean values, standard deviations and 95\%~upper limits at each sampled frequency. \textit{Top}:~The three panels contain the pixelated posterior distributions in the marginalized $(\Acos, \Asin)$~plane based on the MCMC~samples in blue~(light/dark blue: low/high density) for the three frequencies~$\nu = 0.16, 6.3\text{ and }45$. We see the strong and weak correlation of these two amplitudes at small and intermediate/large frequencies, respectively, which is also illustrated by the red contours which enclose 95\%~of the samples. This is also the case for the orange circles which display our main statistic, the phase-marginalized 95\%~(exclusion) limits on the overall amplitude~$A\fnlosc = \sqrt{\Acos^2 + \Asin^2}$, which are derived following~\cite{Beutler:2019ojk} as summarized in the main text. We additionally refer to Appendix~\ref{app:analysis} for further details and these $(\Acos, \Asin)$~panels for all sampled frequencies~$\nu$. \textit{Bottom}:~The standard deviations, positive and negative mean values of~$\Acos$~(blue) and~$\Asin$~(green) are shown as solid, dashed and dotted lines. The gray dots are the mentioned 95\%~exclusion limits on the PNG~amplitude~$A\fnlosc$ for the 19~sampled values of the frequency~$\nu$, with the orange points corresponding to the three panels at the top. We can clearly observe that no non-zero PNG~amplitude is detected and how the degeneracy line between~$\Acos$ and~$\Asin$ rotates for $\nu \lesssim 1$ while these \mbox{amplitudes become nearly uncorrelated for larger frequencies.}}
	\label{fig:constraints_1d_delta-0}
\end{figure}

In order to constrain the oscillatory PNG~amplitude while marginalizing over the phase, we follow the same method introduced in~\cite{Beutler:2019ojk}~(and also applied to our forecasts in~\textsection\ref{sec:Fisher}) to infer the limit on the overall~(phase-independent) oscillation amplitude from an analysis based on the amplitude parametrization. This means that we analyze the marginalized posteriors of the oscillatory amplitudes,~$\Acos$ and~$\Asin$, in the two-dimensional plane defined by these parameters. The 95\%~confidence limit on the total amplitude $A\fnlosc = \sqrt{\Acos^2 + \Asin^2}$, cf.~\eqref{eq:png-amplitude}, is defined as the radius of a circle around the origin that encloses 95\%~of the posterior probability or, equivalently, 95\%~of the Monte Carlo samples which implicitly marginalizes over the phase of the non-Gaussian oscillations. This procedure therefore accounts for the fact that the phase is not independently constrained and that the maximum posterior point may be offset from the origin due to noise. While the posteriors are pixelated in the $(\Acos, \Asin)$~plane, we explicitly verified that pixelation artifacts are negligible. We note that~$\Acos$ and~$\Asin$ are effectively uncorrelated for frequencies~$\nu \gtrsim 1$, but exhibit correlations especially for the smallest frequencies under consideration~(see Fig.~\ref{fig:acos-asin_delta-0} and the top panels of Fig.~\ref{fig:constraints_1d_delta-0}%
for some example illustrations, and Appendix~\ref{app:analysis} for additional details). While our procedure is generally robust and conservative in estimating the PNG~amplitude~$A\fnlosc$, this correlation implies that the inferred limits may not always reflect the actual constraining power of the data. For \mbox{full details and validation of this methodology, we refer to Appendix~C of~\cite{Beutler:2019ojk}.}

The bounds on~$\Acos$ and~$\Asin$ for $\alpha = 0$, along with the derived 95\%~exclusion limits of~$A\fnlosc$, are shown in the bottom panel of Fig.~\ref{fig:constraints_1d_delta-0}. Comparing the top panels with the associated frequency illustrates the relationship between the samples in the Monte Carlo Markov chains and the inferred bounds. The mean values of~$\Acos$ and~$\Asin$ are consistent with zero across all of the frequencies~(and scaling exponents) that we sampled. It is important to keep in mind that the analysis at each individual frequency is correlated with other frequencies, especially for $\alpha = 0$, where the information is at low wavenumbers~$k$. As a result, we do not have an expectation to see several $1\sigma$~fluctuations in these 19~frequencies. Similarly, we had already seen and utilized the broad correlations in the scaling exponent in~\cite{Green:2023uyz}.

\subsection{Bounding Non-Gaussian Oscillations}
\label{sec:analysis_limits}

\begin{figure}
	\centering
	\includegraphics{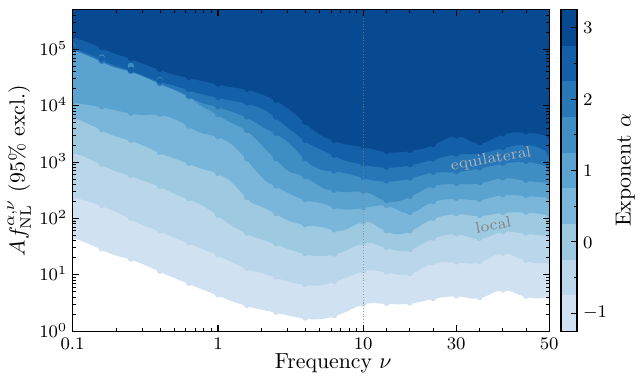}\vspace{-6pt}
	\caption{Exclusion limits at 95\%~c.l.\ on the non-Gaussian oscillatory amplitude~$A\fnlosc$ inferred from the BOSS~DR12~dataset as a function of the oscillatory frequency~$\nu$ of the non-Gaussian signature. The excluded parameter space is indicated by the shaded regions for different scaling exponent~$\alpha$. Note that the horizontal frequency axis is on a logarithmic-linear scale, while the vertical axis for the amplitude is logarithmic. The labels `local' and `equilateral' denote the parameter space excluded for $\alpha = 0\text{ and }2$, respectively, which are the scaling exponents of these standard PNG~types in the $\nu \to 0$ limit. These exclusion limits are inferred from 171~separate Markov chain Monte Carlo analyses at fixed~$(\alpha, \nu)$ using the method introduced in~\cite{Beutler:2019ojk}, which implicitly marginalizes over the oscillatory phase. On this grid, the tightest 95\%~bounds with $A\fnlosc < 1.7$ are inferred for $(\alpha, \nu) = (-1, 4)$, while the weakest 95\%~limits with $A\fnlosc < \num{1.7e5}$ are inferred for $(\alpha, \nu) = (3.0, 0.1)$. We observe that the bounds get weaker for both lower frequencies~$\nu$ and larger exponents~$\alpha$.\vspace{-3pt}}
	\label{fig:data_limits_1d}
\end{figure}

We can straightforwardly infer the posterior distributions for the parameter~$A\fnlosc$ from the Markov chains of the BOSS~DR12~data following the procedures laid out above. For each pair of~($\alpha, \nu$) in our grid of values, we calculate the posterior while marginalizing over the additional bias and EFT~parameters. These results are all consistent with $A\fnlosc = 0$ and, therefore, do not show any evidence for this kind of underlying non-Gaussian signal. We display the 95\%~exclusion limits on the non-Gaussian oscillatory amplitude~$A\fnlosc$ as a function of the frequency~$\nu$ for nine values of the scaling exponent $\alpha \in [-1, 3]$ in Fig.~\ref{fig:data_limits_1d}.%

We can clearly observe the general trends of our inferred exclusion limits as a function of both~$\alpha$ and~$\nu$. We infer the tightest bounds for $\alpha = -1$ which is due to the sharp cubic scaling behavior on large scales. As we increase the scaling exponent~$\alpha$, the limits continuously degrade, being more than three orders of magnitude weaker for $\alpha = 3$. On the other hand, these limits improve by about one to almost two orders of magnitude from $\nu = 0.1$ to $\nu \sim 10$ before remaining roughly at the same level for $\nu \gtrsim 10$. As discussed in~\textsection\ref{sec:Fisher} and~\textsection\ref{sec:analysis_details}, the apparent improvement in constraints from $\nu \ll 1$ to $\nu = O(1)$ is primarily a consequence of marginalizing over the phase of our two-parameter model of the non-Gaussian oscillations. Due to this marginalization, the constraints at small~$\nu$ are much weaker than the constraint for $\nu = 0$ derived in~\cite{Green:2023uyz}. In this sense, the data remains very capable of detecting a signal at low frequency, but the overall sensitivity of the survey is not well captured by our statistic of the exclusion limit of~$A\fnlosc$ when~$\nu \ll 1$ since the sine component becomes poorly constrained.

To better understand the functional dependence of these limits on the frequency~$\nu$, we also included the marginalized constraints on the amplitudes~$\Acos$ and~$\Asin$ for $\alpha = 0$ in Fig.~\ref{fig:constraints_1d_delta-0}. We clearly see that the two amplitudes are highly degenerate for small frequencies $\nu \lesssim 1$, with the degeneracy line rotating in the amplitude plane as a function of~$\nu$. This additionally implies that our procedure to infer a conservative exclusion limit on the overall amplitude~$A$ leads to weaker bounds in this range of parameters. For larger frequencies, the standard deviations of the two amplitudes are very similar and the procedure results in limits that more directly capture the underlying constraining power of the survey. We note that this behavior shown for $\alpha = 0$ is generally similar for other exponents~$\alpha$ as can be assumed from Fig.~\ref{fig:data_limits_1d}~(and Fig.~\ref{fig:data_limits_2d} included in Appendix~\ref{app:analysis}). We also display the two-dimensional histograms for all sampled frequencies at a fixed scaling exponent of $\alpha = 0$ in Fig.~\ref{fig:data_acos-asin_delta-0} in Appendix~\ref{app:analysis}.\medskip

\begin{figure}
	\centering
	\includegraphics{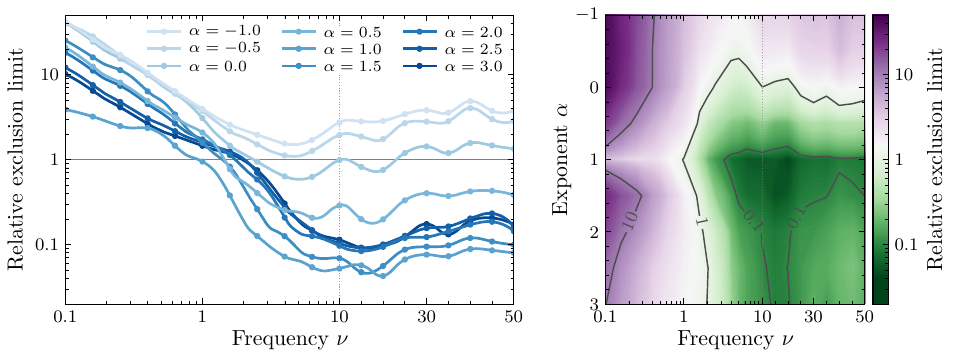}
	\caption{Exclusion limits at 95\%~c.l.\ on the amplitude of the oscillatory primordial non-Gaussianity,~$A\fnlosc$, from BOSS~DR12 relative to the 95\%~limit on the absolute value of the amplitude of the non-oscillatory, simple power-law non-Gaussianity,~$|\fnlalpha|$. The relative limits are shown as a function of the frequency~$\nu$ for the nine scaling exponents~$\alpha$~(\textit{left}), and as a function of both~$\alpha$ and~$\nu$~(\textit{right}). The axes for the exclusion limit, frequency and exponent are on a logarithmic, logarithmic-linear and linear scale, respectively. The oscillatory limits are better than those for the power-law signal for $\alpha \gtrsim 0$ and $\nu \gtrsim 1$, while they are worse otherwise. The improvement reaches factors of five to~23 for $\alpha \geq 0.5$. The relative degradation at small frequencies originates from the phase marginalization and the related strong degeneracy between~$\Acos$ and~$\Asin$ as discussed in the main text. For $\alpha \lesssim 0$, the highly oscillatory limits do not improve over the non-oscillatory constraints since the non-Gaussian signal peaks at the smallest survey wavenumbers and may be aliased. On the other hand, the power of the oscillatory imprint to distinguish the PNG~signal for $\alpha > 0$ from the smooth galaxy power spectrum is clearly evident.}
	\label{fig:data_relative_limits}
\end{figure}

We can additionally compare these new results with the respective limits for the non-oscillatory forms of primordial non-Gaussianity studied in~\cite{Green:2023uyz}. We show the relative exclusion limits in Fig.~\ref{fig:data_relative_limits}.%
To be precise, we present the ratio of the 95\%~exclusion limits on the oscillatory PNG~amplitude~$A\fnlosc$ displayed in Fig.~\ref{fig:data_limits_1d} (and Fig.~\ref{fig:data_limits_2d}) to the bound at 95\%~confidence level~(c.l.)\ on~(the absolute value of) the non-oscillatory, simple power-law PNG~amplitude~$|\fnlalpha|$ that we had studied in detail in~\cite{Green:2023uyz}.\footnote{While we analyzed the BOSS~DR12~dataset for $\alpha \in [0, 2]$ in~\cite{Green:2023uyz}, we performed here an equivalent reanalysis for $\alpha \in [-1, 3]$~(see Appendix~\ref{app:analysis}).} We observe that the oscillatory bounds are weaker than the non-oscillatory bounds for $\nu \lesssim 1-2$ depending on the scaling exponent~$\alpha$ and further weakening towards smaller frequencies. In fact, even for larger frequencies, the oscillatory limits are generally weaker for $\alpha < 0$ and the ratio is about unity for $\alpha = 0$. On the other hand, the oscillatory amplitude is constrained better by a factor of up to five to~23 for $\alpha \geq 0.5$ and frequencies $\nu \gtrsim 10$ on our grid of parameter values.

We can understand the origin of the degraded constraints at small frequencies to come from the degeneracies of the oscillatory signal with the smooth broadband spectrum and of the amplitudes~$A_X$, $X = \sin, \cos$, as discussed above. This also results in a relative limit that is not unity in the limit of very small frequencies~($\nu \to 0$) as one might have initially expected. Relatedly, we do not observe an improvement in the inferred limits compared to the non-oscillatory PNG~results for $\alpha \lesssim 0$ since the signal is dominated by the largest scales of the survey where the non-Gaussian signal is generally not degenerate with the broadband power spectrum. On the other hand, we see a significant improvement for positive exponents~$\alpha$ and larger frequencies since the constraining power is coming from smaller and smaller scales where the additional oscillations break the larger and larger degeneracy of the smooth power-law non-Gaussian signal with the Gaussian broadband power spectrum.\medskip

Comparing these results inferred from the BOSS~data to our Fisher forecasts described in~\textsection\ref{sec:Fisher}, we can clearly see the qualitative similarities as a function of both scaling exponent~$\alpha$ and oscillation frequency~$\nu$. While the precise quantitative limits somewhat differ between the forecasts and data analysis, these may be due to the differences in the underlying modeling and specifications as in~\cite{Green:2023uyz}, especially regarding the specific scale cuts and biases~(with larger~$\alpha$ being particularly sensitive). We in particular note that the bumps for $\nu \approx 1 - 2$ and $\alpha \gtrsim 0$ are as present in the forecasts as the seemingly unexpected behavior for smaller frequencies and $\alpha \gtrsim 1$ in Fig.~\ref{fig:data_relative_limits}. At the same time, the forecasted and inferred limits are also similar for larger frequencies including their roughly logarithmically equidistant values. We additionally observe the same degeneracies and behavior in the two-dimensional $(\Acos, \Asin)$~contours. We therefore conclude that our 95\%~exclusion limits on the oscillatory PNG~amplitude~$A\fnlosc$~(and other results) inferred from the BOSS~DR12~data in this section are broadly consistent with our BOSS~forecasts of~\textsection\ref{sec:Fisher}, with the agreement between the two providing a mutual validation of the forecasting and analysis pipelines.\medskip

To summarize, we have inferred the first constraints from data on the class of oscillatory primordial non-Gaussianity described by the scale-dependent bias~\eqref{eq:b_ng} in the galaxy power spectrum for the wide ranges of $\alpha \in [-1, 3]$ and $\nu \in [0, 50]$. The power law governed by the scaling exponent~$\alpha$ and the oscillations with frequency~$\nu$ decisively change the sensitivity of~BOSS to the PNG~signal compared to the non-oscillatory or standard shapes. We explored this entire and previously unexplored parameter space and explained the origin of the constraining power in the different regimes. This will therefore make it straightforward to augment analyses of~DESI, Euclid, SPHEREx and other future datasets targeting standard local and equilateral non-Gaussianity~(such as the recent~\cite{Chudaykin:2025vdh} for~DESI) to also probe this promising class of oscillatory primordial non-Gaussianity.

\section{Conclusions and Outlook}
\label{sec:conclusions}

Primordial non-Gaussianity offers a unique opportunity to search for new particles and interactions during inflation, and has the potential to probe energy scales many orders of magnitude higher than any terrestrial experiment. Accessing these signatures of ultra-high-energy physics in cosmological data typically requires new templates and dedicated search strategies since the relevant signals may be hidden in more conventional analyses. Given the unparalleled reach of current and near-term surveys, it is timely and important to thoroughly explore the space of models and theoretically motivated signals to the greatest extent possible.\medskip

In this paper, we studied a new theoretically motivated class of models, which was first introduced in~\cite{McAneny:2019epy} and exhibits non-trivial behavior in the squeezed limit of the primordial bispectrum, and their observational imprints in galaxy surveys. These models include a regime with both oscillatory and power-law behavior which gives rise to a hybrid signal between the traditional light- and heavy-field signals that have previously been explored.

The main phenomenological implication of this primordial physics for observations of the large-scale structure of the Universe is that the scale-dependent bias of the galaxy spectra, especially the power spectrum, is modified to include a non-trivial power law and logarithmic oscillations with a wide range of both scaling exponents and frequencies, respectively. We showed that this oscillatory component results in comparatively easy searches in observations and a more straightforward separation of primordial physics from late-time astrophysics. We explicitly demonstrated this in forecasts for current and future galaxy surveys, and in an analysis of the BOSS~DR12~dataset. In this way, we also highlighted the significant discovery potential in this extended parameter space and, since our results are consistent with no such primordial non-Gaussianity, inferred the first constraints on these phenomenological non-Gaussian signals.\medskip

At the same time, it is also this oscillatory nature that typically suppresses the signal beyond the level that intrinsically arises from the inflationary mechanism itself when the primordial bispectrum is translated to the scale-dependent bias in late-time observables. The reason is that this transfer of power involves an average over small-scale (oscillatory)~fluctuations. This suppression, which we absorbed into the inferred PNG~amplitude, becomes increasingly significant at high frequencies. In turn, this suggests that a bispectrum or trispectrum analysis may provide more powerful constraints on the underlying microphysics of these models at least in parts of their parameter space. In fact, the first cosmological analyses for the conventional cosmological collider~(i.e.\ with $\alpha = 3/2$) were only recently performed in~\cite{Cabass:2024wob, Sohn:2024xzd, Suman:2025vuf, Philcox:2025bbo, Suman:2025tpv}. While these measurements offered the first direct constraints on heavy inflationary fields, they are currently much less sensitive than indirect constraints from the equilateral bispectrum~\cite{Planck:2019kim}. This therefore also means that there is still much space to explore not only phenomenologically in extensions of the cosmological collider, but also observationally in pipelines extending our analysis to all observed wavenumbers and involving higher-point statistics in both the cosmic microwave background and the large-scale~structure.\footnote{Our analysis can be extended to all observed wavenumbers for the oscillatory part of the scale-dependent bias similar to~\cite{Beutler:2019ojk}. Apart from higher-point functions of the primary cosmic microwave background and maps of galaxy clustering, which can be jointly analyzed, statistically combined and/or studied at the level of the underlying fields or maps~(cf.~e.g.~\cite{Fergusson:2014hya, Fergusson:2014tza, Meerburg:2015owa, dePutter:2018jqk, Biagetti:2020skr, MoradinezhadDizgah:2020whw, Baumann:2021ykm, Andrews:2022nvv, Coulton:2022qbc, Karagiannis:2023lsj, Euclid:2024ris, Sullivan:2024jxe, Doeser:2025wcb}), other interesting observables in this context include CMB~secondaries, cosmic shear, line intensity mapping and various cross-correlations~(see e.g.~\cite{Camera:2014bwa, Meerburg:2016zdz, Li:2017jnt, Schmittfull:2017ffw, MoradinezhadDizgah:2018lac, CosmicVisions21cm:2018rfq, Munchmeyer:2018eey, PUMA:2019jwd, Karagiannis:2020dpq, Chen:2021vba, Adshead:2024paa, Anbajagane:2025uro, Anbajagane:2025xlt, Zang:2025azh}).}\vskip3pt

More broadly, new signals continue to appear that shift our perspective on what is both allowed and observable, even as techniques ranging from the effective field theory of inflation~\cite{Cheung:2007st} to the cosmological bootstrap~\cite{Arkani-Hamed:2018kmz} have simplified and expanded the exploration of theory space. The signals described in this paper are one such example, but also only represent a small step away from existing searches. In fact, it is important to recognize that the signatures that may be easiest to detect in observations are not necessarily those that may be considered the most generic. This in particular includes the realizations that signals are not limited to low-point statistics~\cite{Flauger:2016idt} and can manifest themselves directly as features in maps of the primordial fluctuations~\cite{Munchmeyer:2019wlh}. A~common aspect of these models, including the ones studied in this paper, is that their novel signatures can make it easier to isolate in the data than many of the~(smooth) conventional shapes of primordial non-Gaussianity. In this sense, there is real discovery potential available from theoretical exploration alone, and the forthcoming wave of cosmological data provides a timely opportunity to convert \mbox{this potential into new signature-driven insights into the primordial~universe.}\vspace{-3pt}

\paragraph{Acknowledgments}
We are grateful to Daniel Baumann, Priyesh Chakraborty, Xingang Chen, Tim Cohen, Kshitij Gupta, Yiwen Huang, Oksana Iarygina, Alec Ridgway, Chia-Hsien Shen, Raman Sundrum, Xi Tong and Zhong-Zhi Xianyu for helpful discussions. D.\,G.~and J.\,H.~acknowledge support from the US~Department of Energy under Grant~\mbox{DE-SC0009919}. B.\,W.~was supported by the Swedish Research Council under Contract No.~\mbox{638-2013-8993}. Nordita was supported in part by NordForsk. B.\,W.~is grateful to INFN~Ferrara, Scuola Normale Superiore di Pisa, University of California San Diego and University of Illinois Urbana-Champaign for their hospitality. Part of this work is based on observations obtained by the Sloan Digital Sky Survey~III (\mbox{SDSS-III}, \href{http://www.sdss3.org/}{http:/\!/www.sdss3.org/}). Funding for SDSS-III has been provided by the Alfred P.~Sloan Foundation, the Participating Institutions, the National Science Foundation and the US~Department of Energy Office of Science. Parts of this work were performed using computing resources provided by the National Academic Infrastructure for Supercomputing in Sweden~(NAISS) under Projects~\mbox{2023/3-21}, \mbox{2023/6-297}, \mbox{2024/5-666} and~\mbox{2024/6-339}, which is partially funded by the Swedish Research Council through Grant~\mbox{2022-06725}. We acknowledge the use of \texttt{CAMB}~\cite{Lewis:1999bs}, \texttt{CLASS}~\cite{Blas:2011rf}, \texttt{CLASS-PT}~\cite{Chudaykin:2020aoj}, \texttt{FAST-PT}~\cite{McEwen:2016fjn}, \texttt{IPython}~\cite{Perez:2007ipy} and \texttt{MontePython}~\cite{Audren:2012wb, Brinckmann:2018cvx}, and the Python packages \texttt{Matplotlib}~\cite{Hunter:2007mat}, \texttt{NumPy}~\cite{Harris:2020xlr} and~\texttt{SciPy}~\cite{Virtanen:2019joe}.

\clearpage
\appendix
\section{Details for the Fisher Forecasts}
\label{app:forecasting}

The forecasts presented in Section~\ref{sec:forecasts} were obtained following standard Fisher methodology. Many of the specific choices and experimental specifications are described in detail in~\cite{Green:2023uyz}. In this appendix, we summarize the implemented changes to address the subtleties induced by high-frequency oscillations that are unique to the models discussed in this paper. We also include additional forecasting results.

\subsection*{Bandpowers}

A lot of useful intuition for measuring primordial oscillations can be derived from time-series analysis. We measure the power spectrum at a finite number of wavenumbers~$k_i$ and can therefore only extract a finite list of frequencies. Unlike most time series, however, we do not actually measure the power spectrum at a unique set of points. Instead, different Fourier modes~$\k$ are binned according to their length~$k$. As the oscillatory frequencies approach the bin size~(or spacing), this detail can significantly impact the results.

\begin{figure}[b!]
	\centering
	\includegraphics{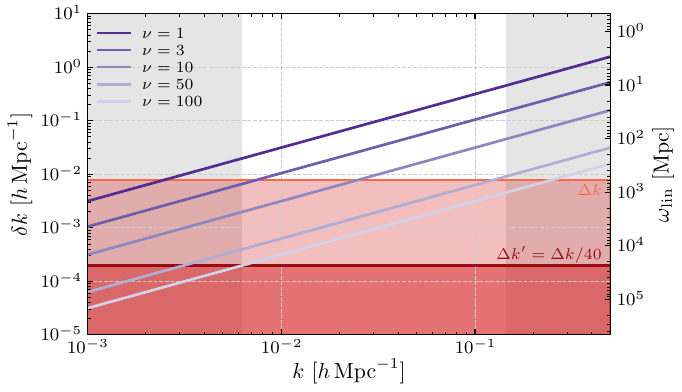}
	\caption{Separation of zeroes~$\delta k$ of the isolated oscillatory signal~(or, alternatively, its effective linear frequency~$\omega_\mathrm{lin}$) for five values of the logarithmic frequency~$\nu$ as a function of wavenumber~$k$. The gray-shaded regions indicate the wavenumbers that are inaccessible in our BOSS~forecasts below~$\kmin$ and above~$\kmax$, respectively, and the red-shaded regions are those in which the survey is partially~(light red) and completely~(dark red) insufficient to resolve the oscillations at~$\kmin$, highlighting the need to use bandpowers in our forecasts.}
	\label{fig:bandpower}
\end{figure}

In order to model the binning procedure, we computed the Fisher matrix by summing over discrete wavenumbers~$k_i$ based on the minimum binning width~$\Delta k$, which is conservatively set by the volume~$V$ of a given redshift bin, and the minimum accessible wavenumber $\kmin = \Delta k$. For a sufficiently small~$\Delta k$, the power spectrum is smooth and we can approximate it in a given band~$[k_i - \Delta/2, k_i + \Delta/2]$ by taking the midpoint value $P_i = P_g(k_i)$. For high-frequency oscillations, there will however be significant variations within each band and our ability to resolve these features will be limited, as depicted in Fig.~\ref{fig:bandpower}.%
If we do not take this into account, our results would suffer from aliasing. This is why we compute the bandpowers~$P_i$ as the average power spectrum in a given bin centered around~$k_i$,
\begin{equation}
	P_i = \frac{1}{\Delta k} \int^{k_i + \Delta k/2}_{k_i - \Delta k/2} \!\d k\, P_g(k) \, .
\end{equation}
When numerically implementing this, it is important to sufficiently sample the theoretical galaxy power spectrum~$P_g$ to resolve the oscillations which requires a sampling step of $\Delta k' < \Delta k$, with $\Delta k' \leq \delta k(\kmin, \nu_\mathrm{max})$, where~$\nu_\mathrm{max}$ is the largest logarithmic frequency considered. We introduced here~$\delta k(k, \nu)$ as the separation of the zeroes of the oscillatory term~$O(k)$ defined in~\eqref{eq:A-cos-sin} which depends on both the wavenumbers~$k$ and the logarithmic frequency~$\nu$, as illustrated in Fig.~\ref{fig:bandpower}. To use our intuition from time series and linear features, we can convert~$\delta k$ to an~(effective) linear frequency~$\omega_\mathrm{lin}$ which naturally decreases with increasing wavenumber for a given logarithmic frequency.

To illustrate the ability to resolve the logarithmic oscillations over the range of wavenumbers used in our forecasts for~BOSS, we also highlight their values of~$\kmin$, $\kmax$ and bandwidth~$\Delta k$~(see also footnote~\ref{fn:bandwidth}). Signals in the white region of Fig.~\ref{fig:bandpower}~($\delta k > \Delta k$) are fully resolved by the survey, those in the light-shaded region ($\Delta k'< \delta k < \Delta k$) are partially affected by the bandpass filtering and those in the dark-shaded region~($\delta k \lesssim \Delta k' = \Delta k/40$) would be entirely affected for frequencies $\nu > 100$. This also explains and the figure visualizes the maximum frequency~$\nu_\mathrm{max} = 55$ for which we forecast the constraining power of~BOSS in~\textsection\ref{sec:Fisher}. As a point of comparison, we found that the BOSS~forecasts using the midpoint approximation lead to approximately the same constraints as those including the bandpowers if $\nu \lesssim 20$. We finally note that our bandpass filtering is additionally serving as an approximate way to include the effects of the survey window function on the oscillations as shown in~\cite{Beutler:2019ojk}.

\subsection*{Phase Dependence}

\begin{figure}
	\centering
	\includegraphics{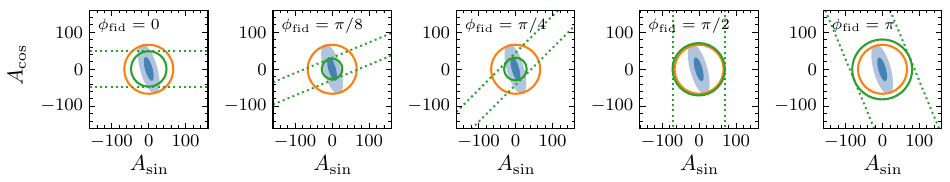}
	\caption{Forecasted posterior distributions for~$(\Acos, \Asin)$ and the derived limits on~$A\fnlosc$ for different fiducial phases~$\phi_\mathrm{fid} = \arctan(-\Asin^\mathrm{fid} / \Acos^\mathrm{fid})$ if we had used the phase instead of the amplitude parametrization to model the logarithmic oscillations. The scaling dimension is set to $\alpha = 0$ and $\nu = 1$, and the phase is varied by setting $\Acos^\mathrm{fid} = \cos\phi_\mathrm{fid}$ and changing~$\Asin^\mathrm{fid}$ accordingly. The light and dark blue contours represent the two-dimensional posterior distributions at~$1\sigma$ and~$2\sigma$, respectively, while the orange circle encloses 95\%~of the probability centered around the origin to indicate the 95\%~upper limit on~$A\fnlosc$. The green dotted lines visualize the constraints projected along the direction specified by the fiducial phase which then translates to the green circle displaying the corresponding limit on~$A\fnlosc$. We can clearly observe the dependence of these projected limits on the fiducial phase due to the highly degenerate nature of~$\Acos$ and~$\Asin$ for the considered scaling dimension, showing why the 95\%~upper limit~(orange circle) is a more robust indicator of the survey sensitivity to the overall non-Gaussian amplitude.}
	\label{fig:sigma-fnl-phase}
\end{figure}

The purpose of forecasting is to anticipate the behavior of our analysis of observed data and estimate the sensitivity of future surveys. One of the downsides of a Fisher approach however is that the uncertainties are effectively implemented as Gaussians around a fiducial value. This is important to keep in mind and one of the reasons why we employ the amplitude instead of the phase parametrization as mentioned in the main text. We explicitly show in Fig.~\ref{fig:sigma-fnl-phase},%
how the forecasted uncertainty for the amplitude(s) is affected by the fiducial value of the phase when characterizing the logarithmic oscillations in terms of an amplitude and a phase. Specifically, we display the two-dimensional contours in the~$(\Acos, \Asin)$ plane for $\alpha = 0$ and $\nu = 1$, as in Fig.~\ref{fig:acos-asin_delta-0}, but for five different values of the fiducial phase. The forecasts in this case are computed by setting the fiducial values $(\Acos^\mathrm{fid}, \Asin^\mathrm{fid})$ as given by the fiducial phase $\phi_\mathrm{fid} = \arctan(-\Asin^\mathrm{fid}/\Acos^\mathrm{fid})$ and overall amplitude. We then calculate the posterior distribution around that~(necessarily non-Gaussian) fiducial point in the~$(\Acos, \Asin)$ plane and convert the two-dimensional distribution to a 95\%~exclusion limit on~$A\fnlosc$ by taking the half-width of the $2\sigma$~contour along the direction specified by the fiducial phase. This is displayed by the green dotted lines and their corresponding circles. While the forecasted two-dimensional posterior for~$(\Acos, \Asin)$ is unaffected, the inferred 95\%~exclusion limits on the overall amplitude clearly vary with the fiducial phase and can therefore obfuscate the survey sensitivity by directly depending on its value. We consequently followed the method developed in~\cite{Beutler:2019ojk} to work in the amplitude parametrization and implicitly marginalize over the phase when deriving the 95\%~upper limits on the non-Gaussian amplitude~$A\fnlosc$~(orange circles). As an additional advantage, this also allows us to fiducially pick the Gaussian cosmology with $\Acos = \Asin = 0$ which would not have been possible with the phase parametrization.

\subsection*{Cosmology Dependence}

Our results depend on cosmology~(and astrophysics) directly through the signal and noise that enter the forecasts, and indirectly through choices in our analyses and parameters we marginalize over. The direct cosmological impact is often self-evident, while the indirect effects may go unnoticed. Given that we scan over a wide range of frequencies, with the signal moving in~$k$ with the scaling exponents, it may be unclear where the changes in sensitivity originate from.

\begin{figure}
	\centering
	\includegraphics{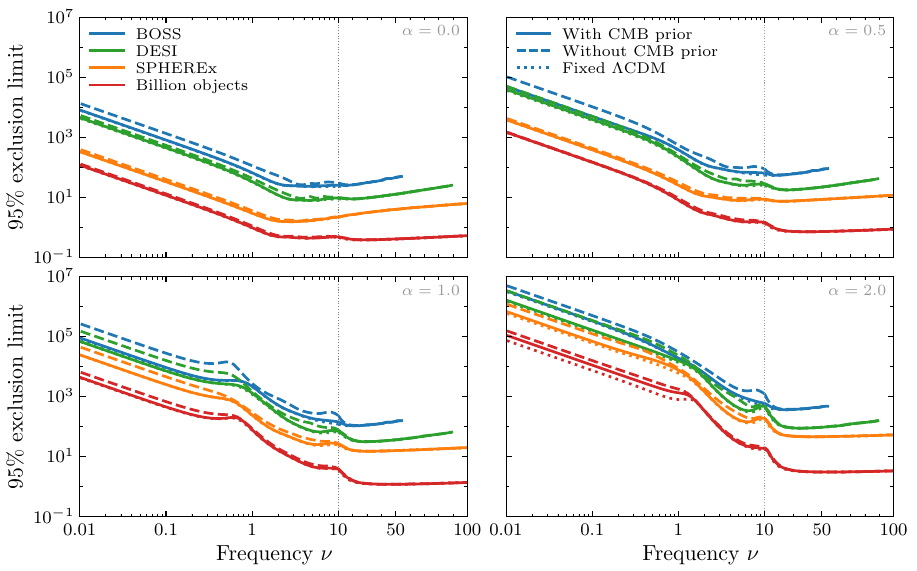}
	\caption{Effect of marginalizing over the cosmological parameters and imposing a CMB~prior on the forecasted 95\%~exclusion limits on the non-Gaussian amplitude~$A\fnlosc$ for four values of~$\alpha$ as a function of~$\nu$. The solid lines, which show the results when marginalizing over the $\Lambda\mathrm{CDM}$~parameters with a Planck~CMB~prior, are the same as in Fig.~\ref{fig:sigma-fnl-experiments}. The dashed and dotted lines display the limits without imposing this prior and when fixing the cosmological parameters to their fiducial values while still marginalizing over the bias parameters. The impact of the CMB~prior on the forecasts amounts to factors of~$O(1)$ for small frequencies~($\nu \lesssim 10$) and percent-level to negligible differences at high frequencies~($\nu \gtrsim 10$) as expected. The difference between imposing a CMB~prior and fixing the cosmological parameters is very small, except for the billion-object survey at large~$\alpha$ and small~$\nu$.}
	\label{fig:prior-impact}
\end{figure}

We illustrate the effects that marginalizing over the cosmological parameters and imposing priors from a Planck-like experiments have on the forecasted non-Gaussian limits in Fig.~\ref{fig:prior-impact}.%
We clearly see that forecasts for $\nu > 10$ are unaffected by marginalizing over~$\Lambda\mathrm{CDM}$ without any priors. This quantifies our expectation that the non-Gaussian signal at high frequencies is not degenerate with the smoother changes of the galaxy power spectrum due to cosmological parameters. Perhaps more surprising is that the impact of the Planck prior for $\nu < 1$ is not dramatic although visible up to factors of~$O(1)$. An important take-away for the analysis of BOSS~data in Section~\ref{sec:analysis} is that fixing the $\Lambda\mathrm{CDM}$~parameters to their fiducial values leads to essentially the same results as varying them with the mentioned CMB~prior. Given the goals of this work and the dramatic speed up in the statistical analysis of the data, we indeed adopted this approach as detailed in the main text. In fact, fixing the cosmological parameters and marginalizing over them with a Planck prior only lead to changes at the level of a few to~20\% depending on~$\alpha$ even for future surveys~(especially around $\nu = 10$ due to the discussed degeneracy with the scale of baryon acoustic oscillations), except for the billion-object survey at large~$\alpha$ and small~$\nu$ for which the forecasted limits are more visibly affected.

\begin{figure}
	\centering
	\includegraphics{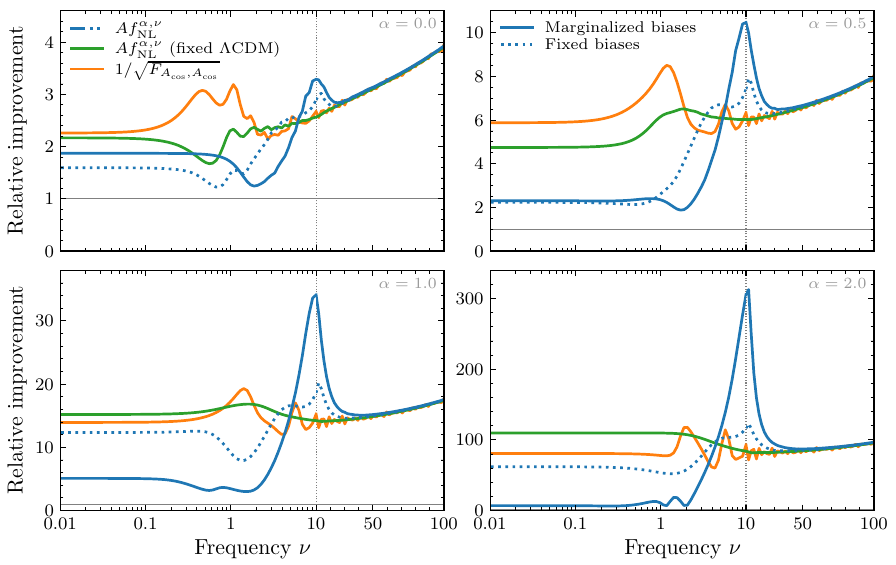}
	\caption{Relative improvement of the 95\%~upper limit on~$A\fnlosc$ for the billion-object survey when increasing the maximum wavenumber from our fiducial and conservative choice of $\kmax = k_\mathrm{halo}$ to the more optimistic $\kmax = k_\mathrm{NL}(z)$ as a function of frequency~$\nu$ for four scaling exponents. The solid blue line is the ratio of the lines displayed in Fig.~\ref{fig:kmax}, while the dotted blue line fixes instead of marginalizes over the bias parameters. For the green and orange curves, the biases were marginalized, but the~$\Lambda\mathrm{CDM}$ parameters and all parameters except for~$\Acos$ fixed, respectively. We can observe that the relative improvement is robust to these variations for $\nu \gtrsim 30$, shows the largest value for $\nu = O(10)$ due to the degeneracy with the baryon acoustic oscillations and exhibits the underlying degeneracies in the small-frequency regime, especially for increasing scaling exponent~$\alpha$.}
	\label{fig:kmax-ratio}
\end{figure}

In Fig.~\ref{fig:kmax-ratio},%
we investigate the impact of a variety of cosmological and astrophysical factors on our forecasts for the billion-object survey through the lens of the employed maximum wavenumber~$\kmax$ by fixing and varying the cosmological and bias parameters, respectively. If our forecasts are limited only by cosmic variance, then increasing the number of modes should dramatically improve the sensitivity. This relative improvement is seen consistently for $\nu > 30$ for all values of~$\alpha$, regardless of the details of the analysis. This is consistent with our expectation that the high-frequency regimes are robust to most degeneracies. In contrast, we notice a significant feature around $\nu = O(10)$ for the full forecasts that is absent when the $\Lambda\mathrm{CDM}$~parameters are held fixed. This feature is consistent with being the result of the degeneracy between~$\nu$ and the scale of baryon acoustic oscillations which corresponds to a linear frequency of~\SI{150}{Mpc}~(cf.~Fig.~\ref{fig:bandpower}). We also see many features around $\nu = O(1)$ due to degeneracies with both biasing and cosmological parameters. Overall, we clearly see that the high-frequency regime is very robust due to its oscillatory nature breaking degeneracies with the other effects that all result in smooth changes to the galaxy power spectrum, as is also evident at small frequencies.

\subsection*{Additional Forecasts}

\begin{figure}
	\centering
	\includegraphics{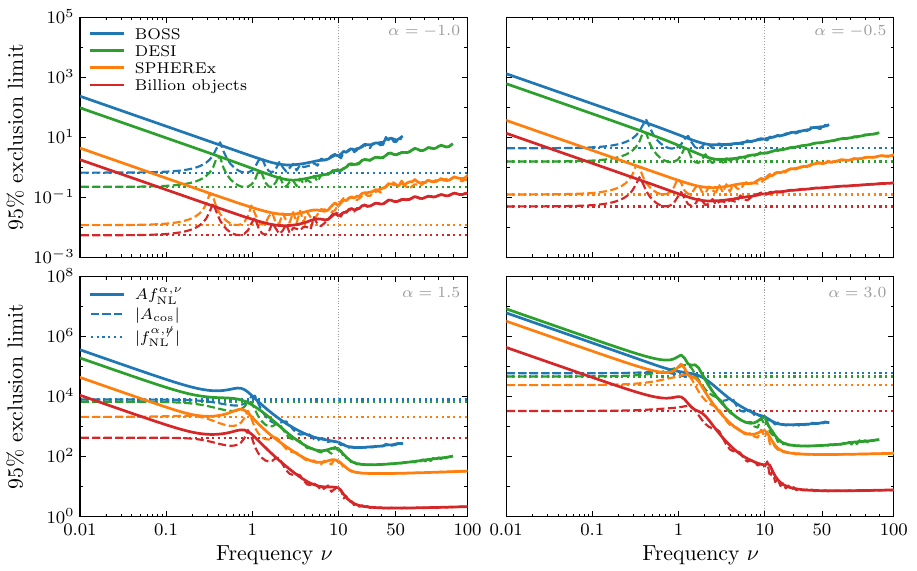}
	\caption{Forecasted constraints on the amplitude of primordial non-Gaussianity in terms of its 95\%~exclusion limit as a function of the oscillation frequency~$\nu$ for four additional values of the scaling exponent~$\alpha$ than in Fig.~\ref{fig:sigma-fnl-experiments}. We refer to the caption in the main text for the description and explanation of the displayed curves, but note that the range of limits differs here between the top and bottom panels by three orders of magnitude. These results extend the values of~$\alpha$ to the entire range investigated in this work and additionally complement the relative exclusion limits of the oscillatory with respect to the non-oscillatory signals presented in Fig.~\ref{fig:sigma-fnl-experiments_relative}.}
	\label{fig:sigma-fnl-experiments2}
\end{figure}

To conclude this appendix, we show the forecasted upper limits on~$A\fnlosc$ for four additional values of the scaling exponent~$\alpha$ in Fig.~\ref{fig:sigma-fnl-experiments2}%
to complement the results displayed in Figures~\ref{fig:sigma-fnl-experiments} and~\ref{fig:sigma-fnl-experiments_relative} in the main text. The negatives values of $\alpha = -1\text{ and }-0.5$ further highlight that the forecasts are generally worse for $\nu \neq 0$ than for $\nu = 0$ when the scale-dependent bias signal is dominant at small wavenumbers. This is compatible with the idea that the low-$k$ information is lost by marginalizing over the phase at small~$\nu$ without sufficient gains from the high-$k$ modes for which the signal vanishes, even at large~$\nu$. At the same time, the overall constraint on the non-Gaussian amplitude is tighter than for positive~$\alpha$, as expected. The cases of $\alpha = 1.5\text{ and }3$ have signals that peak at larger~$k$ where they can be distinguished from nonlinearities when $\nu > 10$. The case $\alpha =1.5$ is of particular note here because it corresponds to the classical cosmological-collider signal of a coupling to a single massive particle. Finally, we see that the improvement in forecasts from~BOSS through a billion-object survey are largely independent of the scaling exponent~$\alpha$. For $\nu > 10$, the improvement by roughly an order of magnitude is effectively independent of~$\alpha$, showing that the improvement in the sensitivity is not strongly affected by the nature of the~signal.

\clearpage
\section{Details for the BOSS~Analysis}
\label{app:analysis}

In this appendix, we collect additional information from our data analysis of the BOSS~DR12 dataset presented in Section~\ref{sec:analysis}. We present three additional figures to illustrate the posterior distributions of the oscillatory amplitudes and our inference of the 95\%~limits, the obtained exclusion limits on the oscillatory PNG~amplitude, and those on the non-oscillatory PNG~amplitude.\medskip

\begin{figure}[b!]
	\centering
	\includegraphics{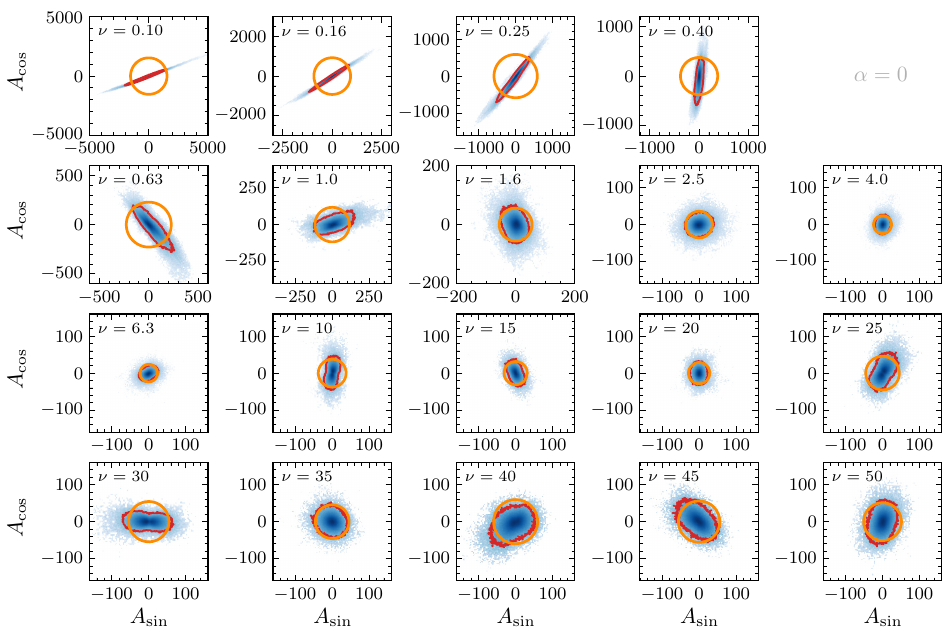}
	\caption{Posterior distributions of~$(\Acos, \Asin)$ for a scaling exponent of $\alpha = 0$ and the 19~frequencies~$\nu$ sampled in our analysis of BOSS~DR12~data. The panels show the pixelated posteriors based on the MCMC~samples marginalized over the bias and EFT~parameters, with their density increasing from light to dark blue. The red contours enclose the pixelated 95\%~confidence regions around the respective maximum posterior points while the orange circles contain 95\%~of the total probability around the origin $A\fnlosc = \Acos = \Asin = 0$ as obtained from the shown pixelated posteriors. As explained in~\textsection\ref{sec:analysis_details}, the radius of the latter therefore is the 95\%~exclusion limit of~$A\fnlosc$ which we employ to quantify and compare our results. We see that this method and our inferred limits well represent the posteriors for frequencies $\nu \gtrsim 1$ when the oscillatory amplitudes~$\Acos$ and~$\Asin$ are~(effectively) uncorrelated, but may underestimate the actual statistical power of~BOSS for smaller frequencies for which this correlation is strong.}
	\label{fig:data_acos-asin_delta-0}
\end{figure}

To provide additional insight into our parameter inference and our procedure for inferring the 95\%~exclusion limits on the overall PNG~amplitude~$A\fnlosc$, Fig.~\ref{fig:data_acos-asin_delta-0}%
shows all posterior distributions directly based on the MCMC~samples for the scaling exponent~$\alpha = 0$ in the plane of the oscillatory amplitudes~$\Acos$ and~$\Asin$. We observe the same rotating degeneracy line with increasing frequency~$\nu$ as predicted in our forecasts of Fig.~\ref{fig:acos-asin_delta-0}. For frequencies $\nu \gtrsim 1$, this degeneracy is effectively absent and the posterior is close to isotropic, while for $\nu \lesssim 1$ the degeneracy becomes very pronounced. As a result, the orange circle, whose radius is the 95\%~exclusion limit on~$A\fnlosc$ since it encloses 95\%~of the Monte Carlo samples~(see~\textsection\ref{sec:analysis_details} for details), provides an accurate summary of the statistical sensitivity of the survey to our PNG~signal at large frequencies, but underestimates the constraining power for small~$\nu$ in the form of weaker reported limits. We also note that the other eight values of~$\alpha$ exhibit qualitatively the same patterns and behaviors as shown here for $\alpha = 0$, with some scaling dimensions exhibiting non-Gaussian posteriors in~$(\Acos, \Asin)$.

\begin{figure}
	\centering
	\includegraphics{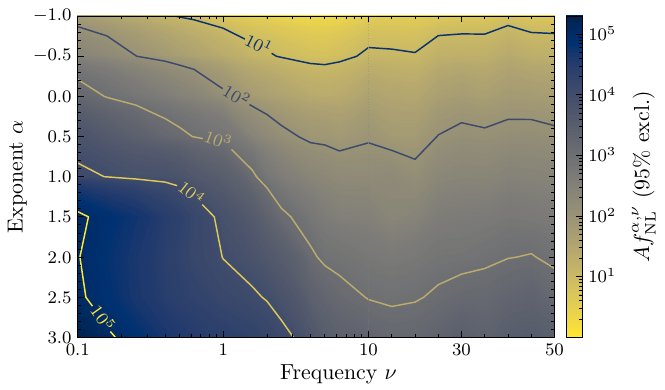}
	\caption{Exclusion limits at 95\%~c.l.\ on the non-Gaussian oscillatory amplitude~$A\fnlosc$ inferred from the BOSS~DR12 dataset as in Fig.~\ref{fig:data_limits_1d}, but shown as a function of both the scaling exponent~$\alpha$ and oscillatory frequency~$\nu$ of the non-Gaussian signature. The axes for~$\alpha$ and~$\nu$, and the colorbar for~$A\fnlosc$ are displayed on a linear, logarithmic-linear and logarithmic scale, respectively. We also show the contours of constant exclusion limit for every decade in the size of the amplitude. We can clearly observe how the constraining power of~BOSS decreases with increasing~$\alpha$ while larger oscillation frequencies help to distinguish and constrain the PNG~amplitude.}
	\label{fig:data_limits_2d}
\end{figure}

We presented the 95\%~exclusion limits on the oscillatory PNG~amplitude~$A\fnlosc$ in Fig.~\ref{fig:data_limits_1d} as a function of the frequency~$\nu$ for the nine sampled exponents~$\alpha$. In Figure~\ref{fig:data_limits_2d},%
we display these limits as a function of both~$\nu$ and~$\alpha$ providing an additional way of interpreting especially their qualitative behavior. The degradation of the inferred limits with decreasing frequency and increasing exponent is clearly visible both in the color gradient and in the contours of constant exclusion limit. This behavior reflects the interplay between the effective wavenumber range probed by the signal at a given signal-to-noise ratio of the survey and the ability of the non-Gaussian oscillations to break degeneracies with smooth bias contributions, as discussed in detail in the main text in the context of both the forecasts~(\textsection\ref{sec:Fisher}) and the data analysis~(\textsection\ref{sec:analysis_limits}).

\begin{figure}
	\centering
	\includegraphics{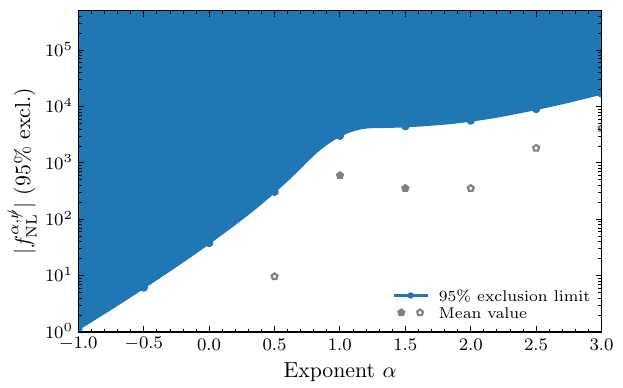}
	\caption{Exclusion limit at 95\%~c.l.\ on the absolute value of the non-Gaussian non-oscillatory amplitude~$|\fnlalpha|$ inferred from the BOSS~DR12 dataset as a function of the scaling exponent~$\alpha$. The filled~(unfilled) diamonds display the inferred positive~(negative) mean values of~$\fnlalpha$. This extends the range of scaling exponents from~$[0, 2]$ as studied in detail in~\cite{Green:2023uyz} to the range~$[-1, 3]$ considered in this work. For ease of comparison, the displayed range of values for~$\fnlalpha$ matches that of Fig.~\ref{fig:data_limits_1d}, with the mean values for $\alpha = -1.0, -0.5\text{ and } 0.0$ being factors of approximately~100, 110 and~220 smaller than their respective 95\%~exclusion limits.}
	\label{fig:data_nonoscillatory}
\end{figure}

Finally, we show the 95\%~exclusion limits on the absolute value of the non-oscillatory PNG~amplitude~$|\fnlalpha|$ as a function of the exponent~$\alpha$ in Fig.~\ref{fig:data_nonoscillatory}.%
Since we extended the range of scaling exponents studied in this work compared to~\cite{Green:2023uyz} from~$[0, 2]$ to~$[-1, 3]$~(see the main text for the phenomenological motivation), we reran the analysis of~\cite{Green:2023uyz} in order to compare the inferred limits for these two classes of primordial non-Gaussianity across the same range of exponents in Fig.~\ref{fig:data_relative_limits}. To be precise, we computed the upper limits on~$|\fnlalpha|$ at 95\%~c.l.\ so that they can be directly compared to the limits on the positive semi-definite oscillatory amplitude~$A\fnlosc$. These limits are essentially the same as the upper and lower $2\sigma$~constraints on~$\fnlalpha$ reported in~\cite{Green:2023uyz}~(with the exception of $\alpha = 3$, for which the difference is approximately~25\%). Across the extended regions $\alpha < 0$ and $\alpha > 2$, the inferred limits and mean values behave as expected from the more detailed and more densely sampled results of~\cite{Green:2023uyz}, with no indication of qualitatively new or unexpected behavior. This in particular includes a large correlation length in the scaling exponent~$\alpha$ and no detection of a nonzero PNG~signal.

\clearpage
\phantomsection
\addcontentsline{toc}{section}{References}
\bibliographystyle{utphys}
\bibliography{references}

\providecommand{\href}[2]{#2}\begingroup\raggedright\begin{thebibliography}{100}

\bibitem{Chang:2022lrw}
C.~Chang {\em et~al.}, ``{Report of the Topical Group on Cosmic Frontier~5 Dark
  Energy and Cosmic Acceleration: Cosmic Dawn and Before for Snowmass~2021}'',
  \href{http://arxiv.org/abs/2209.08265}{{\ttfamily arXiv:2209.08265
  [hep-ex]}}.

\bibitem{Green:2022bre}
D.~Green, ``{TASI Lectures on Cosmic Signals of Fundamental Physics}'',
  \href{http://arxiv.org/abs/2212.08685}{{\ttfamily arXiv:2212.08685
  [hep-ph]}}.

\bibitem{Achucarro:2022qrl}
A.~Ach\'ucarro {\em et~al.}, ``{Inflation: Theory and Observations}'',
  \href{http://arxiv.org/abs/2203.08128}{{\ttfamily arXiv:2203.08128
  [astro-ph.CO]}}.

\bibitem{Chen:2009we}
X.~Chen and Y.~Wang, ``{Large Non-Gaussianities with Intermediate Shapes from
  Quasi-Single-Field Inflation}'',
  \href{http://dx.doi.org/10.1103/PhysRevD.81.063511}{{\em Phys. Rev.~D}
  {\bfseries 81} (2010) 063511},
  \href{http://arxiv.org/abs/0909.0496}{{\ttfamily arXiv:0909.0496
  [astro-ph.CO]}}.

\bibitem{Chen:2009zp}
X.~Chen and Y.~Wang, ``{Quasi-Single-Field Inflation and Non-Gaussianities}'',
  \href{http://dx.doi.org/10.1088/1475-7516/2010/04/027}{{\em JCAP} {\bfseries
  04} (2010) 027}, \href{http://arxiv.org/abs/0911.3380}{{\ttfamily
  arXiv:0911.3380 [hep-th]}}.

\bibitem{Baumann:2011nk}
D.~Baumann and D.~Green, ``{Signatures of Supersymmetry from the Early
  Universe}'', \href{http://dx.doi.org/10.1103/PhysRevD.85.103520}{{\em Phys.
  Rev.~D} {\bfseries 85} (2012) 103520},
  \href{http://arxiv.org/abs/1109.0292}{{\ttfamily arXiv:1109.0292 [hep-th]}}.

\bibitem{Assassi:2012zq}
V.~Assassi, D.~Baumann, and D.~Green, ``{On Soft Limits of Inflationary
  Correlation Functions}'',
  \href{http://dx.doi.org/10.1088/1475-7516/2012/11/047}{{\em JCAP} {\bfseries
  11} (2012) 047}, \href{http://arxiv.org/abs/1204.4207}{{\ttfamily
  arXiv:1204.4207 [hep-th]}}.

\bibitem{Noumi:2012vr}
T.~Noumi, M.~Yamaguchi, and D.~Yokoyama, ``{Effective Field Theory Approach to
  Quasi-Single-Field Inflation and Effects of Heavy Fields}'',
  \href{http://dx.doi.org/10.1007/JHEP06(2013)051}{{\em JHEP} {\bfseries 06}
  (2013) 051}, \href{http://arxiv.org/abs/1211.1624}{{\ttfamily arXiv:1211.1624
  [hep-th]}}.

\bibitem{Arkani-Hamed:2015bza}
N.~Arkani-Hamed and J.~Maldacena, ``{Cosmological Collider Physics}'',
  \href{http://arxiv.org/abs/1503.08043}{{\ttfamily arXiv:1503.08043
  [hep-th]}}.

\bibitem{Lee:2016vti}
H.~Lee, D.~Baumann, and G.~Pimentel, ``{Non-Gaussianity as a Particle
  Detector}'', \href{http://dx.doi.org/10.1007/JHEP12(2016)040}{{\em JHEP}
  {\bfseries 12} (2016) 040}, \href{http://arxiv.org/abs/1607.03735}{{\ttfamily
  arXiv:1607.03735 [hep-th]}}.

\bibitem{Chen:2016uwp}
X.~Chen, Y.~Wang, and Z.-Z. Xianyu, ``{Standard Model Background of the
  Cosmological Collider}'',
  \href{http://dx.doi.org/10.1103/PhysRevLett.118.261302}{{\em Phys. Rev.
  Lett.} {\bfseries 118} (2017) 261302},
  \href{http://arxiv.org/abs/1610.06597}{{\ttfamily arXiv:1610.06597
  [hep-th]}}.

\bibitem{MoradinezhadDizgah:2017szk}
A.~Moradinezhad~Dizgah and C.~Dvorkin, ``{Scale-Dependent Galaxy Bias from
  Massive Particles with Spin during Inflation}'',
  \href{http://dx.doi.org/10.1088/1475-7516/2018/01/010}{{\em JCAP} {\bfseries
  01} (2018) 010}, \href{http://arxiv.org/abs/1708.06473}{{\ttfamily
  arXiv:1708.06473 [astro-ph.CO]}}.

\bibitem{Biagetti:2019bnp}
M.~Biagetti, ``{The Hunt for Primordial Interactions in the Large-Scale
  Structure of the Universe}'',
  \href{http://dx.doi.org/10.3390/galaxies7030071}{{\em Galaxies} {\bfseries 7}
  (2019) 71}, \href{http://arxiv.org/abs/1906.12244}{{\ttfamily
  arXiv:1906.12244 [astro-ph.CO]}}.

\bibitem{Kumar:2019ebj}
S.~Kumar and R.~Sundrum, ``{Cosmological Collider Physics and the Curvaton}'',
  \href{http://dx.doi.org/10.1007/JHEP04(2020)077}{{\em JHEP} {\bfseries 04}
  (2020) 077}, \href{http://arxiv.org/abs/1908.11378}{{\ttfamily
  arXiv:1908.11378 [hep-ph]}}.

\bibitem{Bodas:2020yho}
A.~Bodas, S.~Kumar, and R.~Sundrum, ``{The Scalar Chemical Potential in
  Cosmological Collider Physics}'',
  \href{http://dx.doi.org/10.1007/JHEP02(2021)079}{{\em JHEP} {\bfseries 02}
  (2021) 079}, \href{http://arxiv.org/abs/2010.04727}{{\ttfamily
  arXiv:2010.04727 [hep-ph]}}.

\bibitem{Lu:2021wxu}
Q.~Lu, M.~Reece, and Z.-Z. Xianyu, ``{Missing Scalars at the Cosmological
  Collider}'', \href{http://dx.doi.org/10.1007/JHEP12(2021)098}{{\em JHEP}
  {\bfseries 12} (2021) 098}, \href{http://arxiv.org/abs/2108.11385}{{\ttfamily
  arXiv:2108.11385 [hep-ph]}}.

\bibitem{Chakraborty:2023qbp}
P.~Chakraborty and J.~Stout, ``{Light Scalars at the Cosmological Collider}'',
  \href{http://dx.doi.org/10.1007/JHEP02(2024)021}{{\em JHEP} {\bfseries 02}
  (2024) 021}, \href{http://arxiv.org/abs/2310.01494}{{\ttfamily
  arXiv:2310.01494 [hep-th]}}.

\bibitem{Chakraborty:2023eoq}
P.~Chakraborty and J.~Stout, ``{Compact Scalars at the Cosmological
  Collider}'', \href{http://dx.doi.org/10.1007/JHEP03(2024)149}{{\em JHEP}
  {\bfseries 03} (2024) 149}, \href{http://arxiv.org/abs/2311.09219}{{\ttfamily
  arXiv:2311.09219 [hep-th]}}.

\bibitem{Goldstein:2024bky}
S.~Goldstein, O.~Philcox, J.~C. Hill, and L.~Hui, ``{Intermediate Mass-Range
  Particles from Small Scales: Nonperturbative Techniques for Cosmological
  Collider Physics from Large-Scale Structure Surveys}'',
  \href{http://dx.doi.org/10.1103/PhysRevD.110.083516}{{\em Phys. Rev. D}
  {\bfseries 110} (2024) 083516},
  \href{http://arxiv.org/abs/2407.08731}{{\ttfamily arXiv:2407.08731
  [astro-ph.CO]}}.

\bibitem{Chakraborty:2025myb}
P.~Chakraborty, ``{Primordial Non-Gaussianity from Light Compact Scalars}'',
  \href{http://dx.doi.org/10.1007/JHEP11(2025)023}{{\em JHEP} {\bfseries 11}
  (2025) 023}, \href{http://arxiv.org/abs/2501.07672}{{\ttfamily
  arXiv:2501.07672 [hep-th]}}.

\bibitem{Goldstein:2025eyj}
S.~Goldstein, O.~Philcox, E.~Fondi, and W.~Coulton, ``{Wonderings on Wiggly
  Bispectra: Nonlinear Evolution and Reconstruction of Oscillations in the
  Squeezed Bispectrum}'', \href{http://dx.doi.org/10.1103/q8m3-vspd}{{\em Phys.
  Rev. D} {\bfseries 112} (2025) 083503},
  \href{http://arxiv.org/abs/2505.13443}{{\ttfamily arXiv:2505.13443
  [astro-ph.CO]}}.

\bibitem{Maldacena:2002vr}
J.~Maldacena, ``{Non-Gaussian Features of Primordial Fluctuations in
  Single-Field Inflationary Models}'',
  \href{http://dx.doi.org/10.1088/1126-6708/2003/05/013}{{\em JHEP} {\bfseries
  05} (2003) 013}, \href{http://arxiv.org/abs/astro-ph/0210603}{{\ttfamily
  arXiv:astro-ph/0210603}}.

\bibitem{Creminelli:2004yq}
P.~Creminelli and M.~Zaldarriaga, ``{Single-Field Consistency Relation for the
  Three-Point Function}'',
  \href{http://dx.doi.org/10.1088/1475-7516/2004/10/006}{{\em JCAP} {\bfseries
  10} (2004) 006}, \href{http://arxiv.org/abs/astro-ph/0407059}{{\ttfamily
  arXiv:astro-ph/0407059}}.

\bibitem{Green:2023uyz}
D.~Green, Y.~Guo, J.~Han, and B.~Wallisch, ``{Light Fields during Inflation
  from~BOSS and Future Galaxy Surveys}'',
  \href{http://dx.doi.org/10.1088/1475-7516/2024/05/090}{{\em JCAP} {\bfseries
  05} (2024) 090}, \href{http://arxiv.org/abs/2311.04882}{{\ttfamily
  arXiv:2311.04882 [astro-ph.CO]}}.

\bibitem{Cabass:2024wob}
G.~Cabass, O.~Philcox, M.~Ivanov, K.~Akitsu, S.-F. Chen, M.~Simonovi\'c, and
  M.~Zaldarriaga, ``{BOSS~Constraints on Massive Particles during Inflation:
  The Cosmological Collider in Action}'',
  \href{http://dx.doi.org/10.1103/PhysRevD.111.063510}{{\em Phys. Rev. D}
  {\bfseries 111} (2025) 063510},
  \href{http://arxiv.org/abs/2404.01894}{{\ttfamily arXiv:2404.01894
  [astro-ph.CO]}}.

\bibitem{Sohn:2024xzd}
W.~Sohn, D.-G. Wang, J.~Fergusson, and E.~P.~S. Shellard, ``{Searching for
  Cosmological Collider in the Planck CMB~Data}'',
  \href{http://dx.doi.org/10.1088/1475-7516/2024/09/016}{{\em JCAP} {\bfseries
  09} (2024) 016}, \href{http://arxiv.org/abs/2404.07203}{{\ttfamily
  arXiv:2404.07203 [astro-ph.CO]}}.

\bibitem{Suman:2025vuf}
P.~Suman, D.-G. Wang, W.~Sohn, J.~Fergusson, and E.~P.~S. Shellard, ``{How
  Significant Are Cosmological Collider Signals in the Planck Data?}'',
  \href{http://arxiv.org/abs/2511.17500}{{\ttfamily arXiv:2511.17500
  [astro-ph.CO]}}.

\bibitem{Philcox:2025bbo}
O.~Philcox, K.~Zhong, and S.~S. Sirletti, ``{Separating the Inseparable:
  Constraining Arbitrary Primordial Bispectra with Cosmic Microwave Background
  Data}'', \href{http://arxiv.org/abs/2511.19179}{{\ttfamily arXiv:2511.19179
  [astro-ph.CO]}}.

\bibitem{Suman:2025tpv}
P.~Suman, D.-G. Wang, W.~Sohn, J.~Fergusson, and E.~P.~S. Shellard,
  ``{Searching for Cosmological Collider in the Planck CMB~Data~II: Collider
  Templates and Modal Analysis}'',
  \href{http://arxiv.org/abs/2512.22085}{{\ttfamily arXiv:2512.22085
  [astro-ph.CO]}}.

\bibitem{McAneny:2019epy}
M.~McAneny and A.~Ridgway, ``{New Shapes of Primordial Non-Gaussianity from
  Quasi-Single-Field Inflation with Multiple Isocurvatons}'',
  \href{http://dx.doi.org/10.1103/PhysRevD.100.043534}{{\em Phys. Rev.~D}
  {\bfseries 100} (2019) 043534},
  \href{http://arxiv.org/abs/1903.11607}{{\ttfamily arXiv:1903.11607
  [astro-ph.CO]}}.

\bibitem{Green:2023ids}
D.~Green, Y.~Huang, C.-H. Shen, and D.~Baumann, ``{Positivity from Cosmological
  Correlators}'', \href{http://dx.doi.org/10.1007/JHEP04(2024)034}{{\em JHEP}
  {\bfseries 04} (2024) 034}, \href{http://arxiv.org/abs/2310.02490}{{\ttfamily
  arXiv:2310.02490 [hep-th]}}.

\bibitem{Planck:2018vyg}
{N. Aghanim \textit{et al.} (Planck Collaboration)}, ``{Planck 2018 Results.
  VI.~Cosmological Parameters}'',
  \href{http://dx.doi.org/10.1051/0004-6361/201833910}{{\em Astron. Astrophys.}
  {\bfseries 641} (2020) A6}, \href{http://arxiv.org/abs/1807.06209}{{\ttfamily
  arXiv:1807.06209 [astro-ph.CO]}}.

\bibitem{Planck:2018jri}
{Y. Akrami \textit{et al.} (Planck Collaboration)}, ``{Planck 2018 Results.
  X.~Constraints on Inflation}'',
  \href{http://dx.doi.org/10.1051/0004-6361/201833887}{{\em Astron. Astrophys.}
  {\bfseries 641} (2020) A10},
  \href{http://arxiv.org/abs/1807.06211}{{\ttfamily arXiv:1807.06211
  [astro-ph.CO]}}.

\bibitem{Planck:2019kim}
{Y. Akrami \textit{et al.} (Planck Collaboration)}, ``{Planck 2018 Results.
  IX.~Constraints on Primordial Non-Gaussianity}'',
  \href{http://dx.doi.org/10.1051/0004-6361/201935891}{{\em Astron. Astrophys.}
  {\bfseries 641} (2020) A9}, \href{http://arxiv.org/abs/1905.05697}{{\ttfamily
  arXiv:1905.05697 [astro-ph.CO]}}.

\bibitem{AtacamaCosmologyTelescope:2025blo}
{T. Louis \textit{et al.} (ACT~Collaboration)}, ``{The Atacama Cosmology
  Telescope: DR6~Power Spectra, Likelihoods and $\Lambda$CDM~Parameters}'',
  \href{http://dx.doi.org/10.1088/1475-7516/2025/11/062}{{\em JCAP} {\bfseries
  11} (2025) 062}, \href{http://arxiv.org/abs/2503.14452}{{\ttfamily
  arXiv:2503.14452 [astro-ph.CO]}}.

\bibitem{SPT-3G:2025bzu}
{E. Camphuis \textit{et al.} (SPT-3G~Collaboration)}, ``{SPT-3G~D1:
  CMB~Temperature and Polarization Power Spectra and Cosmology from~2019 and
  2020~Observations of the SPT-3G~Main Field}'',
  \href{http://arxiv.org/abs/2506.20707}{{\ttfamily arXiv:2506.20707
  [astro-ph.CO]}}.

\bibitem{Babich:2004gb}
D.~Babich, P.~Creminelli, and M.~Zaldarriaga, ``{The Shape of
  Non-Gaussianities}'',
  \href{http://dx.doi.org/10.1088/1475-7516/2004/08/009}{{\em JCAP} {\bfseries
  08} (2004) 009}, \href{http://arxiv.org/abs/astro-ph/0405356}{{\ttfamily
  arXiv:astro-ph/0405356}}.

\bibitem{Smith:2006ud}
K.~Smith and M.~Zaldarriaga, ``{Algorithms for Bispectra: Forecasting, Optimal
  Analysis and Simulation}'',
  \href{http://dx.doi.org/10.1111/j.1365-2966.2010.18175.x}{{\em Mon. Not. Roy.
  Astron. Soc.} {\bfseries 417} (2011) 2},
  \href{http://arxiv.org/abs/astro-ph/0612571}{{\ttfamily
  arXiv:astro-ph/0612571}}.

\bibitem{Senatore:2009gt}
L.~Senatore, K.~Smith, and M.~Zaldarriaga, ``{Non-Gaussianities in Single-Field
  Inflation and Their Optimal Limits from the WMAP 5-Year Data}'',
  \href{http://dx.doi.org/10.1088/1475-7516/2010/01/028}{{\em JCAP} {\bfseries
  01} (2010) 028}, \href{http://arxiv.org/abs/0905.3746}{{\ttfamily
  arXiv:0905.3746 [astro-ph.CO]}}.

\bibitem{Flauger:2009ab}
R.~Flauger, L.~McAllister, E.~Pajer, A.~Westphal, and G.~Xu, ``{Oscillations in
  the~CMB from Axion Monodromy Inflation}'',
  \href{http://dx.doi.org/10.1088/1475-7516/2010/06/009}{{\em JCAP} {\bfseries
  06} (2010) 009}, \href{http://arxiv.org/abs/0907.2916}{{\ttfamily
  arXiv:0907.2916 [hep-th]}}.

\bibitem{Flauger:2010ja}
R.~Flauger and E.~Pajer, ``{Resonant Non-Gaussianity}'',
  \href{http://dx.doi.org/10.1088/1475-7516/2011/01/017}{{\em JCAP} {\bfseries
  01} (2011) 017}, \href{http://arxiv.org/abs/1002.0833}{{\ttfamily
  arXiv:1002.0833 [hep-th]}}.

\bibitem{Smith:2015uia}
K.~Smith, L.~Senatore, and M.~Zaldarriaga, ``{Optimal Analysis of the
  CMB~Trispectrum}'', \href{http://arxiv.org/abs/1502.00635}{{\ttfamily
  arXiv:1502.00635 [astro-ph.CO]}}.

\bibitem{Philcox:2025wts}
O.~Philcox, ``{Searching for Inflationary Physics with the CMB~Trispectrum.
  III.~Constraints from Planck}'',
  \href{http://dx.doi.org/10.1103/y81z-g7th}{{\em Phys. Rev. D} {\bfseries 111}
  (2025) 123534}, \href{http://arxiv.org/abs/2502.06931}{{\ttfamily
  arXiv:2502.06931 [astro-ph.CO]}}.

\bibitem{Chang:2008gj}
S.~Chang, M.~Kleban, and T.~Levi, ``{Watching Worlds Collide: Effects on the
  CMB from Cosmological Bubble Collisions}'',
  \href{http://dx.doi.org/10.1088/1475-7516/2009/04/025}{{\em JCAP} {\bfseries
  04} (2009) 025}, \href{http://arxiv.org/abs/0810.5128}{{\ttfamily
  arXiv:0810.5128 [hep-th]}}.

\bibitem{Feeney:2010dd}
S.~Feeney, M.~Johnson, D.~Mortlock, and H.~Peiris, ``{First Observational Tests
  of Eternal Inflation: Analysis Methods and WMAP 7-Year Results}'',
  \href{http://dx.doi.org/10.1103/PhysRevD.84.043507}{{\em Phys. Rev. D}
  {\bfseries 84} (2011) 043507},
  \href{http://arxiv.org/abs/1012.3667}{{\ttfamily arXiv:1012.3667
  [astro-ph.CO]}}.

\bibitem{Munchmeyer:2019wlh}
M.~M\"unchmeyer and K.~Smith, ``{Higher $N$-Point Function Data Analysis
  Techniques for Heavy Particle Production and WMAP~Results}'',
  \href{http://dx.doi.org/10.1103/PhysRevD.100.123511}{{\em Phys. Rev.~D}
  {\bfseries 100} (2019) 123511},
  \href{http://arxiv.org/abs/1910.00596}{{\ttfamily arXiv:1910.00596
  [astro-ph.CO]}}.

\bibitem{Kim:2023wuk}
T.~Kim, J.~H. Kim, S.~Kumar, A.~Martin, M.~M\"unchmeyer, and Y.~Tsai,
  ``{Probing Cosmological Particle Production and Pairwise Hotspots with Deep
  Neural Networks}'', \href{http://dx.doi.org/10.1103/PhysRevD.108.043525}{{\em
  Phys. Rev. D} {\bfseries 108} (2023) 043525},
  \href{http://arxiv.org/abs/2303.08869}{{\ttfamily arXiv:2303.08869
  [hep-ph]}}.

\bibitem{Philcox:2024jpd}
O.~Philcox, S.~Kumar, and J.~C. Hill, ``{Searching for Inflationary Particle
  Production in Planck Data}'',
  \href{http://dx.doi.org/10.1103/PhysRevD.111.103523}{{\em Phys. Rev. D}
  {\bfseries 111} (2025) 103523},
  \href{http://arxiv.org/abs/2405.03738}{{\ttfamily arXiv:2405.03738
  [astro-ph.CO]}}.

\bibitem{DESI:2022lza}
{D. Schlegel \textit{et al.} (DESI Collaboration)}, ``{A Spectroscopic Road Map
  for Cosmic Frontier: DESI, DESI-II, Stage-5}'',
  \href{http://arxiv.org/abs/2209.03585}{{\ttfamily arXiv:2209.03585
  [astro-ph.CO]}}.

\bibitem{Green:2022hhj}
D.~Green {\em et~al.}, ``{Snowmass Theory Frontier: Astrophysics and
  Cosmology}'', \href{http://arxiv.org/abs/2209.06854}{{\ttfamily
  arXiv:2209.06854 [hep-ph]}}.

\bibitem{Baumann:2010tm}
D.~Baumann, A.~Nicolis, L.~Senatore, and M.~Zaldarriaga, ``{Cosmological
  Nonlinearities as an Effective Fluid}'',
  \href{http://dx.doi.org/10.1088/1475-7516/2012/07/051}{{\em JCAP} {\bfseries
  07} (2012) 051}, \href{http://arxiv.org/abs/1004.2488}{{\ttfamily
  arXiv:1004.2488 [astro-ph.CO]}}.

\bibitem{Cabass:2022avo}
G.~Cabass, M.~Ivanov, M.~Lewandowski, M.~Mirbabayi, and M.~Simonovi\'c,
  ``{Snowmass White Paper: Effective Field Theories in Cosmology}'',
  \href{http://dx.doi.org/10.1016/j.dark.2023.101193}{{\em Phys. Dark Univ.}
  {\bfseries 40} (2023) 101193},
  \href{http://arxiv.org/abs/2203.08232}{{\ttfamily arXiv:2203.08232
  [astro-ph.CO]}}.

\bibitem{Ivanov:2022mrd}
M.~Ivanov, ``{Effective Field Theory for Large-Scale Structure}'',
  \href{http://arxiv.org/abs/2212.08488}{{\ttfamily arXiv:2212.08488
  [astro-ph.CO]}}.

\bibitem{Jasche:2012kq}
J.~Jasche and B.~Wandelt, ``{Bayesian Physical Reconstruction of Initial
  Conditions from Large-Scale-Structure Surveys}'',
  \href{http://dx.doi.org/10.1093/mnras/stt449}{{\em Mon. Not. Roy. Astron.
  Soc.} {\bfseries 432} (2013) 894},
  \href{http://arxiv.org/abs/1203.3639}{{\ttfamily arXiv:1203.3639
  [astro-ph.CO]}}.

\bibitem{Ramanah:2018eed}
D.~K. Ramanah, G.~Lavaux, J.~Jasche, and B.~Wandelt, ``{Cosmological Inference
  from Bayesian Forward Modelling of Deep Galaxy Redshift Surveys}'',
  \href{http://dx.doi.org/10.1051/0004-6361/201834117}{{\em Astron. Astrophys.}
  {\bfseries 621} (2019) A69},
  \href{http://arxiv.org/abs/1808.07496}{{\ttfamily arXiv:1808.07496
  [astro-ph.CO]}}.

\bibitem{Villaescusa-Navarro:2021cni}
F.~Villaescusa-Navarro {\em et~al.}, ``{Robust Marginalization of Baryonic
  Effects for Cosmological Inference at the Field Level}'',
  \href{http://arxiv.org/abs/2109.10360}{{\ttfamily arXiv:2109.10360
  [astro-ph.CO]}}.

\bibitem{Stopyra:2023yqm}
S.~Stopyra, H.~Peiris, A.~Pontzen, J.~Jasche, and G.~Lavaux, ``{Towards
  Accurate Field-Level Inference of Massive Cosmic Structures}'',
  \href{http://dx.doi.org/10.1093/mnras/stad3170}{{\em Mon. Not. Roy. Astron.
  Soc.} {\bfseries 527} (2023) 1244},
  \href{http://arxiv.org/abs/2304.09193}{{\ttfamily arXiv:2304.09193
  [astro-ph.CO]}}.

\bibitem{Doeser:2023yzv}
L.~Doeser, D.~Jamieson, S.~Stopyra, G.~Lavaux, F.~Leclercq, and J.~Jasche,
  ``{Bayesian Inference of Initial Conditions from Nonlinear Cosmic Structures
  using Field-Level Emulators}'',
  \href{http://dx.doi.org/10.1093/mnras/stae2429}{{\em Mon. Not. Roy. Astron.
  Soc.} {\bfseries 535} (2024) 1258},
  \href{http://arxiv.org/abs/2312.09271}{{\ttfamily arXiv:2312.09271
  [astro-ph.CO]}}.

\bibitem{Bairagi:2025sux}
A.~Bairagi, B.~Wandelt, and F.~Villaescusa-Navarro, ``{The \textsc{Big Sobol
  Sequence}: How Many Simulations Do We Need for Simulation-Based Inference in
  Cosmology?}'', \href{http://dx.doi.org/10.1051/0004-6361/202554602}{{\em
  Astron. Astrophys.} {\bfseries 703} (2025) A301},
  \href{http://arxiv.org/abs/2503.13755}{{\ttfamily arXiv:2503.13755
  [astro-ph.CO]}}.

\bibitem{McAlpine:2025uzh}
S.~McAlpine, J.~Jasche, M.~Ata, G.~Lavaux, R.~Stiskalek, C.~Frenk, and
  A.~Jenkins, ``{The Manticore Project~I: A Digital Twin of Our Cosmic
  Neighbourhood from Bayesian Field-Level Analysis}'',
  \href{http://dx.doi.org/10.1093/mnras/staf767}{{\em Mon. Not. Roy. Astron.
  Soc.} {\bfseries 540} (2025) 716},
  \href{http://arxiv.org/abs/2505.10682}{{\ttfamily arXiv:2505.10682
  [astro-ph.CO]}}.

\bibitem{Schmidt:2018bkr}
F.~Schmidt, F.~Elsner, J.~Jasche, N.~M. Nguyen, and G.~Lavaux, ``{A Rigorous
  EFT-Based Forward Model for Large-Scale Structure}'',
  \href{http://dx.doi.org/10.1088/1475-7516/2019/01/042}{{\em JCAP} {\bfseries
  01} (2019) 042}, \href{http://arxiv.org/abs/1808.02002}{{\ttfamily
  arXiv:1808.02002 [astro-ph.CO]}}.

\bibitem{Kokron:2021xgh}
N.~Kokron, J.~DeRose, S.-F. Chen, M.~White, and R.~Wechsler, ``{The Cosmology
  Dependence of Galaxy Clustering and Lensing from a Hybrid
  $N$-Body-Perturbation-Theory Model}'',
  \href{http://dx.doi.org/10.1093/mnras/stab1358}{{\em Mon. Not. Roy. Astron.
  Soc.} {\bfseries 505} (2021) 1422},
  \href{http://arxiv.org/abs/2101.11014}{{\ttfamily arXiv:2101.11014
  [astro-ph.CO]}}.

\bibitem{Pellejero-Ibanez:2022efv}
M.~Pellejero-Ibanez, R.~Angulo, M.~Zennaro, J.~Stuecker, S.~Contreras,
  G.~Arico, and F.~Maion, ``{The BACCO~Simulation Project: BACCO~Hybrid
  Lagrangian Bias Expansion Model in Redshift Space}'',
  \href{http://dx.doi.org/10.1093/mnras/stad368}{{\em Mon. Not. Roy. Astron.
  Soc.} {\bfseries 520} (2023) 3725},
  \href{http://arxiv.org/abs/2207.06437}{{\ttfamily arXiv:2207.06437
  [astro-ph.CO]}}.

\bibitem{LSSTDarkEnergyScience:2023qfp}
{A. Nicola \textit{et al.} (LSST~Dark Energy Science~Collaboration)}, ``{Galaxy
  Bias in the Era of~LSST: Perturbative Bias Expansions}'',
  \href{http://dx.doi.org/10.1088/1475-7516/2024/02/015}{{\em JCAP} {\bfseries
  02} (2024) 015}, \href{http://arxiv.org/abs/2307.03226}{{\ttfamily
  arXiv:2307.03226 [astro-ph.CO]}}.

\bibitem{Ivanov:2024xgb}
M.~Ivanov, A.~Obuljen, C.~Cuesta-Lazaro, and M.~Toomey, ``{Full-Shape Analysis
  with Simulation-Based Priors: Cosmological Parameters and the
  Structure-Growth Anomaly}'',
  \href{http://dx.doi.org/10.1103/PhysRevD.111.063548}{{\em Phys. Rev. D}
  {\bfseries 111} (2025) 063548},
  \href{http://arxiv.org/abs/2409.10609}{{\ttfamily arXiv:2409.10609
  [astro-ph.CO]}}.

\bibitem{Dalal:2007cu}
N.~Dalal, O.~Dor\'e, D.~Huterer, and A.~Shirokov, ``{The Imprints of Primordial
  Non-Gaussianities on Large-Scale Structure: Scale-Dependent Bias and
  Abundance of Virialized Objects}'',
  \href{http://dx.doi.org/10.1103/PhysRevD.77.123514}{{\em Phys. Rev.~D}
  {\bfseries 77} (2008) 123514},
  \href{http://arxiv.org/abs/0710.4560}{{\ttfamily arXiv:0710.4560
  [astro-ph]}}.

\bibitem{Meerburg:2019qqi}
P.~D. Meerburg {\em et~al.}, ``{Primordial Non-Gaussianity}'', {\em
  \href{https://baas.aas.org/pub/2020n3i107}{Bull. Am. Astron. Soc.}}
  {\bfseries \href{https://baas.aas.org/pub/2020n3i107}{51}}
  (\href{https://baas.aas.org/pub/2020n3i107}{2019})
  \href{https://baas.aas.org/pub/2020n3i107}{107},
  \href{http://arxiv.org/abs/1903.04409}{{\ttfamily arXiv:1903.04409
  [astro-ph.CO]}}.

\bibitem{SPHEREx:2014bgr}
{O. Dor\'e \textit{et al.} (SPHEREx Collaboration)}, ``{Cosmology with the
  SPHEREX All-Sky Spectral Survey}'',
  \href{http://arxiv.org/abs/1412.4872}{{\ttfamily arXiv:1412.4872
  [astro-ph.CO]}}.

\bibitem{Heinrich:2023qaa}
C.~Heinrich, O.~Dore, and E.~Krause, ``{Measuring~$\fnl^\mathrm{loc}$ with the
  SPHEREx~Multitracer Redshift-Space Bispectrum}'',
  \href{http://dx.doi.org/10.1103/PhysRevD.109.123511}{{\em Phys. Rev. D}
  {\bfseries 109} (2024) 123511},
  \href{http://arxiv.org/abs/2311.13082}{{\ttfamily arXiv:2311.13082
  [astro-ph.CO]}}.

\bibitem{Shiveshwarkar:2023afl}
C.~Shiveshwarkar, T.~Brinckmann, and M.~Loverde, ``{Constraining Multi-Field
  Inflation using the SPHEREx All-Sky Survey Power Spectra}'',
  \href{http://dx.doi.org/10.1088/1475-7516/2024/05/094}{{\em JCAP} {\bfseries
  05} (2024) 094}, \href{http://arxiv.org/abs/2312.15038}{{\ttfamily
  arXiv:2312.15038 [astro-ph.CO]}}.

\bibitem{Chen:2006xjb}
X.~Chen, R.~Easther, and E.~Lim, ``{Large Non-Gaussianities in Single-Field
  Inflation}'', \href{http://dx.doi.org/10.1088/1475-7516/2007/06/023}{{\em
  JCAP} {\bfseries 06} (2007) 023},
  \href{http://arxiv.org/abs/astro-ph/0611645}{{\ttfamily
  arXiv:astro-ph/0611645}}.

\bibitem{Slosar:2019gvt}
A.~Slosar {\em et~al.}, ``{Scratches from the Past: Inflationary Archaeology
  through Features in the Power Spectrum of Primordial Fluctuations}'', {\em
  \href{https://baas.aas.org/pub/2020n3i098}{Bull. Am. Astron. Soc.}}
  {\bfseries \href{https://baas.aas.org/pub/2020n3i098}{51}}
  (\href{https://baas.aas.org/pub/2020n3i098}{2019})
  \href{https://baas.aas.org/pub/2020n3i098}{98},
  \href{http://arxiv.org/abs/1903.09883}{{\ttfamily arXiv:1903.09883
  [astro-ph.CO]}}.

\bibitem{Beutler:2019ojk}
F.~Beutler, M.~Biagetti, D.~Green, A.~Slosar, and B.~Wallisch, ``{Primordial
  Features from Linear to Nonlinear Scales}'',
  \href{http://dx.doi.org/10.1103/PhysRevResearch.1.033209}{{\em Phys. Rev.
  Res.} {\bfseries 1} (2019) 033209},
  \href{http://arxiv.org/abs/1906.08758}{{\ttfamily arXiv:1906.08758
  [astro-ph.CO]}}.

\bibitem{Vlah:2015zda}
Z.~Vlah, U.~Seljak, M.~Y. Chu, and Y.~Feng, ``{Perturbation Theory, Effective
  Field Theory and Oscillations in the Power Spectrum}'',
  \href{http://dx.doi.org/10.1088/1475-7516/2016/03/057}{{\em JCAP} {\bfseries
  03} (2016) 057}, \href{http://arxiv.org/abs/1509.02120}{{\ttfamily
  arXiv:1509.02120 [astro-ph.CO]}}.

\bibitem{Vasudevan:2019ewf}
A.~Vasudevan, M.~Ivanov, S.~Sibiryakov, and J.~Lesgourgues, ``{Time-Sliced
  Perturbation Theory with Primordial Non-Gaussianity and Effects of Large Bulk
  Flows on Inflationary Oscillating Features}'',
  \href{http://dx.doi.org/10.1088/1475-7516/2019/09/037}{{\em JCAP} {\bfseries
  09} (2019) 037}, \href{http://arxiv.org/abs/1906.08697}{{\ttfamily
  arXiv:1906.08697 [astro-ph.CO]}}.

\bibitem{Li:2021jvz}
Y.~Li, H.-M. Zhu, and B.~Li, ``{Nonlinear Reconstruction of Features in the
  Primordial Power Spectrum from Large-Scale Structure}'',
  \href{http://dx.doi.org/10.1093/mnras/stac1544}{{\em Mon. Not. Roy. Astron.
  Soc.} {\bfseries 514} (2022) 4363},
  \href{http://arxiv.org/abs/2102.09007}{{\ttfamily arXiv:2102.09007
  [astro-ph.CO]}}.

\bibitem{Chen:2020ckc}
S.-F. Chen, Z.~Vlah, and M.~White, ``{Modeling Features in the Redshift-Space
  Halo Power Spectrum with Perturbation Theory}'',
  \href{http://dx.doi.org/10.1088/1475-7516/2020/11/035}{{\em JCAP} {\bfseries
  11} (2020) 035}, \href{http://arxiv.org/abs/2007.00704}{{\ttfamily
  arXiv:2007.00704 [astro-ph.CO]}}.

\bibitem{Ballardini:2024dto}
M.~Ballardini and N.~Barbieri, ``{Refining the Nonlinear Modelling of
  Primordial Oscillatory Features}'',
  \href{http://dx.doi.org/10.1088/1475-7516/2025/05/059}{{\em JCAP} {\bfseries
  05} (2025) 059}, \href{http://arxiv.org/abs/2411.02261}{{\ttfamily
  arXiv:2411.02261 [astro-ph.CO]}}.

\bibitem{Ballardini:2022wzu}
M.~Ballardini, F.~Finelli, F.~Marulli, L.~Moscardini, and A.~Veropalumbo,
  ``{New Constraints on Primordial Features from the Galaxy Two-Point
  Correlation Function}'',
  \href{http://dx.doi.org/10.1103/PhysRevD.107.043532}{{\em Phys. Rev. D}
  {\bfseries 107} (2023) 043532},
  \href{http://arxiv.org/abs/2202.08819}{{\ttfamily arXiv:2202.08819
  [astro-ph.CO]}}.

\bibitem{Mergulhao:2023ukp}
T.~Mergulh{\~a}o, F.~Beutler, and J.~Peacock, ``{Primordial Feature Constraints
  from \mbox{BOSS + eBOSS}}'',
  \href{http://dx.doi.org/10.1088/1475-7516/2023/08/012}{{\em JCAP} {\bfseries
  08} (2023) 012}, \href{http://arxiv.org/abs/2303.13946}{{\ttfamily
  arXiv:2303.13946 [astro-ph.CO]}}.

\bibitem{Calderon:2025xod}
R.~Calderon, T.~Simon, A.~Shafieloo, and D.~K. Hazra, ``{Primordial Features in
  Light of the Effective Field Theory of Large-Scale Structure}'',
  \href{http://dx.doi.org/10.1088/1475-7516/2026/01/057}{{\em JCAP} {\bfseries
  01} (2026) 057}, \href{http://arxiv.org/abs/2504.06183}{{\ttfamily
  arXiv:2504.06183 [astro-ph.CO]}}.

\bibitem{Sohn:2019rlq}
W.~Sohn and J.~Fergusson, ``{CMB-S4 Forecast on the Primordial Non-Gaussianity
  Parameter of Feature Models}'',
  \href{http://dx.doi.org/10.1103/PhysRevD.100.063536}{{\em Phys. Rev. D}
  {\bfseries 100} (2019) 063536},
  \href{http://arxiv.org/abs/1902.01142}{{\ttfamily arXiv:1902.01142
  [astro-ph.CO]}}.

\bibitem{Euclid:2023shr}
{M. Ballardini \textit{et al.} (Euclid Collaboration)}, ``{Euclid: The Search
  for Primordial Features}'',
  \href{http://dx.doi.org/10.1051/0004-6361/202348162}{{\em Astron. Astrophys.}
  {\bfseries 683} (2024) A220},
  \href{http://arxiv.org/abs/2309.17287}{{\ttfamily arXiv:2309.17287
  [astro-ph.CO]}}.

\bibitem{Marolf:2010zp}
D.~Marolf and I.~Morrison, ``{The IR Stability of de Sitter: Loop Corrections
  to Scalar Propagators}'',
  \href{http://dx.doi.org/10.1103/PhysRevD.82.105032}{{\em Phys. Rev. D}
  {\bfseries 82} (2010) 105032},
  \href{http://arxiv.org/abs/1006.0035}{{\ttfamily arXiv:1006.0035 [gr-qc]}}.

\bibitem{Hogervorst:2021uvp}
M.~Hogervorst, J.~Penedones, and K.~S. Vaziri, ``{Towards the Non-Perturbative
  Cosmological Bootstrap}'',
  \href{http://dx.doi.org/10.1007/JHEP02(2023)162}{{\em JHEP} {\bfseries 02}
  (2023) 162}, \href{http://arxiv.org/abs/2107.13871}{{\ttfamily
  arXiv:2107.13871 [hep-th]}}.

\bibitem{DiPietro:2021sjt}
L.~Di~Pietro, V.~Gorbenko, and S.~Komatsu, ``{Analyticity and Unitarity for
  Cosmological Correlators}'',
  \href{http://dx.doi.org/10.1007/JHEP03(2022)023}{{\em JHEP} {\bfseries 03}
  (2022) 023}, \href{http://arxiv.org/abs/2108.01695}{{\ttfamily
  arXiv:2108.01695 [hep-th]}}.

\bibitem{Loparco:2023rug}
M.~Loparco, J.~Penedones, K.~Salehi~Vaziri, and Z.~Sun, ``{The
  K{\"a}ll{\'e}n-Lehmann Representation in de Sitter Spacetime}'',
  \href{http://dx.doi.org/10.1007/JHEP12(2023)159}{{\em JHEP} {\bfseries 12}
  (2023) 159}, \href{http://arxiv.org/abs/2306.00090}{{\ttfamily
  arXiv:2306.00090 [hep-th]}}.

\bibitem{Cohen:2024anu}
T.~Cohen, D.~Green, and Y.~Huang, ``{Operator Origin of Anomalous Dimensions in
  de Sitter Space}'', \href{http://dx.doi.org/10.1103/PhysRevD.111.103513}{{\em
  Phys. Rev. D} {\bfseries 111} (2025) 103513},
  \href{http://arxiv.org/abs/2407.08581}{{\ttfamily arXiv:2407.08581
  [hep-th]}}.

\bibitem{Pimentel:2025rds}
G.~Pimentel and C.~Yang, ``{Strongly Coupled Sectors in Inflation: Gapless
  Theories and Unparticles}'',
  \href{http://arxiv.org/abs/2503.17840}{{\ttfamily arXiv:2503.17840
  [hep-th]}}.

\bibitem{Jiang:2025mlm}
Y.~Jiang, G.~Pimentel, and C.~Yang, ``{Strongly Coupled Sectors in Inflation:
  Gapped Theories of Unparticles}'',
  \href{http://arxiv.org/abs/2512.23796}{{\ttfamily arXiv:2512.23796
  [hep-th]}}.

\bibitem{Gleyzes:2016tdh}
J.~Gleyzes, R.~de~Putter, D.~Green, and O.~Dor\'e, ``{Biasing and the Search
  for Primordial Non-Gaussianity Beyond the Local Type}'',
  \href{http://dx.doi.org/10.1088/1475-7516/2017/04/002}{{\em JCAP} {\bfseries
  04} (2017) 002}, \href{http://arxiv.org/abs/1612.06366}{{\ttfamily
  arXiv:1612.06366 [astro-ph.CO]}}.

\bibitem{McCulloch:2024hiz}
C.~McCulloch, E.~Pajer, and X.~Tong, ``{A Cosmological Tachyon Collider:
  Enhancing the Long-Short-Scale Coupling}'',
  \href{http://dx.doi.org/10.1007/JHEP05(2024)262}{{\em JHEP} {\bfseries 05}
  (2024) 262}, \href{http://arxiv.org/abs/2401.11009}{{\ttfamily
  arXiv:2401.11009 [hep-th]}}.

\bibitem{Assassi:2013gxa}
V.~Assassi, D.~Baumann, D.~Green, and L.~McAllister, ``{Planck-Suppressed
  Operators}'', \href{http://dx.doi.org/10.1088/1475-7516/2014/01/033}{{\em
  JCAP} {\bfseries 01} (2014) 033},
  \href{http://arxiv.org/abs/1304.5226}{{\ttfamily arXiv:1304.5226 [hep-th]}}.

\bibitem{Mirbabayi:2015hva}
M.~Mirbabayi and M.~Simonovi{\'c}, ``{Effective Theory of Squeezed Correlation
  Functions}'', \href{http://dx.doi.org/10.1088/1475-7516/2016/03/056}{{\em
  JCAP} {\bfseries 03} (2016) 056},
  \href{http://arxiv.org/abs/1507.04755}{{\ttfamily arXiv:1507.04755
  [hep-th]}}.

\bibitem{Baumann:2012bc}
D.~Baumann, S.~Ferraro, D.~Green, and K.~Smith, ``{Stochastic Bias from
  Non-Gaussian Initial Conditions}'',
  \href{http://dx.doi.org/10.1088/1475-7516/2013/05/001}{{\em JCAP} {\bfseries
  05} (2013) 001}, \href{http://arxiv.org/abs/1209.2173}{{\ttfamily
  arXiv:1209.2173 [astro-ph.CO]}}.

\bibitem{Chen:2012ge}
X.~Chen and Y.~Wang, ``{Quasi-Single Field Inflation with Large Mass}'',
  \href{http://dx.doi.org/10.1088/1475-7516/2012/09/021}{{\em JCAP} {\bfseries
  09} (2012) 021}, \href{http://arxiv.org/abs/1205.0160}{{\ttfamily
  arXiv:1205.0160 [hep-th]}}.

\bibitem{Green:2022fwg}
D.~Green and Y.~Huang, ``{Flat Space Analog for the Quantum Origin of
  Structure}'', \href{http://dx.doi.org/10.1103/PhysRevD.106.023531}{{\em Phys.
  Rev. D} {\bfseries 106} (2022) 023531},
  \href{http://arxiv.org/abs/2203.10042}{{\ttfamily arXiv:2203.10042
  [hep-th]}}.

\bibitem{Creminelli:2012ed}
P.~Creminelli, J.~Nore\~na, and M.~Simonovi\'c, ``{Conformal Consistency
  Relations for Single-Field Inflation}'',
  \href{http://dx.doi.org/10.1088/1475-7516/2012/07/052}{{\em JCAP} {\bfseries
  07} (2012) 052}, \href{http://arxiv.org/abs/1203.4595}{{\ttfamily
  arXiv:1203.4595 [hep-th]}}.

\bibitem{Hinterbichler:2012nm}
K.~Hinterbichler, L.~Hui, and J.~Khoury, ``{Conformal Symmetries of Adiabatic
  Modes in Cosmology}'',
  \href{http://dx.doi.org/10.1088/1475-7516/2012/08/017}{{\em JCAP} {\bfseries
  08} (2012) 017}, \href{http://arxiv.org/abs/1203.6351}{{\ttfamily
  arXiv:1203.6351 [hep-th]}}.

\bibitem{Goldberger:2013rsa}
W.~Goldberger, L.~Hui, and A.~Nicolis, ``{One-Particle-Irreducible Consistency
  Relations for Cosmological Perturbations}'',
  \href{http://dx.doi.org/10.1103/PhysRevD.87.103520}{{\em Phys. Rev. D}
  {\bfseries 87} (2013) 103520},
  \href{http://arxiv.org/abs/1303.1193}{{\ttfamily arXiv:1303.1193 [hep-th]}}.

\bibitem{dePutter:2016moa}
R.~de~Putter, O.~Dor{\'e}, D.~Green, and J.~Meyers, ``{Single-Field Inflation
  and the Local Ansatz: Distinguishability and Consistency}'',
  \href{http://dx.doi.org/10.1103/PhysRevD.95.063501}{{\em Phys. Rev. D}
  {\bfseries 95} (2017) 063501},
  \href{http://arxiv.org/abs/1610.00785}{{\ttfamily arXiv:1610.00785
  [hep-th]}}.

\bibitem{Hui:2018cag}
L.~Hui, A.~Joyce, and S.~Wong, ``{Inflationary Soft Theorems Revisited:
  A~Generalized Consistency Relation}'',
  \href{http://dx.doi.org/10.1088/1475-7516/2019/02/060}{{\em JCAP} {\bfseries
  02} (2019) 060}, \href{http://arxiv.org/abs/1811.05951}{{\ttfamily
  arXiv:1811.05951 [hep-th]}}.

\bibitem{Creminelli:2013mca}
P.~Creminelli, J.~Nore\~na, M.~Simonovi\'c, and F.~Vernizzi, ``{Single-Field
  Consistency Relations of Large-Scale Structure}'',
  \href{http://dx.doi.org/10.1088/1475-7516/2013/12/025}{{\em JCAP} {\bfseries
  12} (2013) 025}, \href{http://arxiv.org/abs/1309.3557}{{\ttfamily
  arXiv:1309.3557 [astro-ph.CO]}}.

\bibitem{Creminelli:2013poa}
P.~Creminelli, J.~Gleyzes, M.~Simonovi\'c, and F.~Vernizzi, ``{Single-Field
  Consistency Relations of Large-Scale Structure. Part~II:~Resummation and
  Redshift Space}'',
  \href{http://dx.doi.org/10.1088/1475-7516/2014/02/051}{{\em JCAP} {\bfseries
  02} (2014) 051}, \href{http://arxiv.org/abs/1311.0290}{{\ttfamily
  arXiv:1311.0290 [astro-ph.CO]}}.

\bibitem{Horn:2014rta}
B.~Horn, L.~Hui, and X.~Xiao, ``{Soft-Pion Theorems for Large-Scale
  Structure}'', \href{http://dx.doi.org/10.1088/1475-7516/2014/09/044}{{\em
  JCAP} {\bfseries 09} (2014) 044},
  \href{http://arxiv.org/abs/1406.0842}{{\ttfamily arXiv:1406.0842 [hep-th]}}.

\bibitem{Esposito:2019jkb}
A.~Esposito, L.~Hui, and R.~Scoccimarro, ``{Non-Perturbative Test of
  Consistency Relations and Their Violation}'',
  \href{http://dx.doi.org/10.1103/PhysRevD.100.043536}{{\em Phys. Rev. D}
  {\bfseries 100} (2019) 043536},
  \href{http://arxiv.org/abs/1905.11423}{{\ttfamily arXiv:1905.11423
  [astro-ph.CO]}}.

\bibitem{Baumann:2011su}
D.~Baumann and D.~Green, ``{Equilateral Non-Gaussianity and New Physics on the
  Horizon}'', \href{http://dx.doi.org/10.1088/1475-7516/2011/09/014}{{\em JCAP}
  {\bfseries 09} (2011) 014}, \href{http://arxiv.org/abs/1102.5343}{{\ttfamily
  arXiv:1102.5343 [hep-th]}}.

\bibitem{Behbahani:2012be}
S.~Behbahani and D.~Green, ``{Collective Symmetry Breaking and Resonant
  Non-Gaussianity}'',
  \href{http://dx.doi.org/10.1088/1475-7516/2012/11/056}{{\em JCAP} {\bfseries
  11} (2012) 056}, \href{http://arxiv.org/abs/1207.2779}{{\ttfamily
  arXiv:1207.2779 [hep-th]}}.

\bibitem{Flauger:2016idt}
R.~Flauger, M.~Mirbabayi, L.~Senatore, and E.~Silverstein, ``{Productive
  Interactions: Heavy Particles and Non-Gaussianity}'',
  \href{http://dx.doi.org/10.1088/1475-7516/2017/10/058}{{\em JCAP} {\bfseries
  10} (2017) 058}, \href{http://arxiv.org/abs/1606.00513}{{\ttfamily
  arXiv:1606.00513 [hep-th]}}.

\bibitem{Behbahani:2011it}
S.~Behbahani, A.~Dymarsky, M.~Mirbabayi, and L.~Senatore, ``{(Small) Resonant
  Non-Gaussianities: Signatures of a Discrete Shift Symmetry in the Effective
  Field Theory of Inflation}'',
  \href{http://dx.doi.org/10.1088/1475-7516/2012/12/036}{{\em JCAP} {\bfseries
  12} (2012) 036}, \href{http://arxiv.org/abs/1111.3373}{{\ttfamily
  arXiv:1111.3373 [hep-th]}}.

\bibitem{Jazayeri:2025vlv}
S.~Jazayeri, X.~Tong, and Y.~Zhu, ``{Every Wrinkle Carries A Memory: An
  Integro-Differential Bootstrap for Features in Cosmological Correlators}'',
  \href{http://arxiv.org/abs/2511.00152}{{\ttfamily arXiv:2511.00152
  [hep-th]}}.

\bibitem{Green:2013rd}
D.~Green, M.~Lewandowski, L.~Senatore, E.~Silverstein, and M.~Zaldarriaga,
  ``{Anomalous Dimensions and Non-Gaussianity}'',
  \href{http://dx.doi.org/10.1007/JHEP10(2013)171}{{\em JHEP} {\bfseries 10}
  (2013) 171}, \href{http://arxiv.org/abs/1301.2630}{{\ttfamily arXiv:1301.2630
  [hep-th]}}.

\bibitem{Desjacques:2016bnm}
V.~Desjacques, D.~Jeong, and F.~Schmidt, ``{Large-Scale Galaxy Bias}'',
  \href{http://dx.doi.org/10.1016/j.physrep.2017.12.002}{{\em Phys. Rept.}
  {\bfseries 733} (2018) 1}, \href{http://arxiv.org/abs/1611.09787}{{\ttfamily
  arXiv:1611.09787 [astro-ph.CO]}}.

\bibitem{Bernardeau:2001qr}
F.~Bernardeau, S.~Colombi, E.~Gaztanaga, and R.~Scoccimarro, ``{Large-Scale
  Structure of the Universe and Cosmological Perturbation Theory}'',
  \href{http://dx.doi.org/10.1016/S0370-1573(02)00135-7}{{\em Phys. Rept.}
  {\bfseries 367} (2002) 1},
  \href{http://arxiv.org/abs/astro-ph/0112551}{{\ttfamily
  arXiv:astro-ph/0112551}}.

\bibitem{McDonald:2009dh}
P.~McDonald and A.~Roy, ``{Clustering of Dark Matter Tracers: Generalizing Bias
  for the Coming Era of Precision~LSS}'',
  \href{http://dx.doi.org/10.1088/1475-7516/2009/08/020}{{\em JCAP} {\bfseries
  08} (2009) 020}, \href{http://arxiv.org/abs/0902.0991}{{\ttfamily
  arXiv:0902.0991 [astro-ph.CO]}}.

\bibitem{Press:1973iz}
W.~Press and P.~Schechter, ``{Formation of Galaxies and Clusters of Galaxies by
  Self-Similar Gravitational Condensation}'',
  \href{http://dx.doi.org/10.1086/152650}{{\em Astrophys. J.} {\bfseries 187}
  (1974) 425}.

\bibitem{Schmidt:2010gw}
F.~Schmidt and M.~Kamionkowski, ``{Halo Clustering with Non-Local
  Non-Gaussianity}'', \href{http://dx.doi.org/10.1103/PhysRevD.82.103002}{{\em
  Phys. Rev.~D} {\bfseries 82} (2010) 103002},
  \href{http://arxiv.org/abs/1008.0638}{{\ttfamily arXiv:1008.0638
  [astro-ph.CO]}}.

\bibitem{Creminelli:2013cga}
P.~Creminelli, A.~Perko, L.~Senatore, M.~Simonovi\'c, and G.~Trevisan, ``{The
  Physical Squeezed Limit: Consistency Relations at Order~$q^2$}'',
  \href{http://dx.doi.org/10.1088/1475-7516/2013/11/015}{{\em JCAP} {\bfseries
  11} (2013) 015}, \href{http://arxiv.org/abs/1307.0503}{{\ttfamily
  arXiv:1307.0503 [astro-ph.CO]}}.

\bibitem{Desjacques:2008vf}
V.~Desjacques, U.~Seljak, and I.~Iliev, ``{Scale-Dependent Bias Induced by
  Local Non-Gaussianity: A~Comparison to $N$-Body Simulations}'',
  \href{http://dx.doi.org/10.1111/j.1365-2966.2009.14721.x}{{\em Mon. Not. Roy.
  Astron. Soc.} {\bfseries 396} (2009) 85},
  \href{http://arxiv.org/abs/0811.2748}{{\ttfamily arXiv:0811.2748
  [astro-ph]}}.

\bibitem{Pillepich:2008ka}
A.~Pillepich, C.~Porciani, and O.~Hahn, ``{Universal Halo Mass Function and
  Scale-Dependent Bias from $N$-Body Simulations with Non-Gaussian Initial
  Conditions}'', \href{http://dx.doi.org/10.1111/j.1365-2966.2009.15914.x}{{\em
  Mon. Not. Roy. Astron. Soc.} {\bfseries 402} (2010) 191},
  \href{http://arxiv.org/abs/0811.4176}{{\ttfamily arXiv:0811.4176
  [astro-ph]}}.

\bibitem{Grossi:2009an}
M.~Grossi, L.~Verde, C.~Carbone, K.~Dolag, E.~Branchini, F.~Iannuzzi,
  S.~Matarrese, and L.~Moscardini, ``{Large-Scale Non-Gaussian Mass Function
  and Halo Bias: Tests on $N$-Body Simulations}'',
  \href{http://dx.doi.org/10.1111/j.1365-2966.2009.15150.x}{{\em Mon. Not. Roy.
  Astron. Soc.} {\bfseries 398} (2009) 321},
  \href{http://arxiv.org/abs/0902.2013}{{\ttfamily arXiv:0902.2013
  [astro-ph.CO]}}.

\bibitem{Scoccimarro:2011pz}
R.~Scoccimarro, L.~Hui, M.~Manera, and K.~C. Chan, ``{Large-Scale Bias and
  Efficient Generation of Initial Conditions for Non-Local Primordial
  Non-Gaussianity}'', \href{http://dx.doi.org/10.1103/PhysRevD.85.083002}{{\em
  Phys. Rev.~D} {\bfseries 85} (2012) 083002},
  \href{http://arxiv.org/abs/1108.5512}{{\ttfamily arXiv:1108.5512
  [astro-ph.CO]}}.

\bibitem{Baldauf:2015vio}
T.~Baldauf, U.~Seljak, L.~Senatore, and M.~Zaldarriaga, ``{Linear Response to
  Long-Wavelength Fluctuations using Curvature Simulations}'',
  \href{http://dx.doi.org/10.1088/1475-7516/2016/09/007}{{\em JCAP} {\bfseries
  09} (2016) 007}, \href{http://arxiv.org/abs/1511.01465}{{\ttfamily
  arXiv:1511.01465 [astro-ph.CO]}}.

\bibitem{Biagetti:2016ywx}
M.~Biagetti, T.~Lazeyras, T.~Baldauf, V.~Desjacques, and F.~Schmidt,
  ``{Verifying the Consistency Relation for the Scale-Dependent Bias from Local
  Primordial Non-Gaussianity}'',
  \href{http://dx.doi.org/10.1093/mnras/stx714}{{\em Mon. Not. Roy. Astron.
  Soc.} {\bfseries 468} (2017) 3277},
  \href{http://arxiv.org/abs/1611.04901}{{\ttfamily arXiv:1611.04901
  [astro-ph.CO]}}.

\bibitem{Barreira:2020kvh}
A.~Barreira, G.~Cabass, F.~Schmidt, A.~Pillepich, and D.~Nelson, ``{Galaxy Bias
  and Primordial Non-Gaussianity: Insights from Galaxy Formation Simulations
  with~IllustrisTNG}'',
  \href{http://dx.doi.org/10.1088/1475-7516/2020/12/013}{{\em JCAP} {\bfseries
  12} (2020) 013}, \href{http://arxiv.org/abs/2006.09368}{{\ttfamily
  arXiv:2006.09368 [astro-ph.CO]}}.

\bibitem{Barreira:2020ekm}
A.~Barreira, ``{On the Impact of Galaxy Bias Uncertainties on Primordial
  Non-Gaussianity Constraints}'',
  \href{http://dx.doi.org/10.1088/1475-7516/2020/12/031}{{\em JCAP} {\bfseries
  12} (2020) 031}, \href{http://arxiv.org/abs/2009.06622}{{\ttfamily
  arXiv:2009.06622 [astro-ph.CO]}}.

\bibitem{Barreira:2021ukk}
A.~Barreira, T.~Lazeyras, and F.~Schmidt, ``{Galaxy Bias from Forward Models:
  Linear and Second-Order Bias of IllustrisTNG Galaxies}'',
  \href{http://dx.doi.org/10.1088/1475-7516/2021/08/029}{{\em JCAP} {\bfseries
  08} (2021) 029}, \href{http://arxiv.org/abs/2105.02876}{{\ttfamily
  arXiv:2105.02876 [astro-ph.CO]}}.

\bibitem{Coulton:2022rir}
W.~Coulton, F.~Villaescusa-Navarro, D.~Jamieson, M.~Baldi, G.~Jung,
  D.~Karagiannis, M.~Liguori, L.~Verde, and B.~Wandelt, ``{Quijote-PNG: The
  Information Content of the Halo Power Spectrum and Bispectrum}'',
  \href{http://dx.doi.org/10.3847/1538-4357/aca7c1}{{\em Astrophys. J.}
  {\bfseries 943} (2023) 178},
  \href{http://arxiv.org/abs/2206.15450}{{\ttfamily arXiv:2206.15450
  [astro-ph.CO]}}.

\bibitem{Barreira:2023rxn}
A.~Barreira and E.~Krause, ``{Towards Optimal and Robust $\fnl$~Constraints
  with Multi-Tracer Analyses}'',
  \href{http://dx.doi.org/10.1088/1475-7516/2023/10/044}{{\em JCAP} {\bfseries
  10} (2023) 044}, \href{http://arxiv.org/abs/2302.09066}{{\ttfamily
  arXiv:2302.09066 [astro-ph.CO]}}.

\bibitem{Sullivan:2023qjr}
J.~Sullivan, T.~Prijon, and U.~Seljak, ``{Learning to Concentrate: Multi-Tracer
  Forecasts on Local Primordial Non-Gaussianity with Machine-Learned~Bias}'',
  \href{http://dx.doi.org/10.1088/1475-7516/2023/08/004}{{\em JCAP} {\bfseries
  08} (2023) 004}, \href{http://arxiv.org/abs/2303.08901}{{\ttfamily
  arXiv:2303.08901 [astro-ph.CO]}}.

\bibitem{Hadzhiyska:2025rez}
B.~Hadzhiyska and S.~Ferraro, ``{Refining Local-Type Primordial
  Non-Gaussianity: Sharpened $b_\phi$~Constraints through Bias Expansion}'',
  \href{http://dx.doi.org/10.1103/PhysRevD.111.103521}{{\em Phys. Rev. D}
  {\bfseries 111} (2025) 103521},
  \href{http://arxiv.org/abs/2501.14873}{{\ttfamily arXiv:2501.14873
  [astro-ph.CO]}}.

\bibitem{Sharma:2025xss}
D.~Sharma, J.~Sullivan, K.~Akitsu, and M.~Ivanov, ``{Equilateral Non-Gaussian
  Bias at the Field Level}'', \href{http://arxiv.org/abs/2511.14677}{{\ttfamily
  arXiv:2511.14677 [astro-ph.CO]}}.

\bibitem{Perez:2026mjt}
L.~Perez, S.~Genel, E.~Krause, and R.~Somerville, ``{The Impact of Galaxy
  Formation on Galaxy Biasing and Implications for Primordial Non-Gaussianity
  Constraints}'', \href{http://arxiv.org/abs/2602.04987}{{\ttfamily
  arXiv:2602.04987 [astro-ph.CO]}}.

\bibitem{Eisenstein:2006nj}
D.~J. Eisenstein, H.-j. Seo, and M.~J. White, ``{On the Robustness of the
  Acoustic Scale in the Low-Redshift Clustering of Matter}'',
  \href{http://dx.doi.org/10.1086/518755}{{\em Astrophys. J.} {\bfseries 664}
  (2007) 660}, \href{http://arxiv.org/abs/astro-ph/0604361}{{\ttfamily
  arXiv:astro-ph/0604361}}.

\bibitem{Eisenstein:2006nk}
D.~Eisenstein, H.-j. Seo, E.~Sirko, and D.~Spergel, ``{Improving Cosmological
  Distance Measurements by Reconstruction of the Baryon Acoustic Peak}'',
  \href{http://dx.doi.org/10.1086/518712}{{\em Astrophys. J.} {\bfseries 664}
  (2007) 675}, \href{http://arxiv.org/abs/astro-ph/0604362}{{\ttfamily
  arXiv:astro-ph/0604362}}.

\bibitem{Crocce:2007dt}
M.~Crocce and R.~Scoccimarro, ``{Nonlinear Evolution of Baryon Acoustic
  Oscillations}'', \href{http://dx.doi.org/10.1103/PhysRevD.77.023533}{{\em
  Phys. Rev. D} {\bfseries 77} (2008) 023533},
  \href{http://arxiv.org/abs/0704.2783}{{\ttfamily arXiv:0704.2783
  [astro-ph]}}.

\bibitem{BOSS:2016wmc}
{S. Alam \textit{et al.} (BOSS Collaboration)}, ``{The Clustering of Galaxies
  in the Completed SDSS-III Baryon Oscillation Spectroscopic Survey:
  Cosmological Analysis of the DR12 Galaxy Sample}'',
  \href{http://dx.doi.org/10.1093/mnras/stx721}{{\em Mon. Not. Roy. Astron.
  Soc.} {\bfseries 470} (2017) 2617},
  \href{http://arxiv.org/abs/1607.03155}{{\ttfamily arXiv:1607.03155
  [astro-ph.CO]}}.

\bibitem{DESI:2016fyo}
{A. Aghamousa \textit{et al.} (DESI Collaboration)}, ``{The DESI Experiment
  Part I: Science, Targeting and Survey Design}'',
  \href{http://arxiv.org/abs/1611.00036}{{\ttfamily arXiv:1611.00036
  [astro-ph.IM]}}.

\bibitem{Euclid:2021icp}
{R. Scaramella \textit{et al.} (Euclid Collaboration)}, ``{Euclid Preparation
  -- I.~The Euclid Wide Survey}'',
  \href{http://dx.doi.org/10.1051/0004-6361/202141938}{{\em Astron. Astrophys.}
  {\bfseries 662} (2022) A112},
  \href{http://arxiv.org/abs/2108.01201}{{\ttfamily arXiv:2108.01201
  [astro-ph.CO]}}.

\bibitem{LSSTScience:2009jmu}
{P. Abell \textit{et al.} (LSST Science \& LSST Project Collaborations)},
  ``{LSST Science Book, Version 2.0}'',
  \href{http://arxiv.org/abs/0912.0201}{{\ttfamily arXiv:0912.0201
  [astro-ph.IM]}}.

\bibitem{Zhao:2024alp}
C.~Zhao {\em et~al.}, ``{MUltiplexed Survey Telescope: Perspectives for
  Large-Scale Structure Cosmology in the Era of Stage-V Spectroscopic
  Survey}'', \href{http://arxiv.org/abs/2411.07970}{{\ttfamily arXiv:2411.07970
  [astro-ph.CO]}}.

\bibitem{Camera:2014bwa}
S.~Camera, M.~Santos, and R.~Maartens, ``{Probing Primordial Non-Gaussianity
  with SKA Galaxy Redshift Surveys: A Fully Relativistic Analysis}'',
  \href{http://dx.doi.org/10.1093/mnras/stv040}{{\em Mon. Not. Roy. Astron.
  Soc.} {\bfseries 448} (2015) 1035},
  \href{http://arxiv.org/abs/1409.8286}{{\ttfamily arXiv:1409.8286
  [astro-ph.CO]}}.

\bibitem{Meerburg:2016zdz}
P.~D. Meerburg, M.~M\"unchmeyer, J.~Mu\~noz, and X.~Chen, ``{Prospects for
  Cosmological Collider Physics}'',
  \href{http://dx.doi.org/10.1088/1475-7516/2017/03/050}{{\em JCAP} {\bfseries
  03} (2017) 050}, \href{http://arxiv.org/abs/1610.06559}{{\ttfamily
  arXiv:1610.06559 [astro-ph.CO]}}.

\bibitem{Li:2017jnt}
Y.-C. Li and Y.-Z. Ma, ``{Constraints on Primordial Non-Gaussianity from Future
  HI Intensity Mapping Experiments}'',
  \href{http://dx.doi.org/10.1103/PhysRevD.96.063525}{{\em Phys. Rev.~D}
  {\bfseries 96} (2017) 063525},
  \href{http://arxiv.org/abs/1701.00221}{{\ttfamily arXiv:1701.00221
  [astro-ph.CO]}}.

\bibitem{MoradinezhadDizgah:2018lac}
A.~Moradinezhad~Dizgah and G.~Keating, ``{Line Intensity Mapping with [CII] and
  CO(1-0) as Probes of Primordial Non-Gaussianity}'',
  \href{http://dx.doi.org/10.3847/1538-4357/aafd36}{{\em Astrophys. J.}
  {\bfseries 872} (2019) 126},
  \href{http://arxiv.org/abs/1810.02850}{{\ttfamily arXiv:1810.02850
  [astro-ph.CO]}}.

\bibitem{CosmicVisions21cm:2018rfq}
{R. Ansari \textit{et al.} (Cosmic Visions \SI{21}{cm} Collaboration)},
  ``{Inflation and Early Dark Energy with a Stage-II Hydrogen Intensity Mapping
  Experiment}'', \href{http://arxiv.org/abs/1810.09572}{{\ttfamily
  arXiv:1810.09572 [astro-ph.CO]}}.

\bibitem{PUMA:2019jwd}
{A. Slosar \textit{et al.} (PUMA Collaboration)}, ``{Packed Ultra-wideband
  Mapping Array~(PUMA): A~Radio Telescope for Cosmology and Transients}'', {\em
  \href{https://baas.aas.org/pub/2020n7i053}{Bull. Am. Astron. Soc.}}
  {\bfseries \href{https://baas.aas.org/pub/2020n7i053}{51}}
  (\href{https://baas.aas.org/pub/2020n7i053}{2019})
  \href{https://baas.aas.org/pub/2020n7i053}{53},
  \href{http://arxiv.org/abs/1907.12559}{{\ttfamily arXiv:1907.12559
  [astro-ph.IM]}}.

\bibitem{Karagiannis:2020dpq}
D.~Karagiannis, J.~Fonseca, R.~Maartens, and S.~Camera, ``{Probing Primordial
  Non-Gaussianity with the Power Spectrum and Bispectrum of Future \SI{21}{cm}
  Intensity Maps}'', \href{http://dx.doi.org/10.1016/j.dark.2021.100821}{{\em
  Phys. Dark Univ.} {\bfseries 32} (2021) 100821},
  \href{http://arxiv.org/abs/2010.07034}{{\ttfamily arXiv:2010.07034
  [astro-ph.CO]}}.

\bibitem{Castorina:2020zhz}
E.~Castorina {\em et~al.}, ``{Packed Ultra-wideband Mapping Array~(PUMA):
  Astro2020 RFI Response}'', \href{http://arxiv.org/abs/2002.05072}{{\ttfamily
  arXiv:2002.05072 [astro-ph.IM]}}.

\bibitem{Lazeyras:2015lgp}
T.~Lazeyras, C.~Wagner, T.~Baldauf, and F.~Schmidt, ``{Precision Measurement of
  the Local Bias of Dark Matter Halos}'',
  \href{http://dx.doi.org/10.1088/1475-7516/2016/02/018}{{\em JCAP} {\bfseries
  02} (2016) 018}, \href{http://arxiv.org/abs/1511.01096}{{\ttfamily
  arXiv:1511.01096 [astro-ph.CO]}}.

\bibitem{Philcox:2021kcw}
O.~Philcox and M.~Ivanov, ``{BOSS~DR12 Full-Shape Cosmology:
  $\Lambda$CDM~Constraints from the Large-Scale Galaxy Power Spectrum and
  Bispectrum Monopole}'',
  \href{http://dx.doi.org/10.1103/PhysRevD.105.043517}{{\em Phys. Rev.~D}
  {\bfseries 105} (2022) 043517},
  \href{http://arxiv.org/abs/2112.04515}{{\ttfamily arXiv:2112.04515
  [astro-ph.CO]}}.

\bibitem{Pullen:2012rd}
A.~Pullen and C.~Hirata, ``{Systematic Effects in Large-Scale Angular Power
  Spectra of Photometric Quasars and Implications for Constraining Primordial
  Non-Gaussianity}'', \href{http://dx.doi.org/10.1086/671189}{{\em Publ.
  Astron. Soc. Pac.} {\bfseries 125} (2013) 705},
  \href{http://arxiv.org/abs/1212.4500}{{\ttfamily arXiv:1212.4500
  [astro-ph.CO]}}.

\bibitem{Foglieni:2023xca}
M.~Foglieni, M.~Pantiri, E.~Di~Dio, and E.~Castorina, ``{Large-Scale Limit of
  the Observed Galaxy Power Spectrum}'',
  \href{http://dx.doi.org/10.1103/PhysRevLett.131.111201}{{\em Phys. Rev.
  Lett.} {\bfseries 131} (2023) 111201},
  \href{http://arxiv.org/abs/2303.03142}{{\ttfamily arXiv:2303.03142
  [astro-ph.CO]}}.

\bibitem{Martinez-Carrillo:2021lcn}
R.~Martinez-Carrillo, J.~Hidalgo, K.~Malik, and A.~Pourtsidou, ``{Contributions
  from Primordial Non-Gaussianity and General Relativity to the Galaxy Power
  Spectrum}'', \href{http://dx.doi.org/10.1088/1475-7516/2021/12/025}{{\em
  JCAP} {\bfseries 12} (2021) 025},
  \href{http://arxiv.org/abs/2107.10815}{{\ttfamily arXiv:2107.10815
  [astro-ph.CO]}}.

\bibitem{Castorina:2020blr}
E.~Castorina and A.~Moradinezhad~Dizgah, ``{Local Primordial Non-Gaussianities
  and Super-Sample Variance}'',
  \href{http://dx.doi.org/10.1088/1475-7516/2020/10/007}{{\em JCAP} {\bfseries
  10} (2020) 007}, \href{http://arxiv.org/abs/2005.14677}{{\ttfamily
  arXiv:2005.14677 [astro-ph.CO]}}.

\bibitem{Shiveshwarkar:2023xjv}
C.~Shiveshwarkar, T.~Brinckmann, M.~Loverde, and M.~McQuinn,
  ``{Post-Inflationary Contamination of Local Primordial Non-Gaussianity in
  Galaxy Power Spectra}'',
  \href{http://dx.doi.org/10.1103/PhysRevD.108.103538}{{\em Phys. Rev. D}
  {\bfseries 108} (2023) 103538},
  \href{http://arxiv.org/abs/2306.07517}{{\ttfamily arXiv:2306.07517
  [astro-ph.CO]}}.

\bibitem{Cabass:2022wjy}
G.~Cabass, M.~Ivanov, O.~Philcox, M.~Simonovi\'c, and M.~Zaldarriaga,
  ``{Constraints on Single-Field Inflation from the BOSS~Galaxy Survey}'',
  \href{http://dx.doi.org/10.1103/PhysRevLett.129.021301}{{\em Phys. Rev.
  Lett.} {\bfseries 129} (2022) 021301},
  \href{http://arxiv.org/abs/2201.07238}{{\ttfamily arXiv:2201.07238
  [astro-ph.CO]}}.

\bibitem{DAmico:2022gki}
G.~D'Amico, M.~Lewandowski, L.~Senatore, and P.~Zhang, ``{Limits on Primordial
  Non-Gaussianities from BOSS~Galaxy-Clustering Data}'',
  \href{http://dx.doi.org/10.1103/PhysRevD.111.063514}{{\em Phys. Rev. D}
  {\bfseries 111} (2025) 063514},
  \href{http://arxiv.org/abs/2201.11518}{{\ttfamily arXiv:2201.11518
  [astro-ph.CO]}}.

\bibitem{Cabass:2022ymb}
G.~Cabass, M.~Ivanov, O.~Philcox, M.~Simonovi\'c, and M.~Zaldarriaga,
  ``{Constraints on Multi-Field Inflation from the BOSS~Galaxy Survey}'',
  \href{http://dx.doi.org/10.1103/PhysRevD.106.043506}{{\em Phys. Rev.~D}
  {\bfseries 106} (2022) 043506},
  \href{http://arxiv.org/abs/2204.01781}{{\ttfamily arXiv:2204.01781
  [astro-ph.CO]}}.

\bibitem{Chudaykin:2025vdh}
A.~Chudaykin, M.~Ivanov, and O.~Philcox, ``{Reanalyzing DESI DR1:
  3.~Constraints on Inflation from Galaxy Power Spectra and Bispectra}'',
  \href{http://arxiv.org/abs/2512.04266}{{\ttfamily arXiv:2512.04266
  [astro-ph.CO]}}.

\bibitem{Philcox:2020vbm}
O.~Philcox, ``{Cosmology Without Window Functions: Quadratic Estimators for the
  Galaxy Power Spectrum}'',
  \href{http://dx.doi.org/10.1103/PhysRevD.103.103504}{{\em Phys. Rev.~D}
  {\bfseries 103} (2021) 103504},
  \href{http://arxiv.org/abs/2012.09389}{{\ttfamily arXiv:2012.09389
  [astro-ph.CO]}}.

\bibitem{Kitaura:2015uqa}
F.-S. Kitaura {\em et~al.}, ``{The Clustering of Galaxies in the SDSS-III
  Baryon Oscillation Spectroscopic Survey: Mock Galaxy Catalogues for the
  BOSS~Final Data Release}'',
  \href{http://dx.doi.org/10.1093/mnras/stv2826}{{\em Mon. Not. Roy. Astron.
  Soc.} {\bfseries 456} (2016) 4156},
  \href{http://arxiv.org/abs/1509.06400}{{\ttfamily arXiv:1509.06400
  [astro-ph.CO]}}.

\bibitem{Ivanov:2019pdj}
M.~Ivanov, M.~Simonovi\'c, and M.~Zaldarriaga, ``{Cosmological Parameters from
  the BOSS~Galaxy Power Spectrum}'',
  \href{http://dx.doi.org/10.1088/1475-7516/2020/05/042}{{\em JCAP} {\bfseries
  05} (2020) 042}, \href{http://arxiv.org/abs/1909.05277}{{\ttfamily
  arXiv:1909.05277 [astro-ph.CO]}}.

\bibitem{DAmico:2019fhj}
G.~D'Amico, J.~Gleyzes, N.~Kokron, K.~Markovic, L.~Senatore, P.~Zhang,
  F.~Beutler, and H.~Gil-Mar\'\i{}n, ``{The Cosmological Analysis of the
  SDSS/BOSS~Data from the Effective Field Theory of Large-Scale Structure}'',
  \href{http://dx.doi.org/10.1088/1475-7516/2020/05/005}{{\em JCAP} {\bfseries
  05} (2020) 005}, \href{http://arxiv.org/abs/1909.05271}{{\ttfamily
  arXiv:1909.05271 [astro-ph.CO]}}.

\bibitem{Carrasco:2012cv}
J.~J. Carrasco, M.~Hertzberg, and L.~Senatore, ``{The Effective Field Theory of
  Cosmological Large-Scale Structure}'',
  \href{http://dx.doi.org/10.1007/JHEP09(2012)082}{{\em JHEP} {\bfseries 09}
  (2012) 082}, \href{http://arxiv.org/abs/1206.2926}{{\ttfamily arXiv:1206.2926
  [astro-ph.CO]}}.

\bibitem{Pajer:2013jj}
E.~Pajer and M.~Zaldarriaga, ``{On the Renormalization of the Effective Field
  Theory of Large-Scale Structure}'',
  \href{http://dx.doi.org/10.1088/1475-7516/2013/08/037}{{\em JCAP} {\bfseries
  08} (2013) 037}, \href{http://arxiv.org/abs/1301.7182}{{\ttfamily
  arXiv:1301.7182 [astro-ph.CO]}}.

\bibitem{Carrasco:2013mua}
J.~J. Carrasco, S.~Foreman, D.~Green, and L.~Senatore, ``{The Effective Field
  Theory of Large-Scale Structure at Two Loops}'',
  \href{http://dx.doi.org/10.1088/1475-7516/2014/07/057}{{\em JCAP} {\bfseries
  07} (2014) 057}, \href{http://arxiv.org/abs/1310.0464}{{\ttfamily
  arXiv:1310.0464 [astro-ph.CO]}}.

\bibitem{Chudaykin:2020aoj}
A.~Chudaykin, M.~Ivanov, O.~Philcox, and M.~Simonovi\'c, ``{Nonlinear
  Perturbation Theory Extension of the Boltzmann Code~CLASS}'',
  \href{http://dx.doi.org/10.1103/PhysRevD.102.063533}{{\em Phys. Rev.~D}
  {\bfseries 102} (2020) 063533},
  \href{http://arxiv.org/abs/2004.10607}{{\ttfamily arXiv:2004.10607
  [astro-ph.CO]}}.

\bibitem{Blas:2011rf}
D.~Blas, J.~Lesgourgues, and T.~Tram, ``{The Cosmic Linear Anisotropy Solving
  System~(CLASS)~II: Approximation Schemes}'',
  \href{http://dx.doi.org/10.1088/1475-7516/2011/07/034}{{\em JCAP} {\bfseries
  07} (2011) 034}, \href{http://arxiv.org/abs/1104.2933}{{\ttfamily
  arXiv:1104.2933 [astro-ph.CO]}}.

\bibitem{Audren:2012wb}
B.~Audren, J.~Lesgourgues, K.~Benabed, and S.~Prunet, ``{Conservative
  Constraints on Early Cosmology: An Illustration of the MontePython
  Cosmological Parameter Inference Code}'',
  \href{http://dx.doi.org/10.1088/1475-7516/2013/02/001}{{\em JCAP} {\bfseries
  02} (2013) 001}, \href{http://arxiv.org/abs/1210.7183}{{\ttfamily
  arXiv:1210.7183 [astro-ph.CO]}}.

\bibitem{Brinckmann:2018cvx}
T.~Brinckmann and J.~Lesgourgues, ``{MontePython~3: Boosted MCMC~Sampler and
  Other Features}'', \href{http://dx.doi.org/10.1016/j.dark.2018.100260}{{\em
  Phys. Dark Univ.} {\bfseries 24} (2019) 100260},
  \href{http://arxiv.org/abs/1804.07261}{{\ttfamily arXiv:1804.07261
  [astro-ph.CO]}}.

\bibitem{Philcox:2020zyp}
O.~Philcox, M.~Ivanov, M.~Zaldarriaga, M.~Simonovi\'c, and M.~Schmittfull,
  ``{Fewer Mocks and Less Noise: Reducing the Dimensionality of Cosmological
  Observables with Subspace Projections}'',
  \href{http://dx.doi.org/10.1103/PhysRevD.103.043508}{{\em Phys. Rev.~D}
  {\bfseries 103} (2021) 043508},
  \href{http://arxiv.org/abs/2009.03311}{{\ttfamily arXiv:2009.03311
  [astro-ph.CO]}}.

\bibitem{Fergusson:2014hya}
J.~Fergusson, H.~Gruetjen, E.~P.~S. Shellard, and M.~Liguori, ``{Combining
  Power Spectrum and Bispectrum Measurements to Detect Oscillatory Features}'',
  \href{http://dx.doi.org/10.1103/PhysRevD.91.023502}{{\em Phys. Rev. D}
  {\bfseries 91} (2015) 023502},
  \href{http://arxiv.org/abs/1410.5114}{{\ttfamily arXiv:1410.5114
  [astro-ph.CO]}}.

\bibitem{Fergusson:2014tza}
J.~Fergusson, H.~Gruetjen, E.~P.~S. Shellard, and B.~Wallisch, ``{Polyspectra
  Searches for Sharp Oscillatory Features in Cosmic Microwave Sky Data}'',
  \href{http://dx.doi.org/10.1103/PhysRevD.91.123506}{{\em Phys. Rev. D}
  {\bfseries 91} (2015) 123506},
  \href{http://arxiv.org/abs/1412.6152}{{\ttfamily arXiv:1412.6152
  [astro-ph.CO]}}.

\bibitem{Meerburg:2015owa}
P.~D. Meerburg, M.~M\"unchmeyer, and B.~Wandelt, ``{Joint Resonant CMB Power
  Spectrum and Bispectrum Estimation}'',
  \href{http://dx.doi.org/10.1103/PhysRevD.93.043536}{{\em Phys. Rev. D}
  {\bfseries 93} (2016) 043536},
  \href{http://arxiv.org/abs/1510.01756}{{\ttfamily arXiv:1510.01756
  [astro-ph.CO]}}.

\bibitem{dePutter:2018jqk}
R.~de~Putter, ``{Primordial Physics from Large-Scale Structure Beyond the Power
  Spectrum}'', \href{http://arxiv.org/abs/1802.06762}{{\ttfamily
  arXiv:1802.06762 [astro-ph.CO]}}.

\bibitem{Biagetti:2020skr}
M.~Biagetti, A.~Cole, and G.~Shiu, ``{The Persistence of Large-Scale
  Structures~I: Primordial Non-Gaussianity}'',
  \href{http://dx.doi.org/10.1088/1475-7516/2021/04/061}{{\em JCAP} {\bfseries
  04} (2021) 061}, \href{http://arxiv.org/abs/2009.04819}{{\ttfamily
  arXiv:2009.04819 [astro-ph.CO]}}.

\bibitem{MoradinezhadDizgah:2020whw}
A.~Moradinezhad~Dizgah, M.~Biagetti, E.~Sefusatti, V.~Desjacques, and
  J.~Nore\~na, ``{Primordial Non-Gaussianity from Biased Tracers: Likelihood
  Analysis of Real-Space Power Spectrum and Bispectrum}'',
  \href{http://dx.doi.org/10.1088/1475-7516/2021/05/015}{{\em JCAP} {\bfseries
  05} (2021) 015}, \href{http://arxiv.org/abs/2010.14523}{{\ttfamily
  arXiv:2010.14523 [astro-ph.CO]}}.

\bibitem{Baumann:2021ykm}
D.~Baumann and D.~Green, ``{The Power of Locality: Primordial Non-Gaussianity
  at the Map Level}'',
  \href{http://dx.doi.org/10.1088/1475-7516/2022/08/061}{{\em JCAP} {\bfseries
  08} (2022) 061}, \href{http://arxiv.org/abs/2112.14645}{{\ttfamily
  arXiv:2112.14645 [astro-ph.CO]}}.

\bibitem{Andrews:2022nvv}
A.~Andrews, J.~Jasche, G.~Lavaux, and F.~Schmidt, ``{Bayesian Field-Level
  Inference of Primordial Non-Gaussianity using Next-Generation Galaxy
  Surveys}'', \href{http://dx.doi.org/10.1093/mnras/stad432}{{\em Mon. Not.
  Roy. Astron. Soc.} {\bfseries 520} (2023) 5746},
  \href{http://arxiv.org/abs/2203.08838}{{\ttfamily arXiv:2203.08838
  [astro-ph.CO]}}.

\bibitem{Coulton:2022qbc}
W.~Coulton, F.~Villaescusa-Navarro, D.~Jamieson, M.~Baldi, G.~Jung,
  D.~Karagiannis, M.~Liguori, L.~Verde, and B.~Wandelt, ``{Quijote-PNG:
  Simulations of Primordial Non-Gaussianity and the Information Content of the
  Matter Field Power Spectrum and Bispectrum}'',
  \href{http://dx.doi.org/10.3847/1538-4357/aca8a7}{{\em Astrophys. J.}
  {\bfseries 943} (2023) 64}, \href{http://arxiv.org/abs/2206.01619}{{\ttfamily
  arXiv:2206.01619 [astro-ph.CO]}}.

\bibitem{Karagiannis:2023lsj}
D.~Karagiannis, R.~Maartens, J.~Fonseca, S.~Camera, and C.~Clarkson,
  ``{Multi-Tracer Power Spectra and Bispectra: Formalism}'',
  \href{http://dx.doi.org/10.1088/1475-7516/2024/03/034}{{\em JCAP} {\bfseries
  03} (2024) 034}, \href{http://arxiv.org/abs/2305.04028}{{\ttfamily
  arXiv:2305.04028 [astro-ph.CO]}}.

\bibitem{Euclid:2024ris}
{A. Andrews \textit{et al.} (Euclid Collaboration)}, ``{Euclid: Field-level
  inference of primordial non-Gaussianity and cosmic initial conditions}'',
  \href{http://arxiv.org/abs/2412.11945}{{\ttfamily arXiv:2412.11945
  [astro-ph.CO]}}.

\bibitem{Sullivan:2024jxe}
J.~Sullivan and S.-F. Chen, ``{Local Primordial Non-Gaussian Bias at the Field
  Level}'', \href{http://dx.doi.org/10.1088/1475-7516/2025/03/016}{{\em JCAP}
  {\bfseries 03} (2025) 016}, \href{http://arxiv.org/abs/2410.18039}{{\ttfamily
  arXiv:2410.18039 [astro-ph.CO]}}.

\bibitem{Doeser:2025wcb}
L.~Doeser, M.~Ata, and J.~Jasche, ``{Learning the Universe: Learning to
  Optimize Cosmic Initial Conditions with Non-Differentiable Structure
  Formation Models}'', \href{http://dx.doi.org/10.1093/mnras/staf1289}{{\em
  Mon. Not. Roy. Astron. Soc.} {\bfseries 1403} (2025) 1422},
  \href{http://arxiv.org/abs/2502.13243}{{\ttfamily arXiv:2502.13243
  [astro-ph.CO]}}.

\bibitem{Schmittfull:2017ffw}
M.~Schmittfull and U.~Seljak, ``{Parameter Constraints from Cross-Correlation
  of CMB~Lensing with Galaxy Clustering}'',
  \href{http://dx.doi.org/10.1103/PhysRevD.97.123540}{{\em Phys. Rev. D}
  {\bfseries 97} (2018) 123540},
  \href{http://arxiv.org/abs/1710.09465}{{\ttfamily arXiv:1710.09465
  [astro-ph.CO]}}.

\bibitem{Munchmeyer:2018eey}
M.~M\"unchmeyer, M.~Madhavacheril, S.~Ferraro, M.~Johnson, and K.~Smith,
  ``{Constraining Local Non-Gaussianities with Kinetic Sunyaev-Zel'dovich
  Tomography}'', \href{http://dx.doi.org/10.1103/PhysRevD.100.083508}{{\em
  Phys. Rev. D} {\bfseries 100} (2019) 083508},
  \href{http://arxiv.org/abs/1810.13424}{{\ttfamily arXiv:1810.13424
  [astro-ph.CO]}}.

\bibitem{Chen:2021vba}
S.-F. Chen, H.~Lee, and C.~Dvorkin, ``{Precise and Accurate Cosmology with
  CMB\texttimes{}LSS Power Spectra and Bispectra}'',
  \href{http://dx.doi.org/10.1088/1475-7516/2021/05/030}{{\em JCAP} {\bfseries
  05} (2021) 030}, \href{http://arxiv.org/abs/2103.01229}{{\ttfamily
  arXiv:2103.01229 [astro-ph.CO]}}.

\bibitem{Adshead:2024paa}
P.~Adshead and A.~Tishue, ``{Probing Beyond Local-Type Non-Gaussianity with
  Kinematic Sunyaev-Zel'dovich Tomography}'',
  \href{http://dx.doi.org/10.1103/PhysRevD.110.103549}{{\em Phys. Rev. D}
  {\bfseries 110} (2024) 103549},
  \href{http://arxiv.org/abs/2407.21094}{{\ttfamily arXiv:2407.21094
  [astro-ph.CO]}}.

\bibitem{Anbajagane:2025uro}
D.~Anbajagane and H.~Lee, ``{Primordial Physics in the Nonlinear Universe:
  Mapping Cosmological Collider Models to Weak-Lensing Observables}'',
  \href{http://arxiv.org/abs/2509.02693}{{\ttfamily arXiv:2509.02693
  [astro-ph.CO]}}.

\bibitem{Anbajagane:2025xlt}
D.~Anbajagane and H.~Lee, ``{Primordial Physics in the Nonlinear Universe:
  Signatures of Inflationary Resonances, Excitations and Scale Dependence}'',
  \href{http://arxiv.org/abs/2509.02695}{{\ttfamily arXiv:2509.02695
  [astro-ph.CO]}}.

\bibitem{Zang:2025azh}
S.-H. Zang and M.~M\"unchmeyer, ``{Squeezed Limit Non-Gaussianity Estimation
  with Cosmic Shear}'', \href{http://arxiv.org/abs/2512.07295}{{\ttfamily
  arXiv:2512.07295 [astro-ph.CO]}}.

\bibitem{Cheung:2007st}
C.~Cheung, P.~Creminelli, A.~Fitzpatrick, J.~Kaplan, and L.~Senatore, ``{The
  Effective Field Theory of Inflation}'',
  \href{http://dx.doi.org/10.1088/1126-6708/2008/03/014}{{\em JHEP} {\bfseries
  03} (2008) 014}, \href{http://arxiv.org/abs/0709.0293}{{\ttfamily
  arXiv:0709.0293 [hep-th]}}.

\bibitem{Arkani-Hamed:2018kmz}
N.~Arkani-Hamed, D.~Baumann, H.~Lee, and G.~Pimentel, ``{The Cosmological
  Bootstrap: Inflationary Correlators from Symmetries and Singularities}'',
  \href{http://dx.doi.org/10.1007/JHEP04(2020)105}{{\em JHEP} {\bfseries 04}
  (2020) 105}, \href{http://arxiv.org/abs/1811.00024}{{\ttfamily
  arXiv:1811.00024 [hep-th]}}.

\bibitem{Lewis:1999bs}
A.~Lewis, A.~Challinor, and A.~Lasenby, ``{Efficient Computation of
  CMB~Anisotropies in Closed FRW~Models}'',
  \href{http://dx.doi.org/10.1086/309179}{{\em Astrophys. J.} {\bfseries 538}
  (2000) 473}, \href{http://arxiv.org/abs/astro-ph/9911177}{{\ttfamily
  arXiv:astro-ph/9911177}}.

\bibitem{McEwen:2016fjn}
J.~McEwen, X.~Fang, C.~Hirata, and J.~Blazek, ``{FAST-PT: A Novel Algorithm to
  Calculate Convolution Integrals in Cosmological Perturbation Theory}'',
  \href{http://dx.doi.org/10.1088/1475-7516/2016/09/015}{{\em JCAP} {\bfseries
  09} (2016) 015}, \href{http://arxiv.org/abs/1603.04826}{{\ttfamily
  arXiv:1603.04826 [astro-ph.CO]}}.

\bibitem{Perez:2007ipy}
F.~P\'{e}rez and B.~Granger, ``{IPython: A System for Interactive Scientific
  Computing}'', \href{http://dx.doi.org/10.1109/MCSE.2007.53}{{\em Comput. Sci.
  Eng.} {\bfseries 9} (2007) 21}.

\bibitem{Hunter:2007mat}
J.~Hunter, ``{Matplotlib: A 2D Graphics Environment}'',
  \href{http://dx.doi.org/10.1109/MCSE.2007.55}{{\em Comput. Sci. Eng.}
  {\bfseries 9} (2007) 90}.

\bibitem{Harris:2020xlr}
C.~Harris {\em et~al.}, ``{Array Programming with NumPy}'',
  \href{http://dx.doi.org/10.1038/s41586-020-2649-2}{{\em Nature} {\bfseries
  585} (2020) 3572}, \href{http://arxiv.org/abs/2006.10256}{{\ttfamily
  arXiv:2006.10256 [cs.MS]}}.

\bibitem{Virtanen:2019joe}
P.~Virtanen {\em et~al.}, ``{SciPy 1.0 -- Fundamental Algorithms for Scientific
  Computing in Python}'',
  \href{http://dx.doi.org/10.1038/s41592-019-0686-2}{{\em Nat. Methods}
  {\bfseries 17} (2020) 261}, \href{http://arxiv.org/abs/1907.10121}{{\ttfamily
  arXiv:1907.10121 [cs.MS]}}.

\end{thebibliography}\endgroup

\end{document}